\documentclass[useAMS,usenatbib]{mn2e}

\usepackage{natbib}

\usepackage[utf8x]{inputenc}
\usepackage{latexsym,graphicx}%,natbib}
\usepackage{color}
\usepackage{fixltx2e}
\usepackage{verbatim} \usepackage{float}
\usepackage{amsmath,amssymb}
\usepackage{times}

\usepackage{setspace}
\usepackage{rotating}
\usepackage{pdflscape}
\usepackage{subfig}

\usepackage[squaren, Gray, cdot]{SIunits}

\newcommand\hii{H\,{\sc ii} \,}

\def\apgt{\ {\raise-.5ex\hbox{$\buildrel>\over\sim$}}\ }
\def\aplt{\ {\raise-.5ex\hbox{$\buildrel<\over\sim$}}\ }

\let\oldhat\hat
\renewcommand{\hat}[1]{\oldhat{\mathbf{#1}}}

\title[\textcolor{black}{On the ALMA observability} of nascent massive multiple systems]{\textcolor{black}{On the ALMA observability} of nascent massive multiple systems formed by gravitational instability}

\author[D. M.-A.~Meyer et al.]
       {D. M.-A.~Meyer$^{1,2}$\thanks{E-mail: dmameyer.astro@gmail.com}, A.~Kreplin$^{2}$, S.~Kraus$^{2}$, E.~I.~Vorobyov$^{3,4}$, L.~Haemmerle$^{5}$ 
       \newauthor and J.~Eisl\" offel$^{6}$ \\
       $^{1}$Astrophysics Group, School of Physics and Astronomy, University of Exeter, Exeter EX4 4QL, United Kingdom \\
       $^{2}$Institute of Physics and Astronomy, University of Potsdam, D-14476, Potsdam, Germany \\
       $^{3}$Research Institute of Physics, Southern Federal University, Stachki 194, Rostov-on-Don, 344090, Russia \\ 
       $^{4}$Department of Astrophysics, The University of Vienna, Vienna, A-1180, Austria \\  
       $^{5}$Observatoire de Gen\` eve, Universit\'e de Gen\`eve, chemin des Maillettes 51, CH-1290 Sauverny, Switzerland \\ 
       $^{6}$Th\" uringer Landessternwarte Tautenburg, Sternwarte 5, D-07778 Tautenburg, Germany \\       
       }

\begin{document}

% Date
\date{Received; accepted}

\maketitle

\label{firstpage}

\begin{abstract} 
Massive young stellar object (MYSOs) form during the collapse of high-mass pre-stellar cores, where infalling molecular 
material is accreted through a centrifugally-balanced accretion disc that is subject to efficient gravitational instabilities. 
In the resulting fragmented accretion disc of the MYSO, gaseous clumps and low-mass stellar companions can form, 
which will influence the future evolution of massive protostars in the Hertzsprung-Russell diagram. 
We perform dust continuum radiative transfer calculations and compute synthetic images of disc structures modelled by the  
gravito-radiation-hydrodynamics simulation of a forming MYSO, in order to investigate the {\it Atacama Large Millimeter/submillimeter Array} 
({\sc alma}) observability of circumstellar gaseous clumps and forming multiple systems. 
Both spiral arms and gaseous clumps located at $\simeq$ a few $100\, \rm au$ from the protostar can be resolved by 
\textcolor{black}{ interferometric {\sc alma} Cycle 7 C43-8 and C43-10 observations at band 6 ($1.2\, \rm mm$),  
using a maximal $0.015^{\prime\prime}$ beam angular resolution and at least $10$-$30\, \rm min$ exposure 
time for sources at distances of $1$-$2\, \rm kpc$. 
Our study shows that substructures are observable regardless of their viewing geometry or can be inferred 
in the case of an edge-viewed disc. 
}
The observation probability of the clumps increases with the gradually increasing efficiency of gravitational instability 
at work as the disc evolves. As a consequence, large discs around MYSOs close to the zero-age-main-sequence 
line exhibit more substructures than at the end of the gravitational collapse. 
Our results motivate further observational campaigns devoted to the close surroundings of the massive protostars 
S255IR-NIRS3 and NGC 6334I-MM1, whose recent outbursts are a probable signature of disc fragmentation and accretion variability.  
\end{abstract}


%%%%%%%%%%%%%%%%%%%%%%%%%%%%%%%%%%%%%%%%%%%%%%%%%%%%%%%%%%%%%%%%%%%%%%%%%%%%%%%%%%%%%%%%%%%
%%%%%%%%%%%%%%%%%%%%%%%%%%%%%%%%%%%%%%%%%%%%%%%%%%%%%%%%%%%%%%%%%%%%%%%%%%%%%%%%%%%%%%%%%%%
%%%%%%%%%%%%%%%%%%%%%%%%%%%%%%%%%%%%%%%%%%%%%%%%%%%%%%%%%%%%%%%%%%%%%%%%%%%%%%%%%%%%%%%%%%%

\section{Introduction}
\label{sect:intro}

\begin{figure*}
        \centering
        \begin{minipage}[b]{ 0.9\textwidth}
                \includegraphics[width=1.0\textwidth]{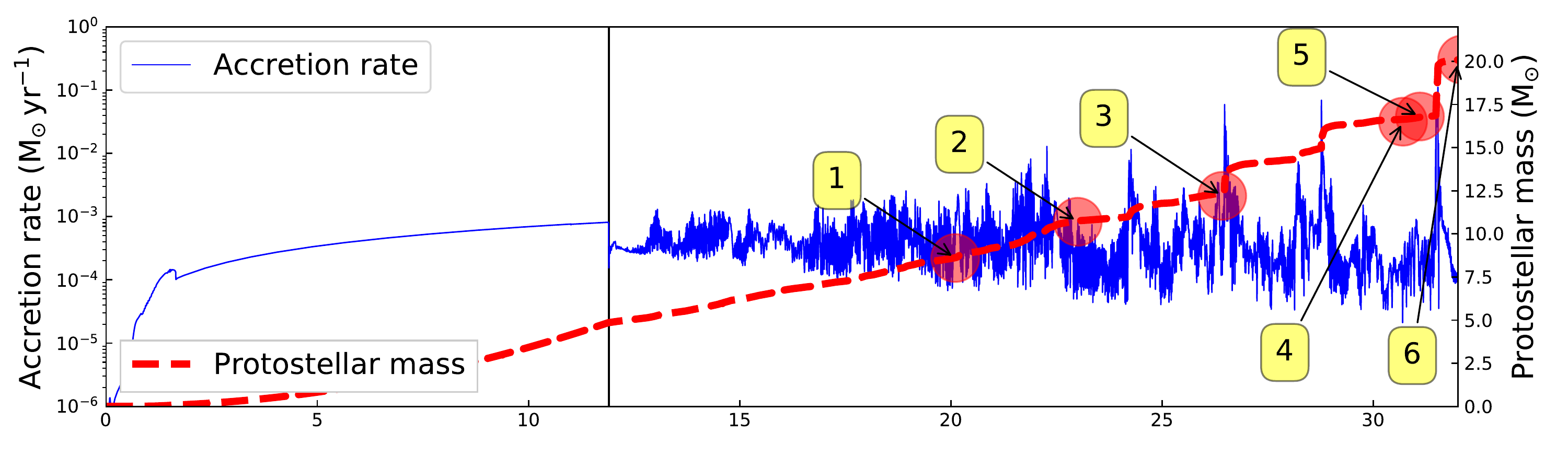}
        \end{minipage}                           
        \caption{ 
        		 Protostellar accretion rate (thin solid blue line, in $\rm M_{\odot}\, \rm yr^{-1}$) and 
        		 \textcolor{black}{protostellar} mass (thick dashed line, in $\rm M_{\odot}$) of model Run-1-hr in~\citet{meyer_mnras_473_2018}.  
				 The labelled red dot represents the 6 selected simulation snapshots for this study and 
				 the thin vertical line is the time instance of the disc formation.  
				 }      
        \label{fig:history}  
\end{figure*}

% Multiplicity in massive star formation ? 
Multiplicity is an indissociable characteristic of massive star evolution. 
Observational studies of OB stars indicate that multiplicity is an intrinsic feature, 
which strongly influences the evolution of massive stars in the Hertzsprung-Russell 
diagram~\citep{sana_sci_337_2012}. 
A wide dispersion of the semi-major axis in hierarchical massive stellar systems, 
which span a range from a small fraction of $\rm au$ for close/spectroscopic 
companions~\citep{2013A&A...550A..27M,2014ApJS..213...34K,chini_424_mnras_2012} to up to a few $100\, \rm au$ in the context of 
massive proto-binaries~\citep{kraus_apj_835_2017}, asks the puzzling question of their formation 
channel. This must be prior to the launching of the strong stellar winds at the zero-age-main-sequence (ZAMS), 
which dissipate the disc and therefore limit the companion formation timescale~\citep{vink_aa_362_2000}. 
On the one hand, investigations of the gravitational capture scenario of stellar systems with N-body simulations 
failed in explaining this high occurence of the small separations observed in spectroscopic companions to 
massive stars~\citep{fujii_sci_334_2011}. 
On the other hand, the disc fragmentation approach by means of multi-dimensional hydrodynamics simulations 
coupled to Lagrangian sink particles that spot secondary star formation in self-gravitating accretion discs suffers 
from the dramatic qualitative algorithm-dependence of the results~\citep{klassen_apj_823_2016}. 
Therefore, understanding the early formation mechanism of hierarchical multiple systems around massive 
stars by confronting numerical models to observations is of prime interests in the study of the 
pre-ZAMS~\citep{zinnecker_araa_45_2007} but also the pre-supernovae evolution of 
massive stars~\citep{langer_araa_50_2012}.

% Self-consistent simulations
State-of-the-art numerical models of the formation of MYSOs predicted the existence of an accretion disc  
connected with bipolar cavities filled by ionized stellar wind and outflow 
material~\citep{seifried_mnras_417_2011,harries_mnras_448_2015,harries_2017}.  
%
%%%However, neither the formation of spiral structures perturbing the accretion flow 
%%%onto massive protostars, nor the possible formation in such discs of a second generation of 
%%%stars had so far been simulated without the use of sink-particles. in accretion discs affected by efficient 
%%%gravitational instability. 
%
%%%This raises the question of possible differences in star-forming mechanisms between the well-studied 
%%%low-mass and the less-studied high-mass regimes of star formation. 
%
The recent gravito-radiation-hydrodynamical simulations of the formation of massive protostars 
in~\citet{meyer_mnras_464_2017} and~\citet{meyer_mnras_473_2018} showed that it is possible to 
obtain self-consistent solutions for multiplicity in the high-mass regime of star formation 
without using artificial sink particles, as long as the spatial resolution of the inner disc 
region is sufficiently high. 
Those high-resolution simulations of $\simeq 100\, \rm M_{\odot}$ collapsing pre-stellar cores 
also demonstrate that even irradiated, massive, self-gravitating accretion discs around massive young objects 
are subject to gravitational fragmentation and generate dense spiral arms in which gaseous clumps 
form. These clumps can contract further and reach the molecular hydrogen dissociation temperature while 
migrating inward onto the central forming massive object. 
The migrating clump can experience a second collapse and become low-mass stars, progenitors 
of the future companions and/or close (spectroscopic) companions to the massive stars.

% Our papers
The series of studies of~\citet{meyer_mnras_464_2017} and~\citet{meyer_mnras_473_2018} provides 
a self-consistent picture unifying disc fragmentation and binary formation in the 
context of massive young stellar objects. The model also predicts that a fraction of the migrating 
gaseous clump material can directly fall onto the protostellar surface and produce 
accretion-driven outbursts as monitored in the MYSOs S255IR-NIRS3 and NGC 6334I-MM1. 
In this picture, accretion bursts are a signature of the presence of a self-gravitating disc in which 
efficient gravitational instability occurs. The bursts happen in series of multiple eruptions, whose 
intensity scales with the quantity of accreted material~\citep{meyer_mnras_482_2019}. 
These episodic events generate modifications in the internal protostellar structure leading to excursions 
in the Hertzsprung-Russel diagram~\citep{2019MNRAS.tmp...10M}. 
Note that this burst mode of massive star formation has an scaled-up equivalent in the low-mass 
regime, as outlined in a series of papers of~\citet{vorobyov_apj_719_2010,vorobyov_apj_768_2013,
dong_apj_823_2016,elbakyan_mnras_484_2019}, \textcolor{black}{as well as in the primordial mass 
regime of star formation, see~\citet{greif_mnras_424_2012} and~\citet{vorobyov_apj_768_2013}.}
The remaining question is therefore, can such self-gravitating effects be directly observed and 
what can we learn from the disc density structure about the high-mass star formation process ?

% Other observations of massive protostars 
Searching for the direct observational signatures of clumpy structures in massive star-forming 
regions is a very active field, as modern facilities are now able to probe the highly-opaque pre-stellar 
cores in which MYSOs grow. 
The last decade saw the report of observations of variabilities in the accretion flow onto massive 
protostars~\citep{keto_apj_637_2006,stecklum_2017a} and in the pulsed bipolar outflows which release 
ionized gas~\citep{Cunningham_apj_692_2009,cesaroni_aa_509_2010,caratti_aa_573_2015,purser_mnras_460_2016,
reiter_mnras_470_2017,burns_mnras_467_2017,burns_iaus_336_2018,purser_mnras_475_2018,samal_mnras_477_2018}, 
together with proofs of disc-jet associations such as G023.01-00.41~\citep{2018arXiv180509842S}. 
Direct observation of Keplerian discs around MYSOs~\citep{johnston_apj_813_2015,ilee_mnras_462_2016,
forgan_mnras_463_2016,2018arXiv180410622G,maud_aa_620_2018} and evidence of filamentary spirals feeding 
the candidate disc W33A MM1-Main~\citep{maud_467_mnras_2017} and of infalling clumps in the systems 
G350.69-0.49~\citep{chen_apj_835_2017} and G11.92-0.61 MM1~\citep{2018ApJ...869L..24I} revealed the existence 
of substructures in discs that are similar to those modelled in~\citet{meyer_mnras_473_2018}. 
Furthermore, the investigations for molecular emission from collapsing clouds and from the close 
surroundings of high-mass protostars confirm the possibility of fragmentation and the presence of 
inhomogeneities in their discs~\citep{2018arXiv181110245B,ahmadi_aa_618_2018}, without excluding the 
possibility of magnetic self-regulation of the inner disc fragmentation~\citep{beuther_aa_614_2018} 
as predicted by analytic and numerical studies~\citep{hennebelle_apj_830_2016}.

\begin{table*}
	\centering
	\caption{
	Accretion disc models selected from~\citet{meyer_mnras_473_2018}. The table gives the simulation time (in $\rm kyr$), 
	protostellar mass (in $\rm M_{\odot}$), the total protostellar luminosity, the protostellar radius and a brief description of 
	the disc structure for each of the considered disc models.   
	}
	\begin{tabular}{lcccccr}
	\hline
	${\rm {Model}}$  &  $t$ ($\rm kyr$)  &  $M_{\star}$ ($\rm M_{\odot}$)  &  $L_{\star}$ ($\rm L_{\odot}$)  &  $R_{\star}$ ($\rm R_{\odot}$)  &  $T_{\rm eff}$ ($\rm K$)  &  description  \\ 
	\hline    
	1       & 20.1                        &  $~8.6$             &      $~~~~~~283.0$     &  $~~~22.3$   &   $~5020.0$     &      young, stable disc   \\  
	2       & 23.0                        &  $10.7$             &      $~~~~~~650.0$     &  $~~~30.6$   &   $~5275.0$     &      stable disc with first gravitational perturbation    \\  
	3       & 26.4                        &  $12.1$             &      $~~~~8359.0$      &  $~~~43.6$   &   $~8374.0$     &      disc with spiral arms and migrating gaseous clump    \\  	
	4       & 30.7                        &  $16.6$             &      $~~27973.0$       &  $~~~32.0$   &   $13210.0$     &      disc with multiple ring-like spiral arms hosting clumps    \\
	5       & 31.1                        &  $16.7$             &      $~~28625.0$       &  $~~~30.6$   &   $13586.0$     &      strongly fragmenting disc with multiple clumps       \\  	
	6       & 32.1                        &  $20.0$             &      $226454.0$        &  $~389.0$    &   $~~6393.0$    &      extended strongly fragmented disc with filamentary spiral arms and clumps               \\	
	\hline    
%	\hline 
	\end{tabular}
\label{tab:models}\\
\end{table*}

\begin{figure}
        \centering
        \begin{minipage}[b]{ 0.45\textwidth}  \centering
                \includegraphics[width=1.0\textwidth]{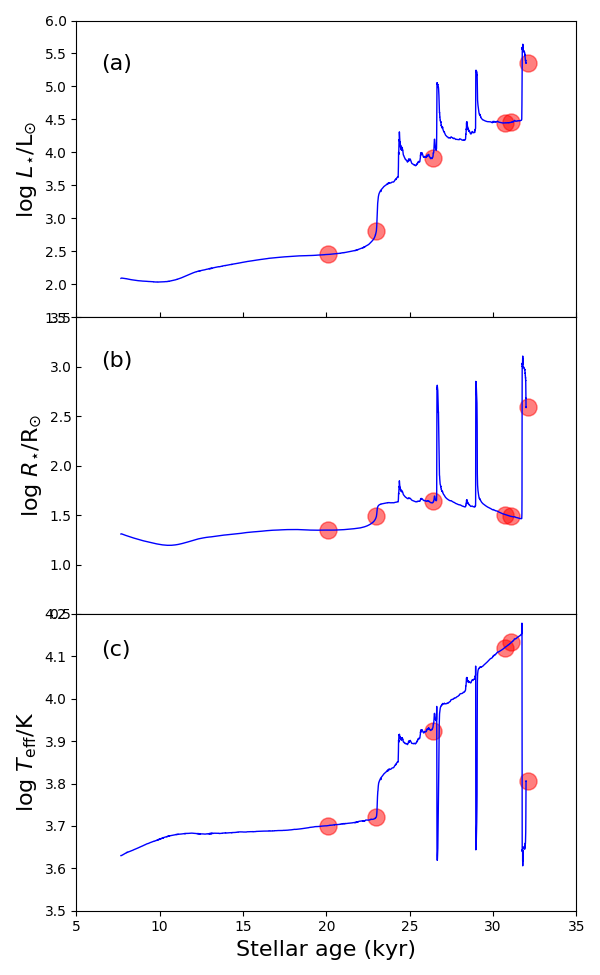}
        \end{minipage}     
        \caption{ 
        		 Evolution as a function of time (in kyr) of the stellar surface lu-
                 minosity (a), stellar radius (b) and effective temperature (c) of our MYSOs
                 experiencing variable disc accretion, episodically interspersed by strong accretion events 
                 responsible for luminous bursts. 
                 %The vertical grey dashed lines are isochrones at the times of the selected disc models, 
                 The protostellar properties of the different disc models are marked with red dots. 
                 }      
        \label{fig:star}  
\end{figure}

\begin{figure*}
        \centering
        \begin{minipage}[b]{ 0.8\textwidth}  \centering
                \includegraphics[width=1.0\textwidth]{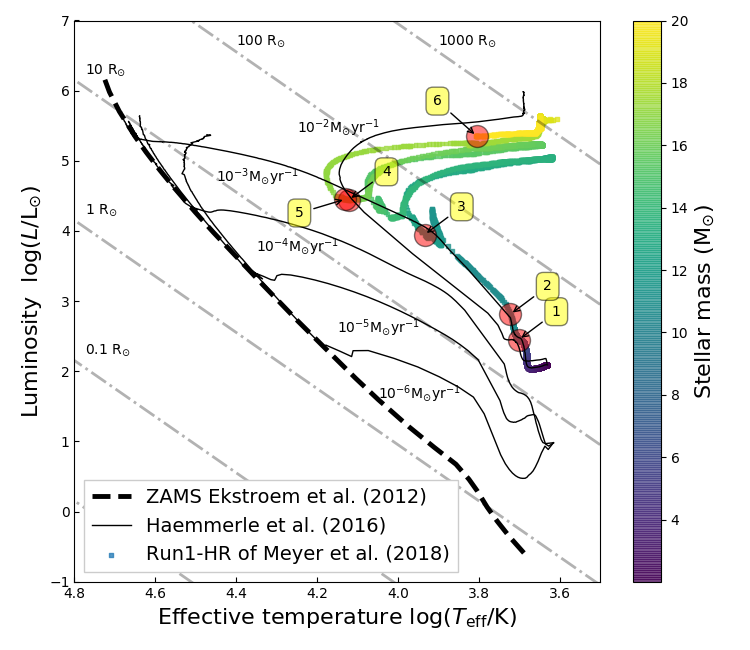}
        \end{minipage}     
        \caption{ 
        		 Comparison between the massive protostar in Run-1-hr (thick coloured line) 
        		 of~\citet{meyer_mnras_473_2018} and the evolutionary tracks for massive protostars accreting 
        		 with constant rates (thin black lines) of~\citet{haemmerle_585_aa_2016}. 
        		 The color coding of our simulated track indicates the stellar mass (in $\rm M_{\odot}$).  
        		 Grey dashed-dotted lines are isoradius and the thick black dashed line is the 
        		 zero-age-main-sequence (ZAMS) track of~\citet{ekstroem_aa_537_2012}. 
        		 The numbered red dots are the selected simulation snapshots from~\citet{meyer_mnras_473_2018} 
        		 that we consider in this study. 
                 }      
        \label{fig:hdr}  
\end{figure*}

% What we do here ?
This study aims at testing the observability of the overdensities and various substructures such 
as spiral arms and gaseous clumps that are predicted by~\citet{meyer_mnras_464_2017} 
and~\citet{meyer_mnras_473_2018}. We adopt the standard approach of radiative transfer 
calculations against dust opacities, performed on the basis of the accretion disc density and temperature 
fields from numerical simulations of~\citet{meyer_mnras_473_2018} and further post-processed in order 
to obtain synthetic interferometric {\it Atacama Large Millimeter/submillimeter Array} ({\sc alma}) 
band 6 ($1.2\, \rm mm$) images. This corresponds to the typical waveband at which the surroundings of 
MYSOs are observed, as it offers a good compromise between high spatial resolution and phase calibration 
problems occurring at longer interferometric baselines. 
This approach has been adopted to predict the observability of gravitational instabilities in discs 
around low-mass stars~\citep{vorobyov_mnras_433_2013,dong_apj_823_2016,seifried_mnras_571_2016}.  
Only very few similar studies have been conducted in the context of high-mass stars, either 
using artificial sink particles as tracers of multiplicity in the disc~\citep{krumholz_apj_665_2007} 
or on the basis of simplistic, analytic disc models~\citep{jankovic_mnras_482_2019}.  
Our synthetic images predict what disc morphologies should be observable with {\sc alma} 
and constitute prognostications that are directly comparable with future high angular 
resolution observations campaigns.

% Plan of the paper...
Our work is organized as follows. In Section~\ref{sect:methods}, we review the methods 
used to perform numerical hydrodynamical simulations, radiation transfer calculations 
and to compute synthetic images of the fragmented circumstellar medium of a massive protostar. 
%
%%%The disc around our young high-mass star develops substructures such as spiral arms and 
%%%gaseous clumps, as the disc grows in size and gains mass, where the $1.2\, \rm mm$ surface brightness 
%%%is of prime interest for observing disc fragmentation around massive protostellar 
%%%objects in the future. 
%
In Section~\ref{sect:results}, we present our results as a series of synthetic images of 
accretion discs one should expect to observe with {\sc alma} in the surroundings of forming 
MYSOs. We consider several viewing angles of the objects with respect to the plane of the sky. 
Our synthetic images are further discussed in Section~\ref{sect:discussion}. Finally, we provide 
our conclusions in Section~\ref{sect:cc}.

%%%%%%%%%%%%%%%%%%%%%%%%%%%%%%%%%%%%%%%%%%%%%%%%%%%%%%%%%%%%%%%%%%%%%%%%%%%%%%%%%%%%%%%%%%%
%%%%%%%%%%%%%%%%%%%%%%%%%%%%%%%%%%%%%%%%%%%%%%%%%%%%%%%%%%%%%%%%%%%%%%%%%%%%%%%%%%%%%%%%%%%
%%%%%%%%%%%%%%%%%%%%%%%%%%%%%%%%%%%%%%%%%%%%%%%%%%%%%%%%%%%%%%%%%%%%%%%%%%%%%%%%%%%%%%%%%%%

\section{Method}
\label{sect:methods}

In the following paragraphs, we remind the reader about the numerical methods used to 
carry out our hydrodynamics disc models. The structure and accretion properties of the 
discs are extracted from the simulation of the collapse of a solid-body-rotating 
pre-stellar core that produces a central, single massive young stellar object~\citep{meyer_mnras_473_2018}. 
The accretion history onto the MYSO is used to feed a stellar evolution code, in order to 
self-consistently derive its protostellar surface properties. 
Then, we present the radiation transfer methods utilised to produce  
dust continuum millimeter images of the discs, by post-processing the density and 
temperature fields of the hydrodynamical circumstellar disc model. 
Finally, we review our method for retrieving synthetic {\sc alma} images, allowing us to transform the radiation 
transfer calculations into realistic {\sc alma} synthetic observations and to discuss the observability of 
the clumps in the disc as a function of, e.g. the protostellar formation phases.

\subsection{Hydrodynamical simulations} 
\label{sect:hydro}

This study focuses on the high-resolution gravito-radiation-hydrodynamics simulation Run-1-hr 
of~\citet{meyer_mnras_473_2018}. It is a numerical model of the gravitational collapse of a 
$100\, \rm M_{\odot}$ solid-body-rotating cold pre-stellar core of uniform initial temperature 
$T_{\rm c}=10\, \rm K$ and of density distribution $\rho(r)\propto r^{-3/2}$, with $r$ the radial 
coordinate. The initial ratio of kinetic-by-gravitational energy of the pre-stellar core was set to 
$\beta=4\, \%$. 
The spherical midplane-symmetric computational domain onto which the run has been performed 
has an inner radius $r_{\rm in}=10\, \rm au$ and an outer radius $R_{\rm c}=0.1\, \rm pc$, respectively. 
A grid maps the domain $[r_{\rm in},R_{\rm c}]\times[0,\pi/2]\times[0,2\pi]$ with 
$N_{\rm r}=256\times\,N_{\rm \phi}=41\times\,N_{\rm \theta}=256$ grid cells, expanding logarithmically 
along the radial direction $r$, going as a cosine in the polar direction $\phi$ and uniformly spaced 
along the azimuthal direction $\theta$. 
The inner hole is a semi-permeable sink cell attached onto the origin of the domain whereas outflow 
boundary conditions are assigned to the outer radius. Therefore, the material $\dot{M}$ that is lost 
through the sink cell allows us to calculate the accretion rate onto the protostar, while the protostellar 
properties such as its stellar radius and its photospheric luminosity are time-dependently estimated 
using the pre-calculated protostellar evolutionary tracks of~\citet{hosokawa_apj_691_2009}. 
Because of time-step restrictions due to the high-resolution of the grid, we computed the collapse of 
the core together with the initial formation phase of the circumstellar disc of the protostar up 
to $t_{\rm end}=\, 32.1\rm kyr$. 
%
%This approach has the advantage to reach high-spatial resolution in the inner region of the disc midplane,  
%while reducing the total number of total grid zones compared to a simulation with a uniform grid, which  
%allows to save computing resources. However, the smallest grid zones close to the inner boundary dramatically 
%affect the simulation time-step, preventing us to reach longer integration timescales and restricting our 
%simulated time to $t_{\rm end}=\, 32.1\rm kyr$. 

We integrate the system by solving the equations of gravito-radiation-hydrodynamics with the 
{\sc pluto} code\footnote{http://plutocode.ph.unito.it/}~\citep{mignone_apj_170_2007,migmone_apjs_198_2012}, 
that has been modified to account for (i) the protostellar feedback and (ii) the self-gravity of the gas. 
The direct proto-stellar irradiation feedback of the central star as well as the inner disc radiative transport 
are taken into account using the gray approximation which has been adapted from the publicly-available scheme 
of~\citet{kolb_aa_559_2013}, see method section of~\citet{meyer_mnras_473_2018}.  
This two-fold algorithm intelligently ray-traces photon packages from the protostellar atmosphere and then 
diffuses their energy by flux-limited propagation into the accretion disc. However, this scheme allows us 
to accurately treat both the inner heating and the outer cooling of the irradiated disc surrounding our 
MYSO~\citep{vaidya_apj_742_2011}. 
We use \textcolor{black}{a constant gas opacity} of $0.01\, \rm  cm^{2}\, g^{−1}$ and the tabulated dust opacities of~\citet{laor_apj_402_1993} in order 
to account for the attenuation of the radiation field. Gas and dust temperatures are calculated assuming the equilibrium 
between the temperature of the silicate dust grains and the total protostellar irradiation and disc radiation fields. 
Any turbulent viscosity is neglected in the computational domain and we assume that the most efficient mechanism for 
angular momentum transport are the gravitational torques generated in the self-gravitating accretion disc that 
surrounds the protostar. 
The gravity of the central star is taken into account by calculating its total gravitational potential and our 
simulations include the self-gravity of the circumstellar gas by directly solving the Poisson equation. 
Finally, we refer the reader who is interested in further details about the numerical method to~\citet{meyer_mnras_473_2018}.

\begin{figure*}
        \centering
        \begin{minipage}[b]{ 0.32\textwidth}
                \includegraphics[width=1.0\textwidth]{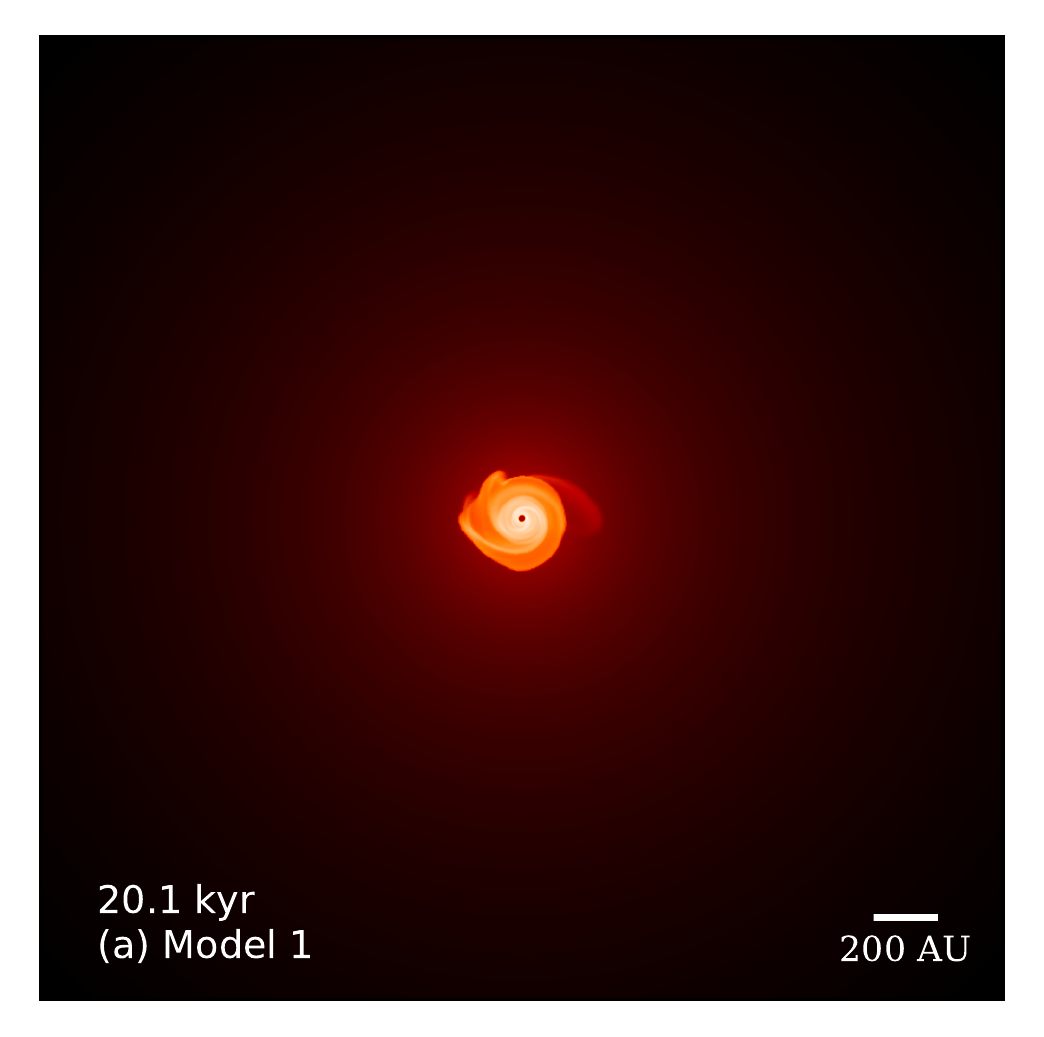}
        \end{minipage}    
        \centering
        \begin{minipage}[b]{ 0.32\textwidth}
                \includegraphics[width=1.0\textwidth]{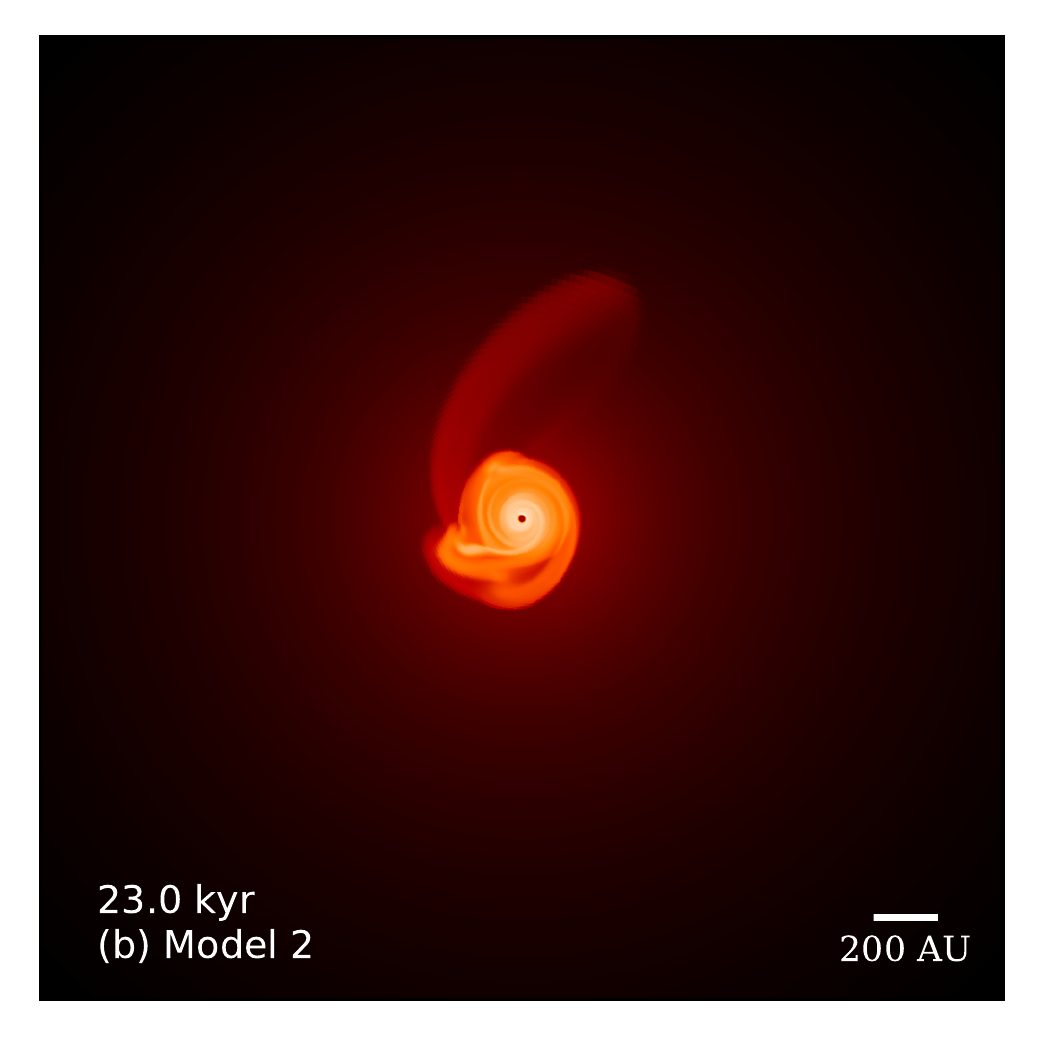}
        \end{minipage}  
        \centering                    
        \begin{minipage}[b]{ 0.32\textwidth}
                \includegraphics[width=1.0\textwidth]{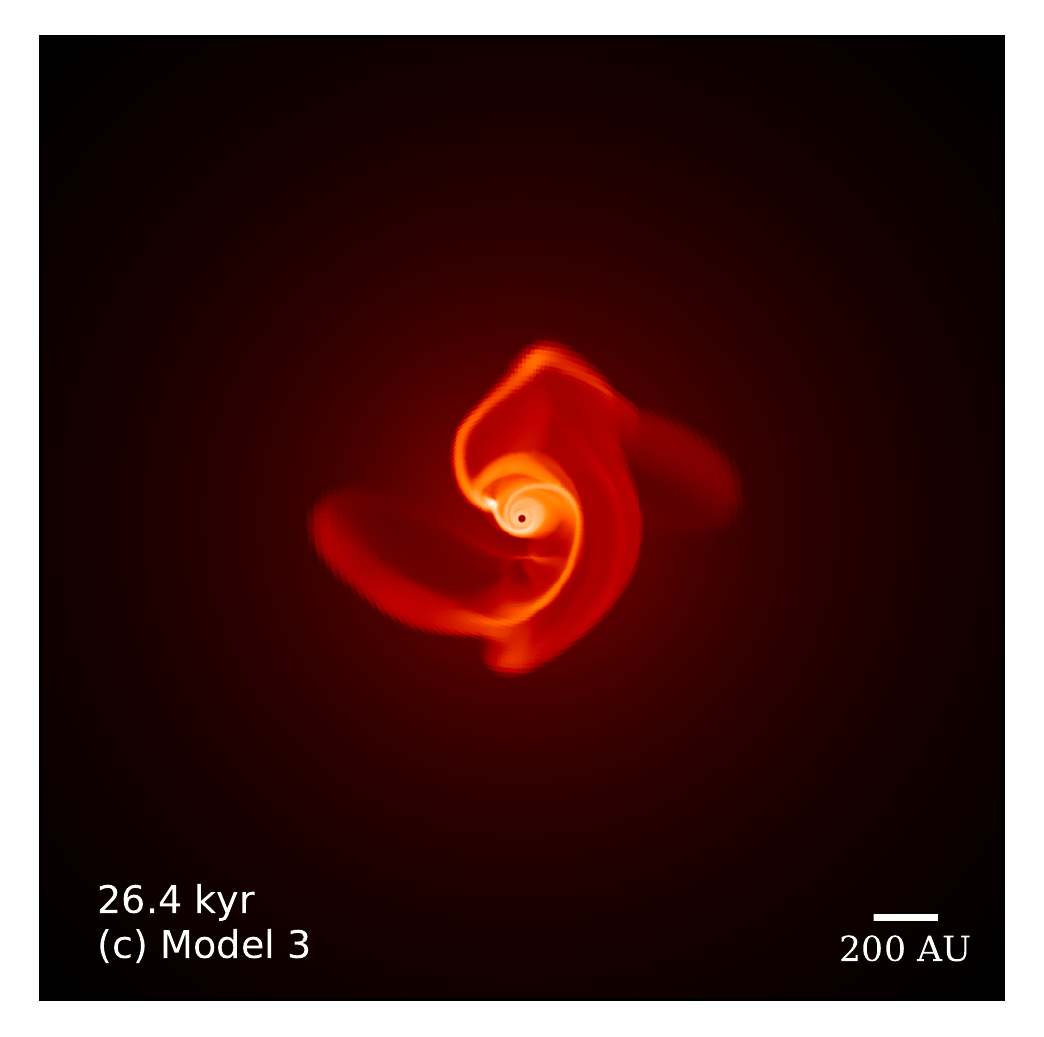}
        \end{minipage}   
        \centering	\\
        \begin{minipage}[b]{ 0.32\textwidth}
                \includegraphics[width=1.0\textwidth]{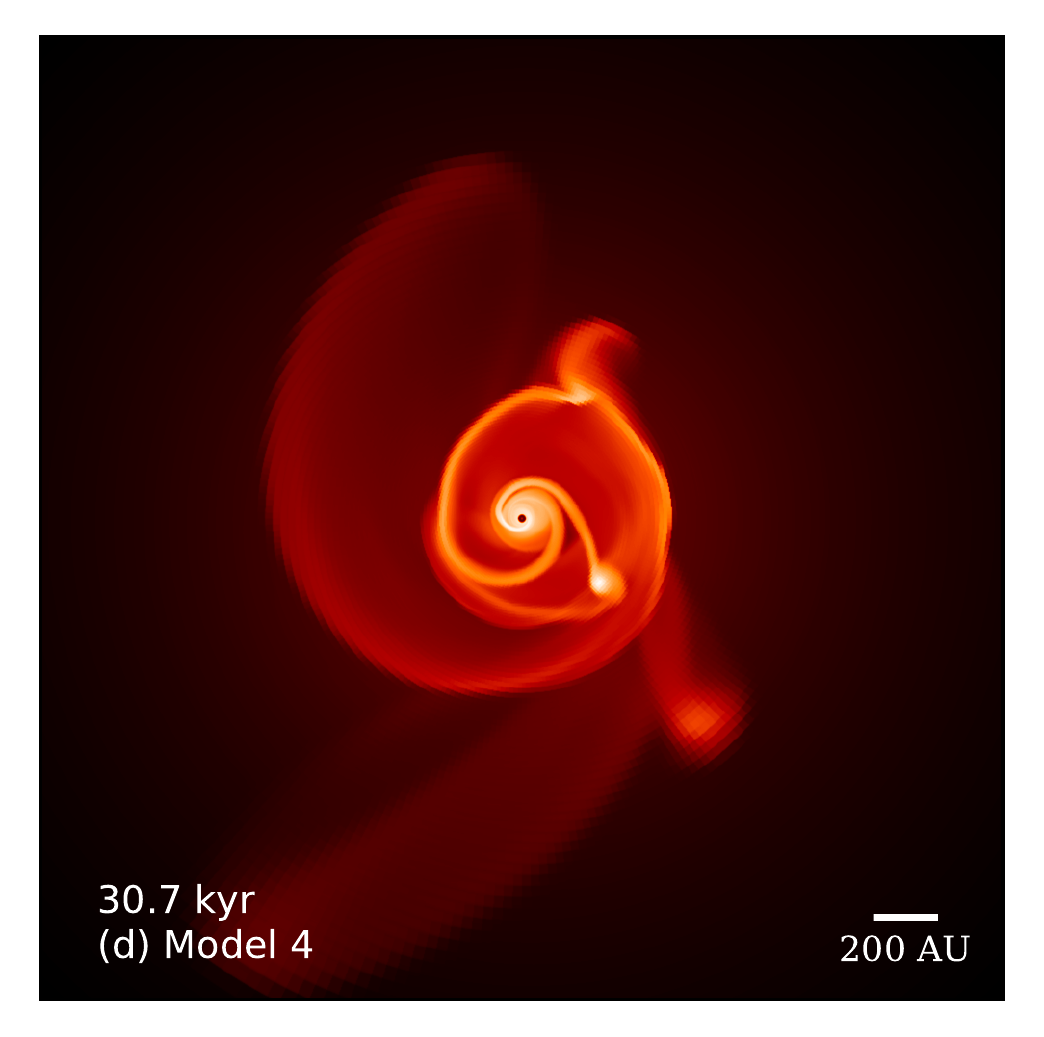}
        \end{minipage}   
        \centering      
        \begin{minipage}[b]{ 0.32\textwidth}
                \includegraphics[width=1.0\textwidth]{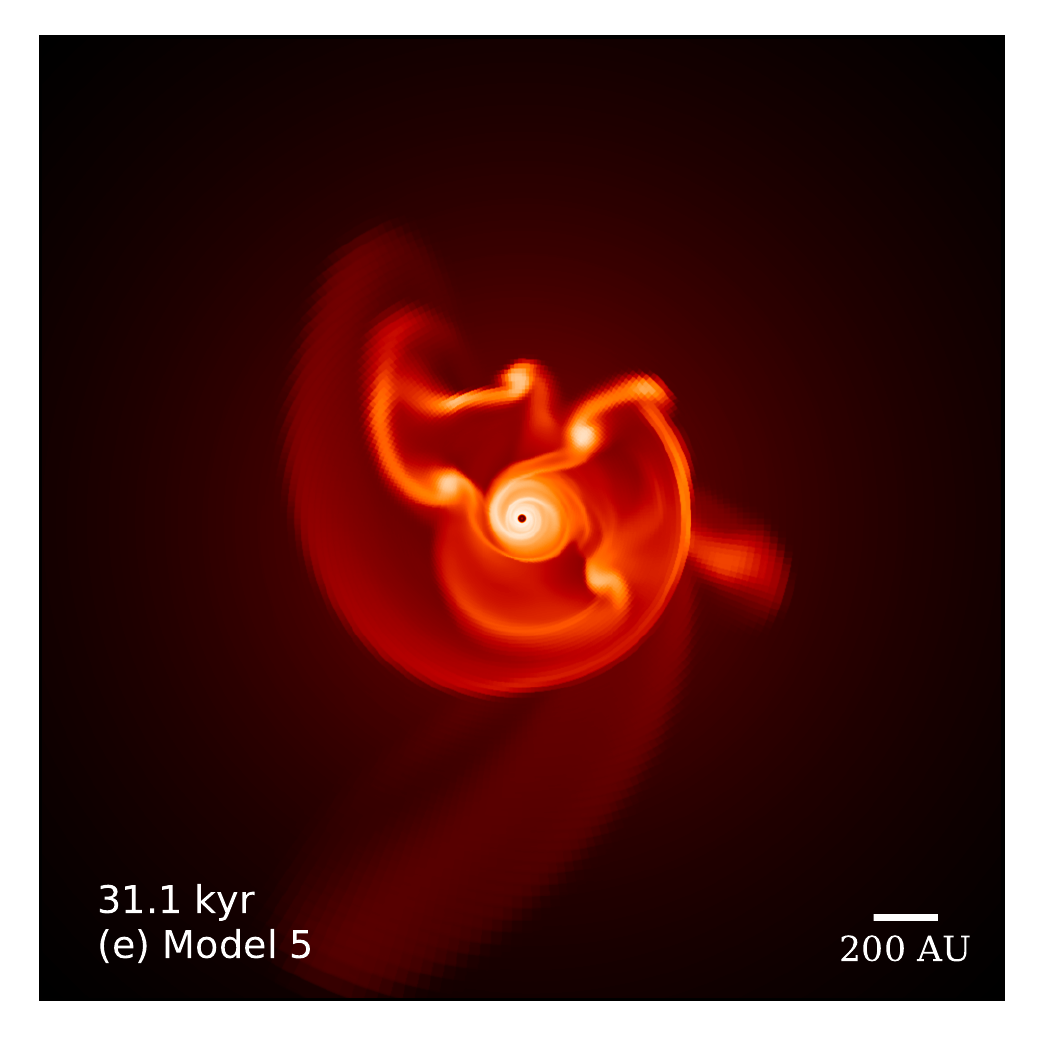}
        \end{minipage}    
        \centering
        \begin{minipage}[b]{ 0.32\textwidth}
                \includegraphics[width=1.0\textwidth]{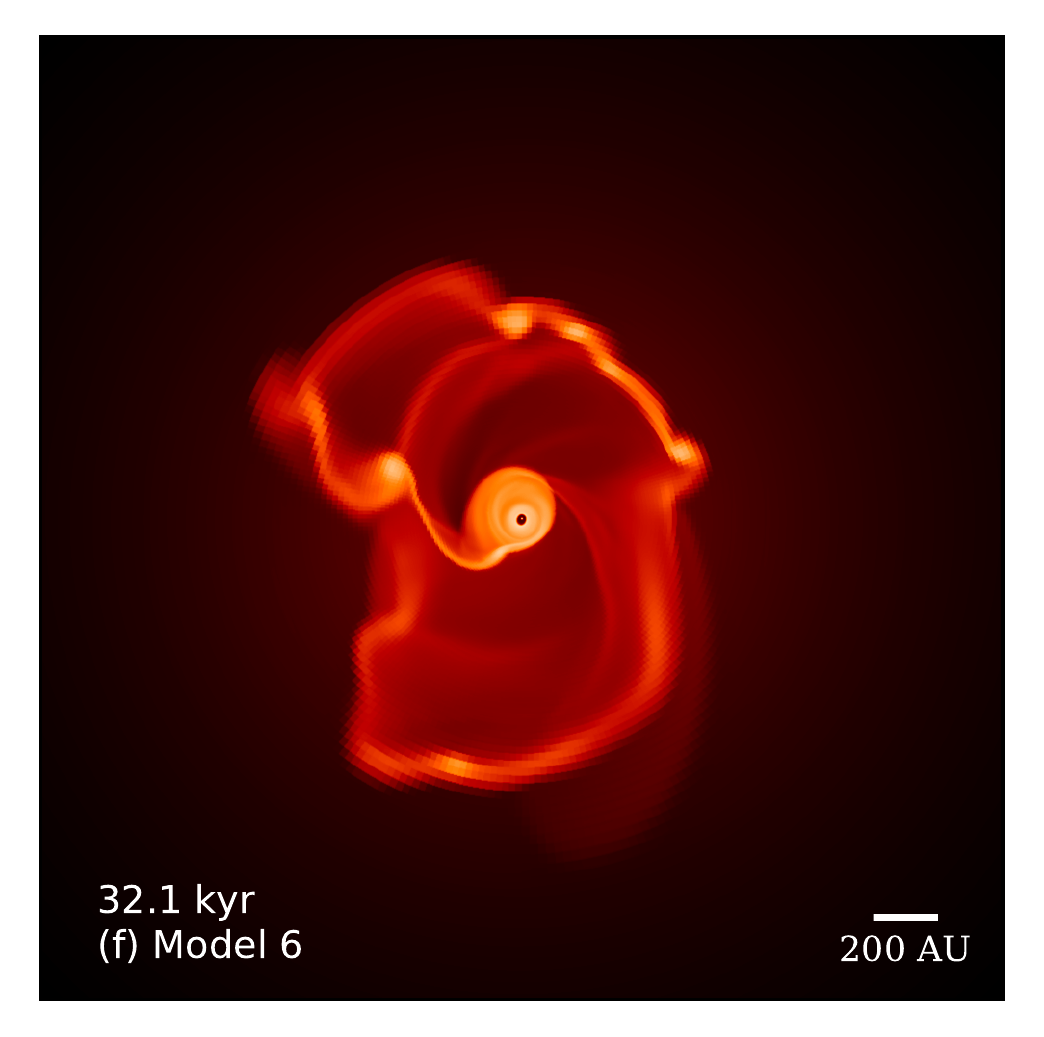}
        \end{minipage}                   
        \caption{ \textcolor{black}{
				 Rendering of the dust-rich, infrared-emitting regions in the structure of our accretion discs.  
				 The figures show the growing disc for different selected time instances throughout the hydrodynamical simulation and no inclination angle is assumed.   
				 Each image is plotted on the logarithmic scale, the most diluted part of the discs are in black color, and the densest disc regions are in white color, respectively. 
				 }
        		 }      
        \label{fig:theory_plot_2}  
\end{figure*}

\begin{figure*}    
        \centering
        \includegraphics[width=1.0\textwidth]{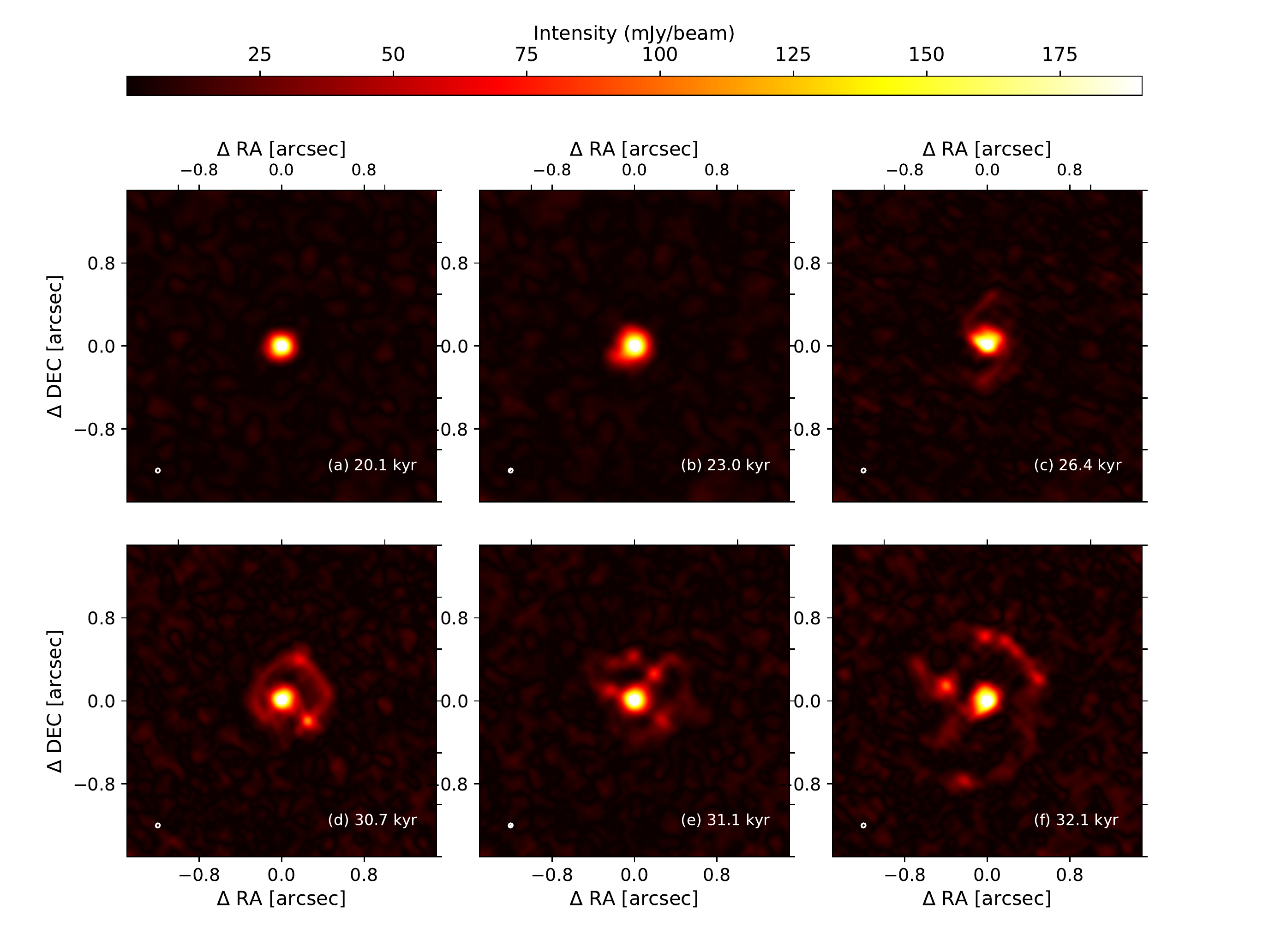}  
        \caption{ 
                 \textcolor{black}{
                 {\sc alma} C43-8 images simulated images at band 6 ($1.2\, \rm mm$) of the 
                 accretion disc growing around our protostar, assuming an antenna configuration 8 and $10\, \rm min$ exposure time.  
                 }
        		 The images are shown for the face-on viewing angle. 
                 The beam size is $0.029^{\prime\prime}$ and it is given in the lower left corner of each panel.
                 The source distance is assumed to be $1\, \rm kpc$. 
                 %The $XXX\, \sigma$ detection limit in simulated {\sc alma} images is 0.034 mJy beam is marked by the green tick on the color bar. 
                }     
        \label{fig:allmodels_68}  
\end{figure*}

\subsection{ Stellar evolution calculations } 
\label{sect:starevol}

The modelled accretion rate history is used as initial conditions to compute the evolutionary tracks in 
the Hertzsprung-Russel diagram of the corresponding episodically-accreting massive protostar, following the 
method of~\citet{2019MNRAS.tmp...10M}. 
To this end, a one-dimensional stellar evolution calculation is carried out with the hydrostatic {\sc genec} 
(i.e. {\sc geneva} code) code~\citep{eggenberger_apss_316_2008}, which was updated with respect to disc accretion physics for 
the study of pre-main-sequence high-mass stars~\citep{haemmerle_phd_2014,haemmerle_585_aa_2016}. 
The accretion mechanism is therein treated within the so-called cold disc accretion approximation~\citep{palla_apj_392_1992}. 
The inner disc region is assumed to be geometrically thin as the accreted material falls onto the surface of the protostar and 
the implementation of~\citet{haemmerle_phd_2014} showed full consistency with the original results of~\citet{hosokawa_apj_721_2010}. 
Any entropy excess is immediately radiated away and the circumstellar material is advected inside the protostar, i.e. the simulation 
considers that the thermal properties of the inner edge of the disc are similar to those of the suface layer of the MYSOs. 
This approach is the lower limit on the entropy attained by the protostar throughout the accretion mechanism, the upper limit 
being the so-called spherical (or hot) accretion scenario~\citep{hosokawa_apj_721_2010}. 
The calculation of the stellar structure is performed with the Henyey method, using the Lagrangian formulation~\citep{haemmerle_phd_2014}  
at solar metallicity (Z=0.014) and making use of the abundances of~\citet{asplund_ASPC_2005} and~\citet{cunha_apj_647_2006}, together with the 
deuterium mass fractions described in the studies of~\citet{norberg_aa_159_2000} and~\citet{behrend_aa_373_2001}.  
The stellar structure calculation is started with a fully convective stellar embryo~\citep{haemmerle_458_mnras_2016} and make use of overshooting and of the 
Schwarzschild criterion in the treatment of the internal stellar convection. 
The outcomes of the calculation are evolution of $L_{\odot}$, $R_{\star}$ and $T_{\rm eff}$ as a function of time.

\begin{figure*}    
        \centering
        \includegraphics[width=1.0\textwidth]{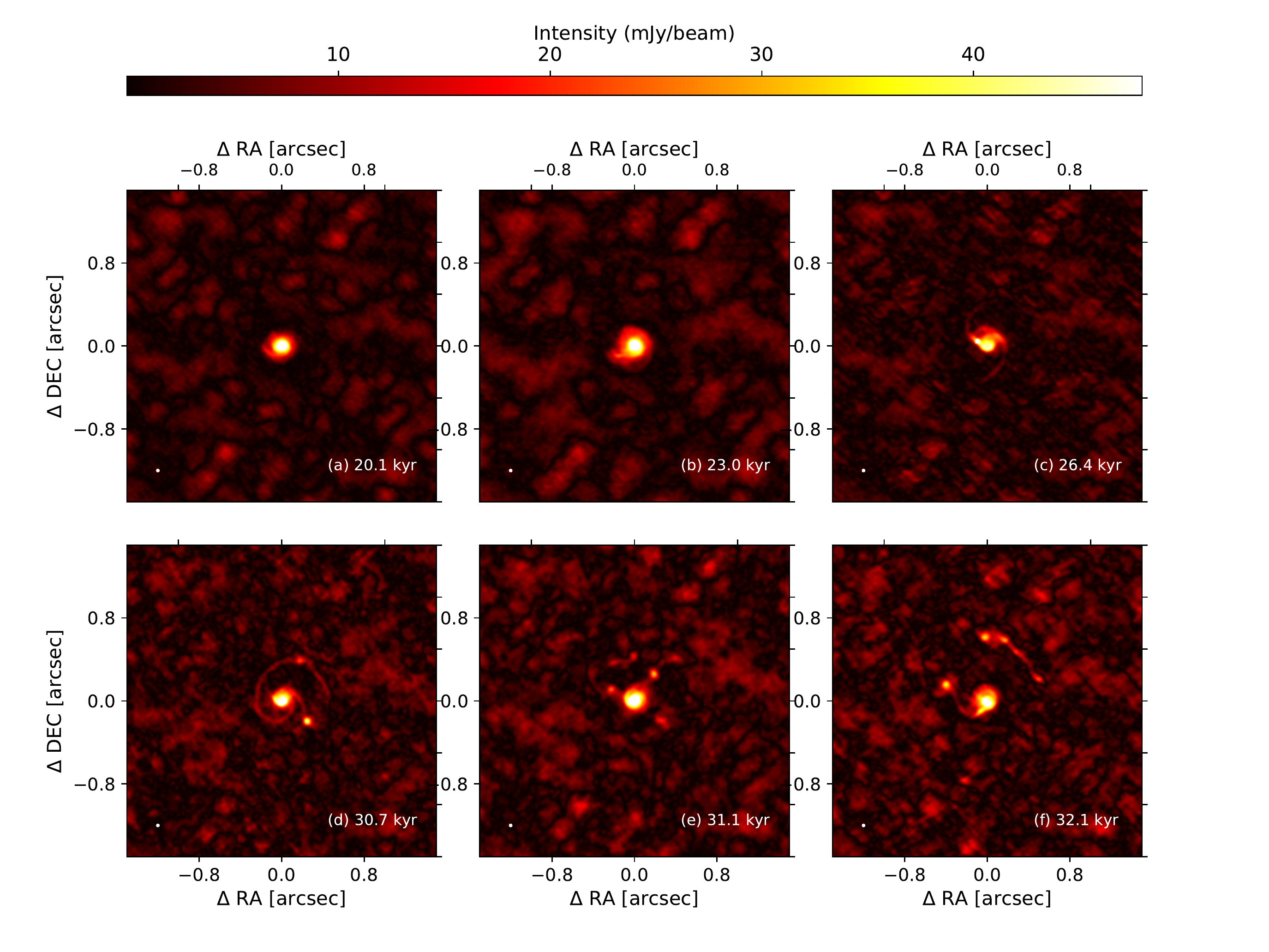}  
        \caption{ 
                 \textcolor{black}{
                 {\sc alma} C43-10 images simulated images at band 6 ($1.2\, \rm mm$) of the 
                 accretion disc growing around our protostar, assuming an antenna configuration 10 and $10\, \rm min$ exposure time.  
                 }
        		 The images are shown for the face-on viewing angle. 
                 The beam size is $0.015^{\prime\prime}$ and it is given in the lower left corner of each panel.
                 The source distance is assumed to be $1\, \rm kpc$.
                }      
        \label{fig:allmodels_610}  
\end{figure*}

\subsection{Radiative transfer calculations} 
\label{sect:transfer}

A series of characteristic simulation snapshots is selected from the simulation Run-1-hr of~\citet{meyer_mnras_473_2018} 
and they are post-processed by performing radiative transfer calculations against dust opacity with the code 
{\sc radmc-3d}\footnote{http://www.ita.uni-heidelberg.de/~dullemond/software/radmc-3d/}~\citep{dullemond_2012}. 
We make use of the~\citet{laor_apj_402_1993} dust mixture based of silicates 
crystals. Hence, our simulated emission is calculated with the same dust opacity that was used in the gravito-radiation-hydrodynamical 
simulations. In our approach, {\sc radmc-3d} directly uses dust density and temperature fields imported from the {\sc pluto} outputs and 
we do not perform any Monte-Carlo simulation prior to the image building, see comparison of both methods in Section~\ref{sect:caveats}. 
The interface between the {\sc pluto} and {\sc radmc-3d} codes is an interpolation between the above described non-uniform 
mesh of {\sc pluto} and the uniform spherically-symmetric grid of {\sc radmc-3d} which reconstructs the full three-dimensional 
structure of the disc taking into account the midplane-symmetry of the hydrodynamical models. 
The protostar is located at the origin of both grids and the {\sc radmc-3d} calculations is then effectuated 
concentrating onto the inner $1000$-$3000\, \rm au$ regions of the disc, such that the corresponding 
map has $2000 \times 2000$ cells along the horizontal and vertical directions of the sky plane, respectively.

The accretion discs are irradiated by packages of $5 \times 10^{6}$ photons ray-traced from the protostellar 
atmosphere of radius $r_{\rm in}$ to the outer disc region, at a radius of a few $1000\, \rm au$. The protostar is assumed to 
be a black body radiator of effective temperature $T_{\rm eff}$, bolometric luminosity $L_{\star}$ and stellar radius 
$R_{\star}$ with values taken from the outcomes of the stellar evolution calculation at the time of the 
selected snapshots. 
The radiative transfer calculations are performed within the anisotropic scattering approximation using the 
Henyey-Greenstein formula. For each of the selected simulation snapshots, we build $1.2\, \rm mm$ 
emission maps by projecting the emitted flux. This is the typical waveband at which observation campaigns for the study 
of accretion discs around MYSOs are currently conducted, see e.g.~\citet{maud_aa_620_2018}, as this waveband offers a good compromise 
between high angular resolution and atmosphere phase stability.
The distance between the source and the observer is set to $1.0\, \rm kpc$, that is of the order of the distance to 
the closest massive star-forming regions such as Orion ($0.41\, \rm kpc$), Cep A ($0.70\, \rm kpc$) and VY CMa 
($1.14\, \rm kpc$), see Table~4 in~\citet{reid_apj_700_2009}, as the angular resolution reached by {\sc alma} when 
observing them is the highest possible. Additionally, we also investigate the observability of disc 
substructures for MYSOs located at a distance of $2.0\, \rm kpc$ which corresponds to further afield 
star-forming regions such as W3(OH), see Section~\ref{sect:distance}. 
For each of the 6 selected simulation shapshots, we use the possibility of {\sc radmc-3d} 
to compute $1.2\, \rm mm$ synthetic images for three different system inclination 
angles, namely face-on ($\phi=0^{\circ}$), intermediate inclination ($\phi=45^{\circ}$), and edge-on ($\phi=90^{\circ}$).
The disc models and their emission properties are presented in detail in Section~\ref{sect:results}.

\subsection{Synthetic observables} 
\label{sect:image}

We further post-process the {\sc radmc-3d} radiation transfer calculations with the Common 
Astronomy Software Applications {\sc casa}\footnote{https://casa.nrao.edu/}~\citep{McMullin_aspc_376_2007} in order to obtain synthetic 
fluxes of the disc models as they would look like if seen by {\sc alma} observations. 
\textcolor{black}{
The successive technical updates and enhancements of {\sc alma} are labelled as "Cycles" and to each of them corresponds a call for 
proposals followed by a series of observations. For each cycle, {\sc alma} operates at different wavebands (the "bands"). 
The one of interest in our study is band 6, corresponding to $1.2\, \rm mm$~\citep{brown_AdSpR_2004,wootten_IEEEP_2009}. Observations are performed with a given 
spatial configuration of all the antennas constituting the {\sc alma} interferometer. The more compact the configuration the larger the beam size probing the sky, 
while the more extended the antennas configuration the better the spatial resolution in the observed images. The antennas are 
available in two categories, distinguishable by the diameter of their parabola, i.e. $7\, \rm m$ and $12\, \rm m$, respectively. The 
maximum number of $12\, \rm m$-antenna that can be used simultaneously is 43. 
We are interested in the $12\, \rm m$ antenna arrays because these 43 antennas can be spread over a wider surface on the Llano de Chajnantor plateau, 
offering a larger maximal separation between two antennas (the maximum baseline) and hence a smaller beam size to probe the internal 
structure of our accretion discs. 
The {\sc alma} Cycle 7 offers ten $12\, \rm m$ antenna array configurations, labelled from 1 (most compact) to 10 (most extended). 
We perform synthetic images at two of the most extended configurations (8 and 10) of the 43 different $12\, \rm m$-antenna, also called configurations C43-8 and C43-10. 
We therefore model images of our discs as {\sc alma} Cycle 7 configuration 8 and 10 observations at band 6 ($1.2\, \rm mm$), respectively. 
}

The {\sc radmc-3d} to {\sc casa} interface is effectuated via the {\sc radmc3dPy} python package available 
with the current distribution of {\sc radmc-3d}. It allows the user to directly output the radiative transfer 
calculations as standard FITS files, which header can be read by the {\sc casa} software. 
We make use the {\sc simalma} task of {\sc casa} with a central frequency of the combined continuum emission of 
$249.827\, \rm Ghz$ ($1.2\, \rm mm$, {\sc alma} band 6) and a channelwidth of $50.0\, \rm Mhz$. 
\textcolor{black}{
The different used long-baseline configurations result in angular resolutions of 
$0.029^{\prime\prime}$ (antenna configuration 8 with maximal baseline of $8.5\, \rm km$) and $0.015^{\prime\prime}$ 
(antenna configuration 10 with maximal baseline of $16.2\, \rm km$), respectively.
%, and, for source located at a distance 
%of $2\, \rm kpc$, this correspond to an angular resolution of $0.058^{\prime\prime}$ and $0.030^{\prime\prime}$, respectively. 
}
The observation integration time is chosen to be $600\, \rm s$ ($10\, \rm min$) of exposition time, as longer durations do not provide 
better images (see Section~\ref{sect:discussion}). The precipitable water vapor (pwv) parameter is chosen 
to be $0.5\, \rm mm$, which represent the amount of pwv at the Llano de Chajnantor plateau in the Chilean Andes during $25\, \%$ of the time
during which the {\it Atacama Large Millimeter/submillimeter Array} ({\sc alma}) is operated. 
%
%The {\sc casa} outputs are exported as FITS files which can easily be plotted with the python package 
%{\sc matplotlib} as series of figures with normalised intensity. 

%%%%%%%%%%%%%%%%%%%%%%%%%%%%%%%%%%%%%%%%%%%%%%%%%%%%%%%%%%%%%%%%%%%%%%%%%%%%%%%%%%%%%%%%%%%
%%%%%%%%%%%%%%%%%%%%%%%%%%%%%%%%%%%%%%%%%%%%%%%%%%%%%%%%%%%%%%%%%%%%%%%%%%%%%%%%%%%%%%%%%%%
%%%%%%%%%%%%%%%%%%%%%%%%%%%%%%%%%%%%%%%%%%%%%%%%%%%%%%%%%%%%%%%%%%%%%%%%%%%%%%%%%%%%%%%%%%%

\section{Results}
\label{sect:results}

In this section, we begin by presenting the samples of the hydrodynamical simulation 
which we post-process to obtain synthetic $1.2\, \rm mm$ interferometric images of dust continuum 
emission. 
Then, we discuss the time effects due to the disc evolution and the effects of the disc inclination 
angle with respect to the plane of sky.

\subsection{Hydrodynamical model and stellar properties} 
\label{sect:disc}

The simulation model Run1-hr of~\citet{meyer_mnras_473_2018} begins with the gravitational collapse 
of $100\, \rm M_{\odot}$ of pre-stellar rotating molecular material onto the stellar embryo. At the 
end of the free-fall collapse phase, the infalling material ends on a centrifugally-balanced disc, 
from which the gas is subsequently transferred to the growing protostar. 
Fig.~\ref{fig:history} plots the accretion rate history onto the central massive protostar 
(solid blue line, in $\rm M_{\odot}\, \rm yr^{-1}$) together with the evolution 
of the protostellar mass (dashed red line, in $\rm M_{\odot}$) that is calculated 
as the integrated disc-to-star mass transfer rate through the sink cell. 
The vertical thin black line indicates the onset of disc formation, when the free-fall 
collapse of the envelope material onto the protostar stops and the star begins to gain its 
mass exclusively via accretion from its surrounding disc. 
After the initial infall of material, the collapse of the parent 
pre-stellar core material generates an initial increase of the mass flux through the inner 
boundary at a time $\approx 2\, \rm kyr$. The accretion rate then reaches the standard value predicted 
for MYSOs of $10^{-3}\, \rm M_{\odot}\, \rm yr^{-1}$~\citep{hosokawa_apj_691_2009} up to the onset 
of the disc formation happening at $\approx 12\, \rm kyr$. 
Variabilities in the accretion flow begin right after the disc formation and the accretion 
rate history exhibits numerous peaks of growing intensity as the MYSO becomes heavier. 
\textcolor{black}{ 
This is not caused by different inner boundary conditions, but simply reflects the time-dependent 
azimuthal anisotropies induced in the accretion flow by the disc evolution.   
}
The efficient gravitational instabilities produce complex substructures in the disc, 
such as overdense spiral arms in which gaseous clumps of various morphologies form at radii of 
$\sim 100\, \rm au$ and inward-migrate down to the central protostar. This produces luminosity 
outbursts via the mechanism revealed in~\citet{meyer_mnras_464_2017} \textcolor{black}{for massive 
stars.} 
These outbursts are responsible for step-like increases in the stellar mass evolution (see thick dotted red line) 
due to the fast accretion of dense circumstellar material inside the sink cell. A more detailed 
description of the evolution of circumstellar discs around young massive stars irradiating their 
self-gravitating discs can be found in our precedent study~\citep{meyer_mnras_473_2018}. 
In Fig.~\ref{fig:history}, several magenta dots mark the time instances of the chosen simulation snapshots 
considered in this study. Note that the young star becomes, by definition, a massive object when 
$M_{\star}=8\, \rm M_{\odot}$. Hence, our selected disc models are all in the high-mass regime.

Fig.~\ref{fig:star} plots the stellar surface properties obtained by post-processing the accretion 
rate history displayed in Fig.~\ref{fig:history} with the {\sc genec} evolution code, with the 
evolution of the protostellar iternal photospheric luminosity (a), radius (b), and effective 
temperature (c) as a function of the stellar age of the MYSO.  
No notable variations happen up to the end of the free-fall collapse and during the early 
phase of the disc formation at times $\approx 20\, \rm kyr$. A slight monotonical increase 
of the photospheric luminosity $L_{\star}$ and of the protostellar radius $R_{\star}$ happens 
as the deuterieum burning keeps the core temperature almost constant. 
When the evolution of the accretion rate exhibits sudden variations in response to the accretion of 
gaseous circumstellar clumps formed in the fragmented accretion disc, it generates changes in the 
internal structure resulting in the bloating of the protostellar radius~\citep{haemmerle_585_aa_2016}. 
At this stage, the entropy of the external layers is high and any episodic deposit of mass on it consequently induces an 
augmentation of $T_{\rm eff}$ and $L_{\star}$, as described in~\citet{2019MNRAS.tmp...10M}. 
During each strong accretion burst, the MYSOs episodically experiences changes in the photospheric 
properties by becoming bigger and cooler. Fig.~\ref{fig:hdr} shows the evolutionary track 
of our MYSO in the Hertzsprung-Russell diagram and compares it with the pre-zero-age-main-sequence (pre-ZAMS) tracks calculated 
with constant accretion rates (thin solid black lines) of~\citet{haemmerle_585_aa_2016} and 
with the ZAMS track (thick dashed black line) of~\citet{ekstroem_aa_537_2012}. 
The grey solid lines are isoradii. 
As detailed in~\citet{2019MNRAS.tmp...10M}, each strong accretion event reaching rates 
$\ge 10^{-2}\, \rm M_{\odot}\, \rm yr^{-1}$ induces important rapid changes in the pre-main-sequence 
evolutionary track of the MYSOs. Following an internal redistribution of entropy in the upper 
layer of the stellar structure, $R_{\star}$ increases. Successive evolutionary loops toward the 
red part of the Hertzsprung-Russell diagram occur during these bloating periods. 
The protostellar properties, required for the radiative transfer calculations, are reported for 
the 6 selected simulation snapshots in our Table~\ref{tab:models}.

\begin{figure} 
        \centering
        \begin{minipage}[b]{ 0.45\textwidth}
                \centering
                \includegraphics[width=1.0\textwidth]{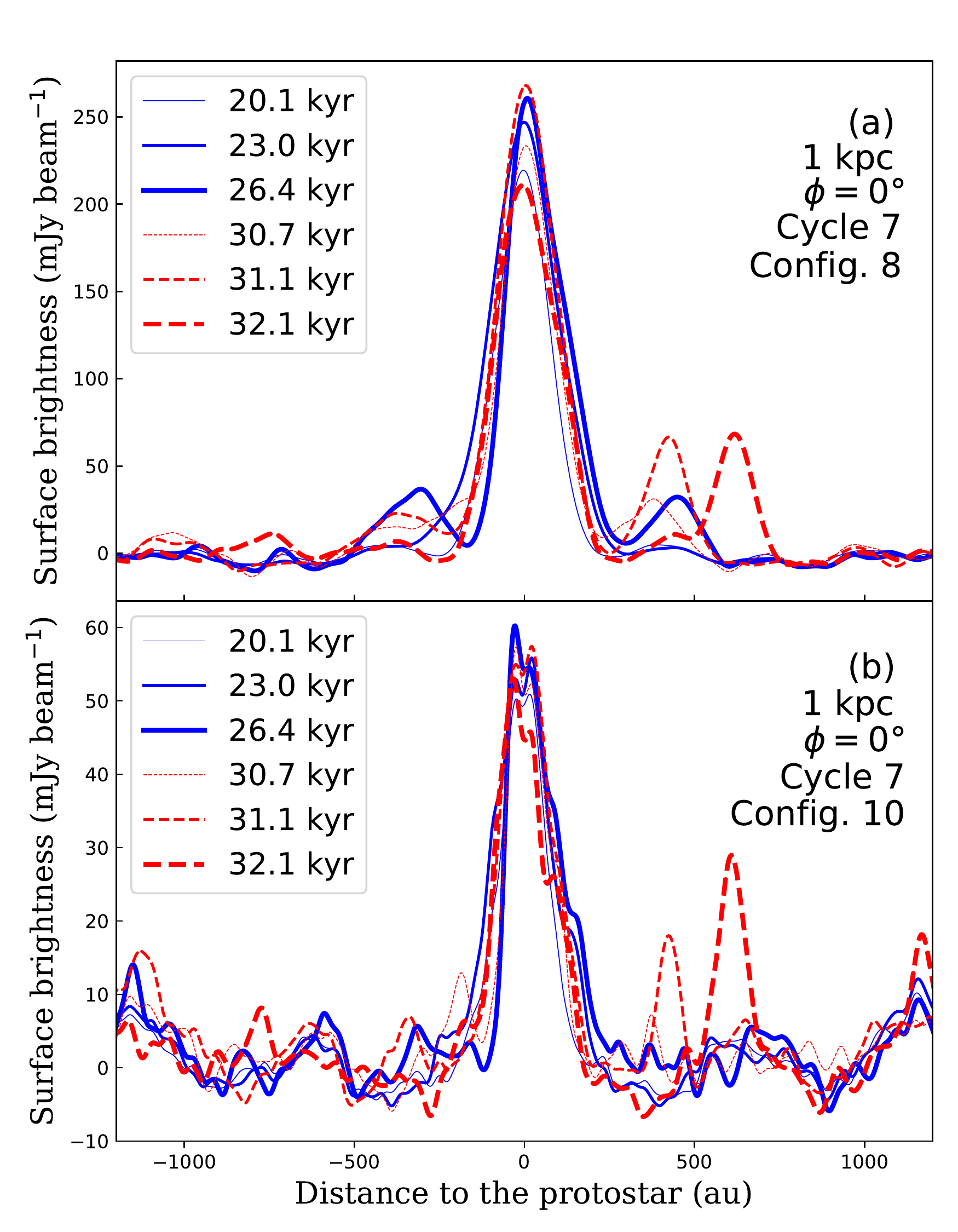}
        \end{minipage} 
        \caption{ 
                 \textcolor{black}{
        		 Cross sections taken in our {\sc alma} C43-8 (a) and C43-10 (b) images simulated images with $10\, \rm min$ exposure time 
        		 at band 6 ($1.2\, \rm mm$) of our accretion discs.
        		 } 
        		 The cuts are taken through images with face-on viewing angle and the source distance is $1\, \rm kpc$. 
        		 The cut are taken along the vertical direction through the origin of the images. 
                 }      
        \label{fig:cut_all_models}  
\end{figure}

\subsection{ Emission maps at $1.2\, \rm mm$} 
\label{sect:mm1200}

Fig.~\ref{fig:theory_plot_2} plots the projected $1.2\, \rm mm$ emission maps 
of our models $1$-$6$. The emission intensity is normalised, the black color representing 
the faintest pixels and the white one the brightest ones, respectively. The white bar in 
the bottom right corner of each figure indicates the physical size of each images.  
The top series of panels represent the young disc at times $20.1\, \rm yr$ (Fig.~\ref{fig:theory_plot_2}a), 
$23.0\, \rm yr$ (Fig.~\ref{fig:theory_plot_2}b) and $26.4\, \rm yr$ (Fig.~\ref{fig:theory_plot_2}c) after 
the beginning of the simulation, respectively. These snapshots correspond to time instances before the MYSO 
experiences its first accretion-driven burst. 
The disc model 1 in Fig.~\ref{fig:theory_plot_2}a is a circular and stable disc that does no 
show any noticeable sign of either fragmentation by gravitational instability or substructures in 
it. This shape persists up to at least $23.0\, \rm yr$ (model 2, Fig.~\ref{fig:theory_plot_2}b). 
During this time interval, the protostellar mass evolves from $8.6, \rm M_{\odot}$ to $10.7\, \rm M_{\odot}$, 
but its circumstellar medium fails in developing a clear spiral structure, although a faint trailing 
structure begins to appear at radii $\ge 200$-$600\, \rm au$. The brightest region is the inner 
$50\, \rm au$ of the disc. 
Our model 3 (Fig.~\ref{fig:theory_plot_2}b) starts exhibiting a two-arms pattern extending up to 
radii $\sim 400\, \rm au$ from the MYSO. Moreover, a dense gaseous clump formed in the spiral arm 
inward-migrates and fast-moves onto the protostar. The clump keeps on contracting during its migration 
and its core increases in density and temperature, constituting the maximal-emitting region of 
the disc. 
The successive, rapid accretion of several migrating clumps from different parent arms, the 
merging of clumps into inhomogeneous gaseous structures or even the migration of clumps 
that separate an inner portion of spiral arms into two segments make the accretion pattern more 
and more complex and strengthen the accretion variability (Fig.~\ref{fig:history}).

The bottom series of panels of Fig.~\ref{fig:theory_plot_2} shows emission maps of our disc 
models 4 to 6, corresponding to the young disc at times $30.7\, \rm yr$ (Fig.~\ref{fig:theory_plot_2}c), 
$31.1\, \rm yr$ (Fig.~\ref{fig:theory_plot_2}e) and $32.1\, \rm yr$ (Fig.~\ref{fig:theory_plot_2}f), 
respectively. 
The disc model 4 (Fig.~\ref{fig:theory_plot_2}d) has the typical morphology of the circumstellar medium 
of a protostellar disc undergoing efficient gravitational instability. 
In addition to clear enrolled spiral arms starting close to the star and fading away at $\ge 600$-$800\, \rm au$, 
several gaseous clumps have formed inside the spiral arms, at distances $\ge 200$-$400\, \rm au$ from 
the protostar. 
The spiral arms and clumps appear very prominently in the images, with a higher integrated surface 
brightness than the homogenous parts of the disk.
The disc model 5 (Fig.~\ref{fig:theory_plot_2}e) is more complex than the rather standard ring-like 
clump-hosting structure of model 4. Its strongly fragmented disc experiences a complex dynamics
in which multiple Toomre-instable gaseous clumps~\citep{meyer_mnras_473_2018} exert torques with each other, 
with the neighbouring spiral arms and with the central MYSO of $16.7\, \rm M_{\odot}$, respectively. 
%The bright dense clumps bend the spiral arms as they move and inward-migrate, suggesting a 
%self-rotation of the fragments~\citep{}. 
%
Finally, the disc model 6 (Fig.~\ref{fig:theory_plot_2}f) exhibits an extended, strongly fragmented 
disc with a filamentary spiral arm hosting several bright gaseous clumps. This morphological configuration 
of the disc substructures is typical of the time following an accretion-driven burst, when the tail of 
a migrating clump is swung away and pushes the circumstellar material to the larger radii as described 
in~\citet{meyer_mnras_464_2017}. The subsequently forming dense ring is the location of the formation of 
new clumps, whose integrated surface brightness can be similar to that of the inner disc.

 \begin{figure*}  
        \begin{minipage}[b]{ 1.0\textwidth}   
        \centering
                \includegraphics[width=1.0\textwidth]{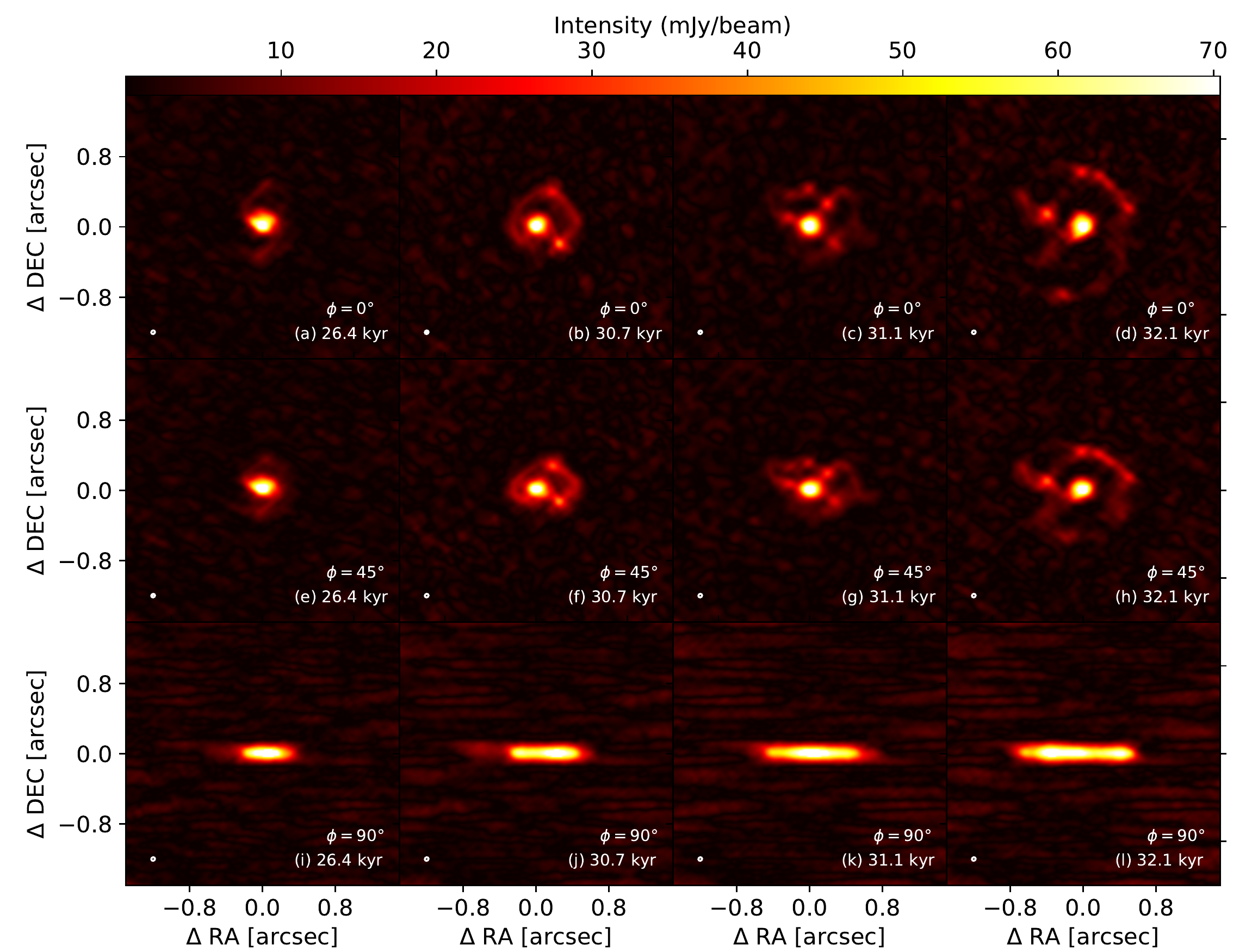}
        \end{minipage}    
        \caption{
                \textcolor{black}{
                 {\sc alma} C43-8 images simulated images at band 6 ($1.2\, \rm mm$) of the 
                 accretion disc growing around our protostar at times $26.4\, \rm kyr$ (left), $30.7\, \rm kyr$ (middle left), $31.1\, \rm kyr$ (middle right) 
                 and $32.1\, \rm kyr$ (right), assuming $10\, \rm min$ exposure time.
                 }
                 The images are shown with a viewing angle of $0\degree$ (top), $45\degree$ (middle) and $90\degree$ (bottom) with respect to the plane of sky. 
                 The beam size is $0.029^{\prime\prime}$ and it is given in the lower left corner of each panel. 
                 The source distance is assumed to be $1\, \rm kpc$.   
                 }      
        \label{fig:maps_angle1}  
\end{figure*}

\subsection{Observability of evolving discs around growing protostars}      
\label{sect:growth}

Figs.~\ref{fig:allmodels_68} and~\ref{fig:allmodels_610} show emission maps of our disc models 
as seen by {\sc alma} if located at a distance of $1\, \rm kpc$ from the observer and monitored with 
antenna configuration 8 and 10, respectively. Both figures show $1.2\, \rm mm$ simulated 
images of the selected snapshots at times $20.1\, \rm kyr$ (a), $23.0\, \rm kyr$ (b), $26.4\, \rm kyr$ (c), 
$30.7\, \rm kyr$ (d), $31.1\, \rm kyr$ (e) and $32.1\, \rm kyr$ (f), respectively. The images assume 
a face-on viewing geometry. On each panel the beam size is given in the bottom left 
panel. The map scale is in units of $\rm arcsec$ and the surface brightness intensity is in 
$\rm mJy\, \rm beam^{-1}$. 
%At this distance, $1000\, \rm au$ represents an angular separation of $1.0^{\prime\prime}$. 
%
The synthetic images of stable disc model 1 exhibit a bright circular shape for both antenna configuration 8 
(Fig.~\ref{fig:allmodels_68}a) and the most extended configuration 10 (Fig.~\ref{fig:allmodels_610}a). 
The same is the case for our disc model 2 ($23.0\, \rm kyr$), as the current beam resolution of {\sc alma} is not 
able to spatially resolve the early forming overdensities in its principal spiral arm (Fig.~\ref{fig:allmodels_68}b 
and Fig.~\ref{fig:allmodels_610}b). 
The $1.2\, \rm mm$ signature of the effects of gravitational instability begins to appear at time $26.4\, \rm kyr$ 
with the antenna configuration 10 (Fig.~\ref{fig:allmodels_610}c), when the protostar has reached $12.1\, \rm M_{\odot}$ 
and a series of twin spiral arms in which a
detached migrating clump falls onto the protostar. 
The images at time $30.7\, \rm kyr$ clearly reveal the fragmenting characteristic of our disc, i.e. a 
self-gravitating spiral arm that includes a few bright gaseous clumps (Fig.~\ref{fig:allmodels_68}d) 
potentially on the way to low-mass star formation~\citep{meyer_mnras_473_2018}. The accretion disc 
and its substructures can be resolved with the {\sc alma} antenna configuration 8. A more extended 
antenna configuration (configuration 10) yields similar results, see Fig.~\ref{fig:allmodels_610}d. 
At time $\ge 31.1\, \rm kyr$ the synthetic observations of models 5 and 6 reveal a nascent multiple, 
hierarchical massive system with clumps and filamentary structures organised around a hot, young massive star 
of mass $16.7\, \rm M_{\odot}$ (Figs.~\ref{fig:allmodels_68}e,f) and $20.0\, \rm M_{\odot}$ 
(Figs.~\ref{fig:allmodels_610}e,f), respectively.

\begin{figure*}  
        \begin{minipage}[b]{ 1.0\textwidth}   
        \centering
                \includegraphics[width=1.0\textwidth]{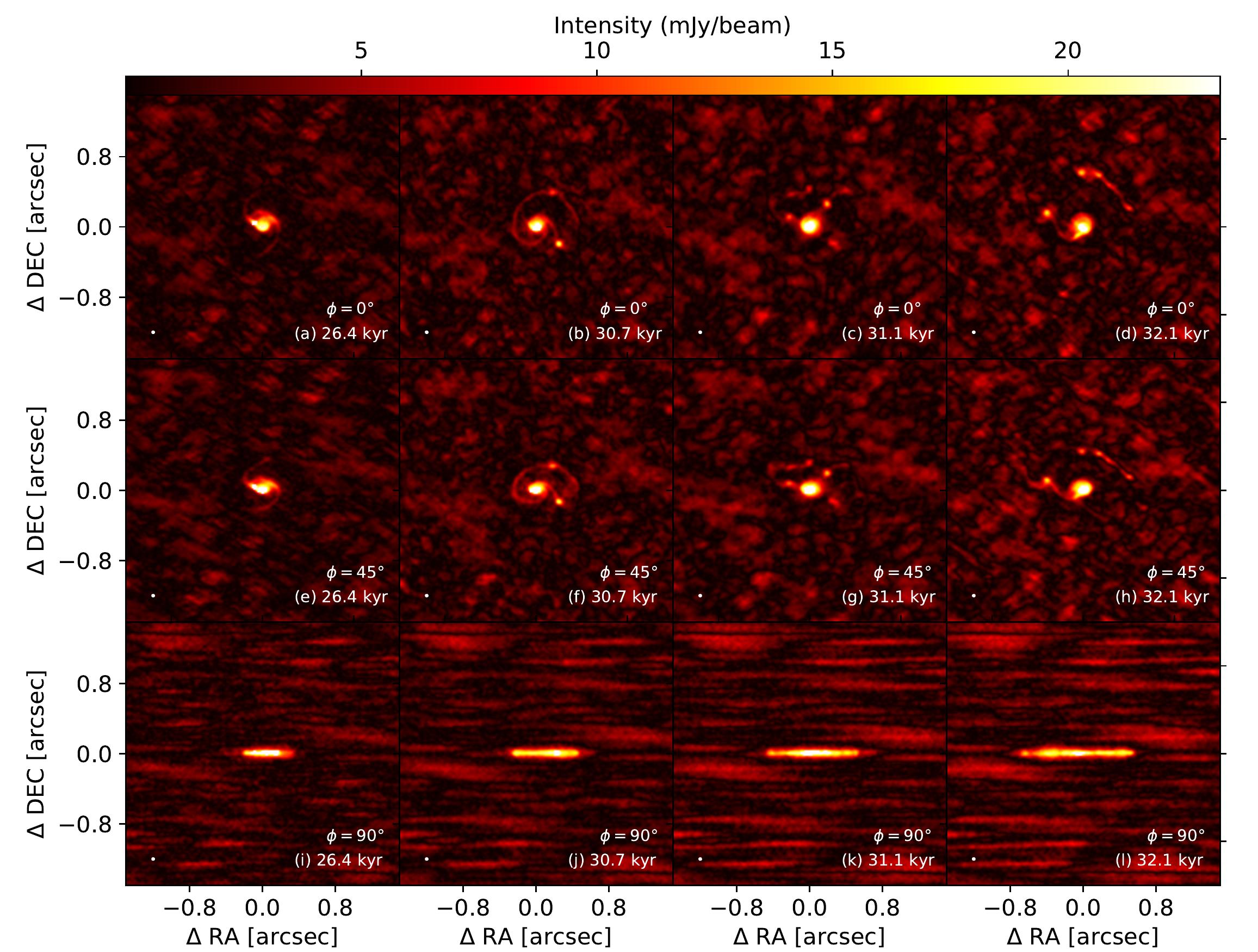}
        \end{minipage}    
        \caption{ 
                \textcolor{black}{
                 {\sc alma} C43-10 images simulated images at band 6 ($1.2\, \rm mm$) of the 
                 accretion disc growing around our protostar at times $26.4\, \rm kyr$ (left), $30.7\, \rm kyr$ (middle left), $31.1\, \rm kyr$ (middle right) 
                 and $32.1\, \rm kyr$ (right), assuming $10\, \rm min$ exposure time.
                 }
                 The images are shown with a viewing angle of $0\degree$ (top), $45\degree$ (middle) and $90\degree$ (bottom) with respect to the plane of sky. 
                 The beam size is $0.015^{\prime\prime}$ and it is given in the lower left corner of each panel. 
                 The source distance is assumed to be $1\, \rm kpc$.    
                 }      
        \label{fig:maps_angle2}  
\end{figure*}

Shown in Fig.~\ref{fig:cut_all_models} are cross-sections extracted from the {\sc alma} cycle 6 maps with 
antenna configuration 8 \textcolor{black}{(Fig.~\ref{fig:allmodels_68})} and antenna configuration 10 
\textcolor{black}{(Fig.~\ref{fig:allmodels_610})}, respectively. 
The figures compare the emission intensity at times $20.1\, \rm kyr$ (thin solid blue line), 
$23.0\, \rm kyr$ (solid blue line), $26.4\, \rm kyr$ (thick solid blue line), $30.7\, \rm kyr$ 
(thin dashed red line), $31.1\, \rm kyr$ (dashed red line) and $32.1\, \rm kyr$ (thick dashed red line). 
These profiles assume a face-on viewing geometry ($\phi=0\degree$). The source distance 
is $1\, \rm kpc$ and the cuts is taken along the vertical direction through the center of the synthetic images.      
The shape of \textcolor{black}{the cross-sections (solid blue lines) corresponding to Fig.~\ref{fig:allmodels_68}} 
further illustrates that the surface brightness of our modelled accretion 
disc does not exhibit signs of substructured circumstellar disc at times $\le 23.0\, \rm kyr$, when the 
protostar is $\le 10.7\, \rm M_{\odot}$, although the intensity cut extracted from the image at 
$26.4\, \rm kyr$ (thick solid blue line) has two additional bright emission peaks corresponding 
to the two spiral arms. 
\textcolor{black}{
The slightly negative fluxes in some places in the interarm region of the discs result from the noise of 
the substracted background thermal emission \textcolor{black}{arising from the incomplete u-v coverage} in the simulated observations. 
}
The cuts taken in the models 4-6 clearly highlight the presence of a dense ring-like spiral arm that  
surrounds the MYSO and which produces intensity peaks of $\approx 75\, \rm mJy/beam$ above the background of 
$\approx 10\, \rm mJy/beam$. 
\textcolor{black}{
The emission cuts in Fig.~\ref{fig:cut_all_models}b have a higher spatial resolution but a lower sensitivity than 
those of Fig.~\ref{fig:cut_all_models}a, i.e. the intensity differences between the close stellar surroundings 
and the arms/clumps in the disc are less pronounced (see also Fig.~\ref{fig:config}). 
}
For example, the northern ring of model 6 ($32.1\, \rm kyr$) peaks at $\approx 50$ and 
$\approx 30\, \rm mJy/beam$, which represents about $25\%$ and $60\%$ of the disc central emission, respectively. 
The images in Fig.~\ref{fig:allmodels_68}a-f are complementary to that of Fig.~\ref{fig:allmodels_610}a-f 
in the observation and analysis of the discs, \textcolor{black}{and observations of the circumstellar medium of a high-mass star  
should be performed with several antenna configurations of {\sc alma} for a given waveband.}

\subsection{Effect of the inclination angle}  
\label{sect:angle}

Figs.~\ref{fig:maps_angle1} and~\ref{fig:maps_angle2} show synthetic {\sc alma} cycle 6 ($1.2\, \rm mm$) images 
of our disc models. \textcolor{black}{In the figure, each row represents an inclination, and each column a time, and the images 
are calculated for the {\sc alma} C43-8 configuration, assuming a distance to the source of $1\, \rm kpc$. }
%
%at times $26.4\, \rm kyr$ (left column, Figs.~\ref{fig:maps_angle2}a,e,i), $30.7\, \rm kyr$ 
%(middle left column, Figs.~\ref{fig:maps_angle2}b,f,j), $31.1\, \rm kyr$ (middle right column, 
%Figs.~\ref{fig:maps_angle2}c,g,k) 
%and $32.1\, \rm kyr$ (right column, Figs.~\ref{fig:maps_angle2}d,h,l) for inclination angles of 
%$0\degree$ (top row, Figs.~\ref{fig:maps_angle2}a-d), $45\degree$ (middle row, Figs.~\ref{fig:maps_angle2}e-h) 
%and $90\degree$ (bottom row, Figs.~\ref{fig:maps_angle2}i-l) with respect to the plane of sky, assuming 
%a distance to the source of $1\, \rm kpc$. 
% 
The beam size is given in the lower left corner of each panel. Again, the map scale is in units of $\rm arcsec$ and 
the surface brightness intensity is in $\rm mJy\, \rm beam^{-1}$. 
%At this distance, $1000\, \rm au$ represents an angular separation of $1.0^{\prime\prime}$.  
%
For $\phi=45\degree$, the emission maps do not exhibit important changes in the disc shapes, except that their projected size is 
squeezed by a factor $\cos(\pi/4)$ along the direction normal of the disc plane (the vertical direction of the maps). 
This is consistent with the studies on discs from low-mass stars of~\citet{dong_apj_823_2016} which 
demonstrated that moderate inclination angles with respect to the plane of sky have little incidence 
in the disc emission maps compare to that with $\phi=0\degree$. 
%
%Importantly, the disc inclination does not affect the observability of the bright gaseous clumps 
%in it, in the sense that they can not be observed at time $\le 26.4\, \rm kyr$ but become clearly visible afterwards. 
%
In the case of edge-on accretion discs, the antenna configuration 8 does not allow the observer to clearly 
distinguish clumpy structures in the accretion discs, even if they can be inferred from the meridional emission 
profile, e.g. at time $32.1\, \rm kyr$ (Fig.~\ref{fig:maps_angle1}l). 
Similar trends in the morphological signatures of the accretion discs happen with a higher resolution of the beam, when the 
antenna configuration is more extended (Fig.~\ref{fig:maps_angle2}). 
%
%Let also notice the better resolution of the inner disc organisation for edge-viewed accretion discs for the most 
%extended antenna configuration giving the finest beam size (Fig.~\ref{fig:maps_angle2}j-l). 
%
%Again, as stated in Fig.~\ref{sect:growth}, in addition to a finer spatial resolution that naturally highlights 
%better disc structures, the extended antenna configuration 10 has a smaller contrast between the disc and 
%the background (Fig.~\ref{fig:maps_angle2}) which may make the clump detection more difficult, while it greatly facilitate their study %once they are already localised, e.g. with a more compact antenna configuration (Fig.~\ref{fig:maps_angle1}). 

\begin{figure*}  
        \centering
        \begin{minipage}[b]{ 1.0\textwidth}
                \centering
                \includegraphics[width=1.0\textwidth]{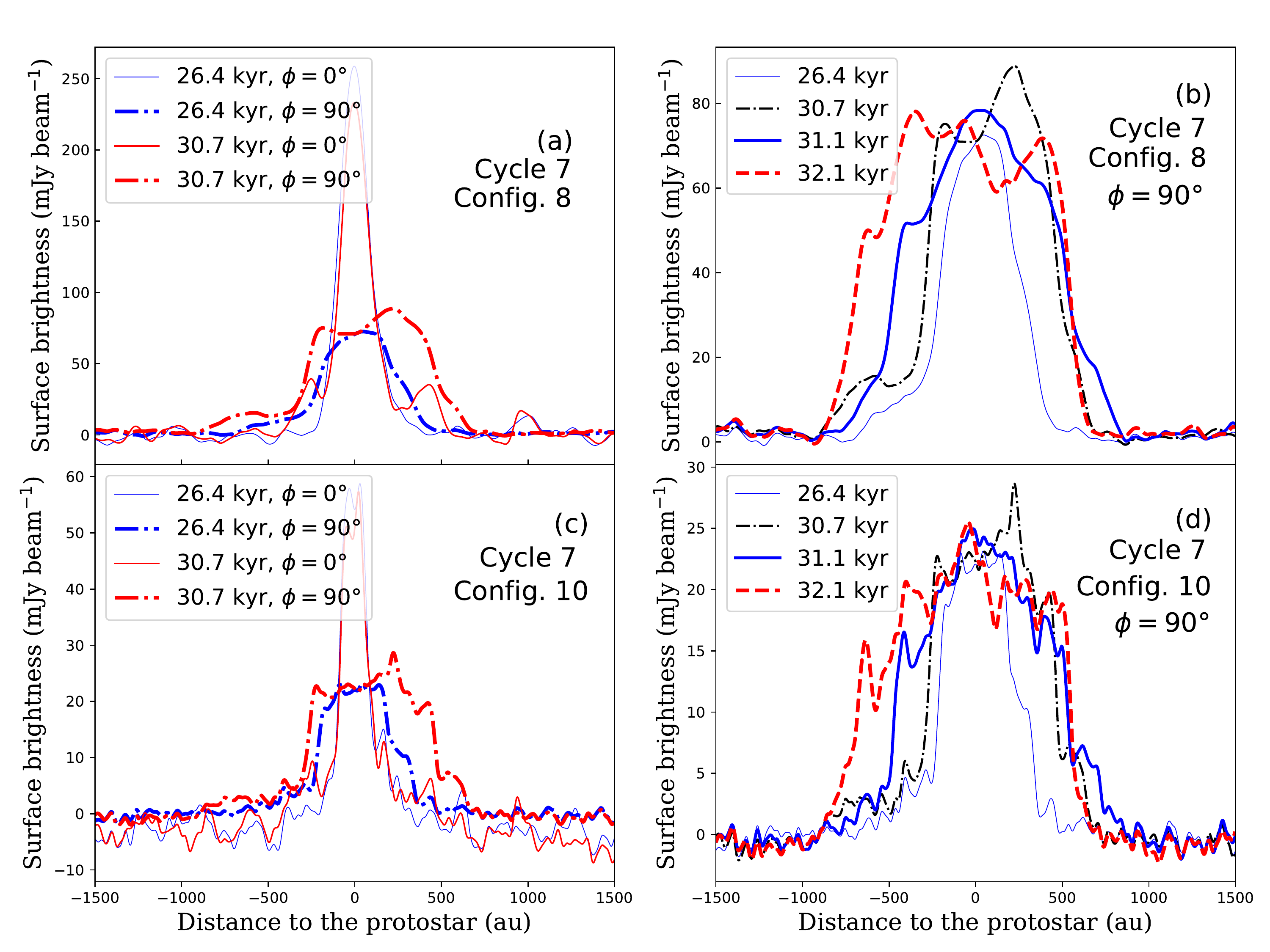}
        \end{minipage}       
        \caption{ 
                \textcolor{black}{        
        		 Comparison between the cross sections taken in our simulated {\sc alma} images simulated at band 6 
        		 ($1.2\, \rm mm$) and assuming antenna configurations 8 (C43-8 observations) and 10 (C43-10 observations), respectively.
        		 }
        		 Left panels compare the surface brightnesses of the discs at times $20.1\ \rm kyr$
        		 and $23.0\ \rm kyr$ considered with a viewing angles of $\phi=0\degree$ 
        		 and $\phi=90\degree$, respectively. 
        		 Right panels compare the surface brightnesses of the discs at times $26.4\ \rm kyr$,
        		 to $30.7\ \rm kyr$ (blue) considered with a viewing angles of $\phi=0\degree$ 
        		 and $\phi=90\degree$, respectively.         		 
        		 The source distance is $1\, \rm kpc$. 
        		 The cut are taken in East-West direction through the origin of the images.         
                 }      
        \label{fig:angle_cut}  
\end{figure*}

Fig.~\ref{fig:angle_cut} compares cross sections taken from our simulated {\sc alma} cycle 6 
($1.2\, \rm mm$) images, with antenna configuration 8 (compact, coarser resolution) and 10 (extended, higher resolution). 
%
%%%The left panels compare the surface brightnesses of the discs at times $26.4\ \rm kyr$ (blue) 
%%%and $30.7\ \rm kyr$ (red) considered with a viewing angles of $\phi=0\degree$ (thin solid lines) 
%%%and $\phi=90\degree$ (thick dashed lines), for antenna configurations 8 (a) and 10 (c), respectively. 
%
%%%Right panels compare the surface brightnesses of the discs at times $26.4\ \rm kyr$ (thin solid 
%%%blue line), $30.7\ \rm kyr$ (dotted-dashed black line), $31.1\ \rm kyr$ (thick blue line) and 
%%%$32.1\ \rm kyr$ (dashed red line) considered with a viewing angle of $\phi=90\degree$ with respect 
%%%to the plane of sky, for antenna configurations 8 (b) and 10 (d), respectively,  
%%%and $\phi=90\degree$ (thick dashed lines), for antenna configurations 8 (a) and 10 (b), respectively.   
%
The source distance is $1\, \rm kpc$ and all cuts are taken along the horizontal direction through 
the location of the star. 
The first panel Fig.~\ref{fig:angle_cut}a illustrates the main difference between face-on and edge-viewed 
accretion discs during the early phase, at times $\le 26.4\, \rm kyr$. While the face-on disc ($\phi=0\degree$) 
has first an emission profile that does not exhibit clear substructures, its edge-view ($\phi=90\degree$) indicates 
the presence of its large, developing spiral arm. At later time and by projection, it gives the disc a symmetric 
morphology but not a uniform surface brightness with respect to its axis of rotation. 
%
%More details are seen initially with $\phi=90\degree$, the intensity is nevertheless smaller than with $\phi=0\degree$. 

Similar conclusions are drawn with the antenna configuration 10 (Fig.~\ref{fig:angle_cut}c). The disc projected 
intensity deceases as a function of the inclination, as other circumstellar structures around more evolved 
massive stars do~\citep{meyer_459_mnras_2016}.  
The time evolution of the surface brightnesses of the edge-viewed accretion discs is as follows: the 
intensity peaks on the projection plane at the location of the star. The emission signature gradually increases 
in size as the disc grows, by accumulating molecular gas from the surrounding infalling envelope, up to reaching 
a radius of $\simeq 800$-$1000\, \rm au$ (thick red curves of Fig.~\ref{fig:angle_cut}b,d). 
As a result of the antenna configuration 10 and smaller beam size, the meridional cuts through the disc reveal 
more details in the projected disc emission. 
Interestingly, some intensity peaks such as at time $30.7\, \rm kyr$ (thin black dotted dashed line 
of Fig.~\ref{fig:angle_cut}d) reveal the presence of a couple of aligned clumps, while the other 
snapshots do not. 
%
%Hence, particular clump spatial disposition may produce intensity enhancements when the disc 
%is inclined, and their presence inferred by means of {\sc alma} observation. 
%
Polarization measurements in the near-infrared waveband may be useful to further investigate this phenomenon.

\subsection{Effect of the distance of the source}  
\label{sect:distance}

\begin{figure*}  
        \begin{minipage}[b]{ 1.0\textwidth}   
        \centering
                \includegraphics[width=1.0\textwidth]{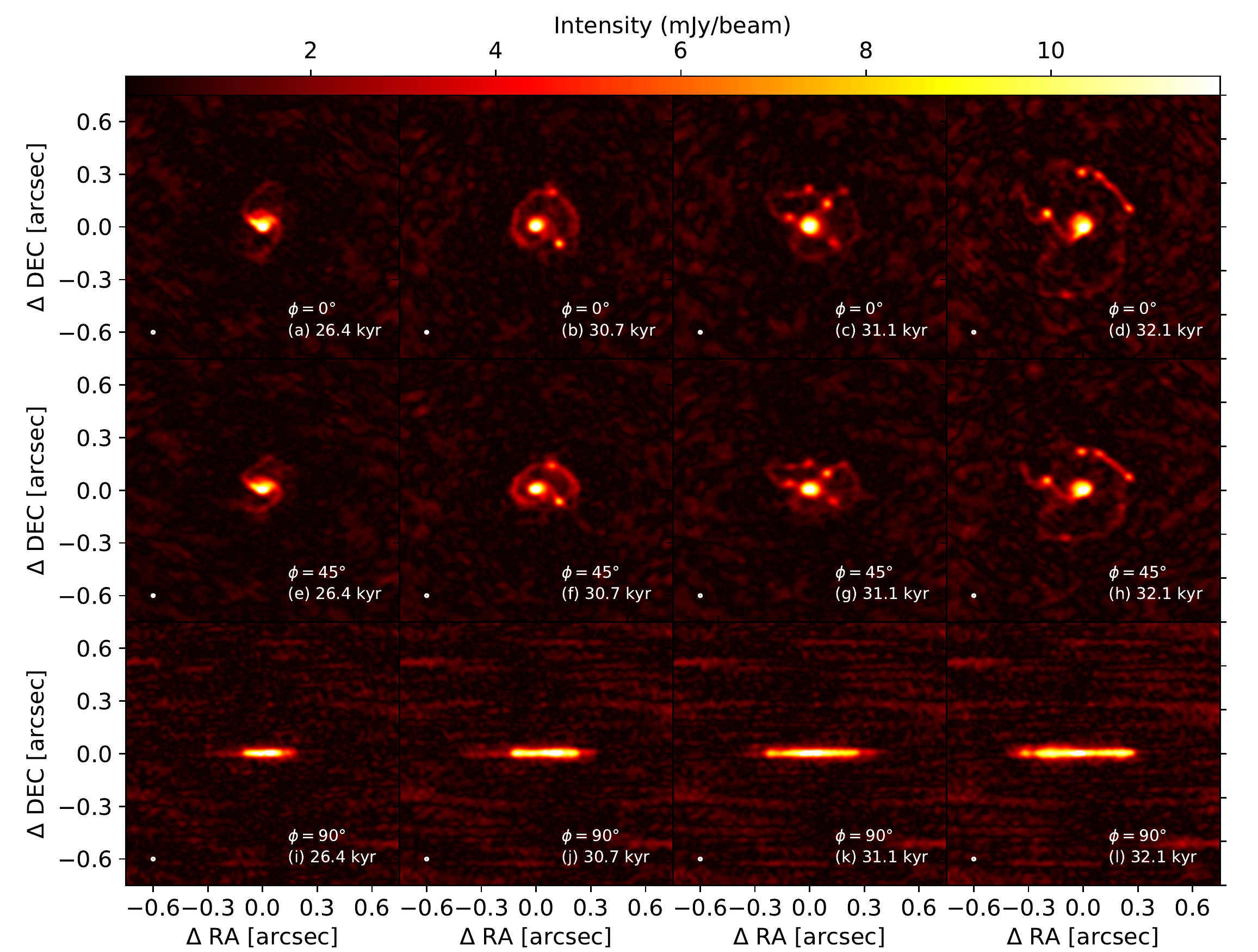}
        \end{minipage}    
        \caption{ 
                 \textcolor{black}{
                 Similar to Fig.~\ref{fig:maps_angle2}, but for a distance to the source of $2\, \rm kpc$. 
                 It displays {\sc alma} Cycle 7 band 6 ($1.2\, \rm mm$) images simulated with an antenna configuration 10 (C43-10 images).
                 The beam size is $0.015^{\prime\prime}$ and it is given in the lower left corner of each panel. 
                 Note that the angular scales are different than for the models with a distance to the source of $1\, \rm kpc$ (see Fig.~\ref{fig:maps_angle2}).  
                 }
        }      
        \label{fig:maps_distance}  
\end{figure*}

Although the Earth's closest high-mass star-forming regions are located at distances 
$\le 1\, \rm kpc$~\citep{reid_apj_700_2009}, numerous giant molecular regions with massive protostars 
are more distant. We check the observability of the substructures in our accretion disc models 
assuming a distance of $\le 2\, \rm kpc$. The radiative transfer methods and the computations of synthetic images 
are identical as above described for sources located at $1\, \rm kpc$, the only difference being the exposure time that 
we derive to be $30\, \rm min$ (Section~\ref{sect:effects}). 
Fig.~\ref{fig:maps_distance} shows synthetic {\sc alma} cycle 6 ($1.2\, \rm mm$) images 
of our disc models. 
\textcolor{black}{
In the figure, each row represents an inclination, and each column a time, and the images 
are calculated for the {\sc alma} C43-10 configuration, assuming a distance to the source of $2\, \rm kpc$.  
} 
The beam size is given in the lower left corner of each panel. Again, the map scale is in units of $\rm arcsec$ and 
the surface brightness intensity is in $\rm mJy\, \rm beam^{-1}$. 
In accretion discs located at $2\, \rm kpc$ rather than \textcolor{black}{at $1\, \rm kpc$}, it is more difficult to observe clearly 
detached clumps for young discs such at times $\le 26.4\, \rm kyr$ whatever their viewing geometry is 
(Fig.~\ref{fig:maps_distance}a,e,i). Synthetic images of later time instances ($>26.4\, \rm kyr$) lead to similar conclusions regarding 
the possible observability of clumps and spiral arms as those derived on the basis of the synthetic images computed 
assuming a distance of $1\, \rm kpc$. In the case of edge-on discs, inferring clumps is somewhat less evident 
(Fig.~\ref{fig:maps_distance}i-k), even though the far West clump developing at time $32.1\, \rm kyr$ can be 
distinguished from the rest of the circumstellar structure in Fig.~\ref{fig:maps_distance}l. 
We conclude that the outcomes of gravitational instabilities in accretion discs around MYSOs can be detected with 
{\sc alma}, with observations performed during the Cycle 7 observational run and using its most extended 
available antenna configuration.

%%%%%%%%%%%%%%%%%%%%%%%%%%%%%%%%%%%%%%%%%%%%%%%%%%%%%%%%%%%%%%%%%%%%%%%%%%%%%%%%%%%%%%%%%%%
%%%%%%%%%%%%%%%%%%%%%%%%%%%%%%%%%%%%%%%%%%%%%%%%%%%%%%%%%%%%%%%%%%%%%%%%%%%%%%%%%%%%%%%%%%%
%%%%%%%%%%%%%%%%%%%%%%%%%%%%%%%%%%%%%%%%%%%%%%%%%%%%%%%%%%%%%%%%%%%%%%%%%%%%%%%%%%%%%%%%%%%
%%%%%%%%%%%%%%%%%%%%%%%%%%%%%%%%%%%%%%%%%%%%%%%%%%%%%%%%%%%%%%%%%%%%%%%%%%%%%%%%%%%%%%%%%%%
%%%%%%%%%%%%%%%%%%%%%%%%%%%%%%%%%%%%%%%%%%%%%%%%%%%%%%%%%%%%%%%%%%%%%%%%%%%%%%%%%%%%%%%%%%%
%%%%%%%%%%%%%%%%%%%%%%%%%%%%%%%%%%%%%%%%%%%%%%%%%%%%%%%%%%%%%%%%%%%%%%%%%%%%%%%%%%%%%%%%%%%

\begin{figure*}  
        \centering
        \begin{minipage}[b]{ 0.85\textwidth}   
                \centering
                \includegraphics[width=1.0\textwidth]{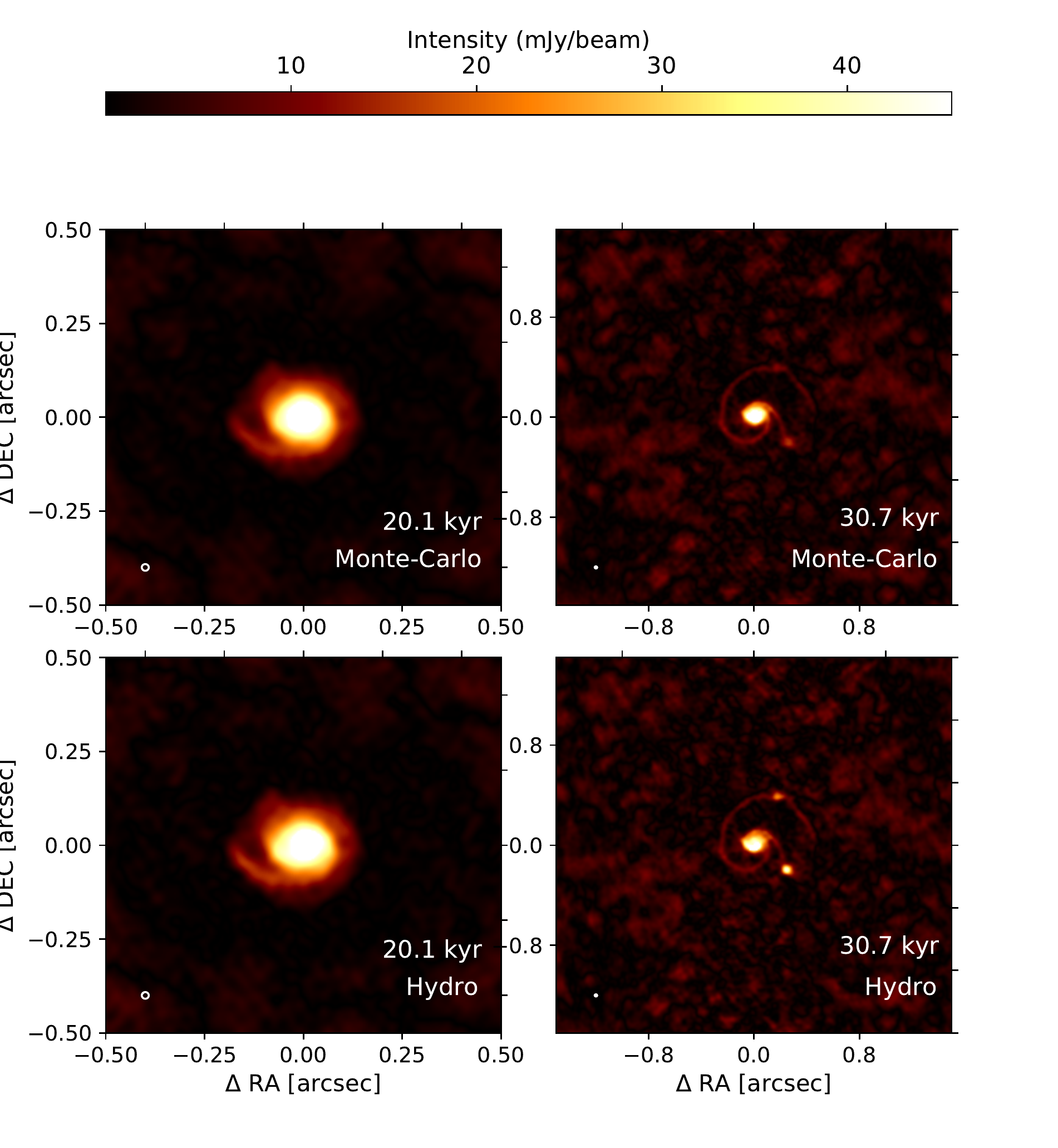}
        \end{minipage}    
        \caption{ 
        		 Comparison between simulated {\sc alma} Cycle 7 band 6 ($1.2\, \rm mm$) images with antenna 
        		 configuration 10 of stable (left) and fragmented (right) disc models, respectively. 
        		 Differences lie in the dust temperature field calculation method, either 
        		 by Monte-Carlo simulation (top) or by importing the dust temperatures from the 
        		 hydrodynamics simulations (bottom). 
        		 The images are shown for face-on viewing geometry.  
                 The beam size is indicated in the lower left corner of each panel and  
                 the source distance is located at $1\, \rm kpc$. 
                 }      
        \label{fig:tdust}  
\end{figure*}

\section{Discussion}
\label{sect:discussion}

This last section is devoted to the discussion of our synthetic observations of fragmenting 
circumstellar discs around MYSOs. We first review the limitations and caveats of our methods, 
investigate how the effects of the integration time of the interferometric observations and of 
the {\sc alma} facility beam resolution can affect the detection of substructures in the disc. 
We briefly compare our results with precedent studies and we discuss future prospects for observing 
gravitational instability and disc fragmentation around MYSOs. 
Finally, we discuss our results in the context of high-mass protostellar disc-hosting 
candidates.

\subsection{Limitations of the model}   
\label{sect:caveats}

Our models of synthetic images are affected by two kinds of caveats: on the one hand the numerical
hydrodynamics simulations of our forming accretion discs are intrinsically subject to limitations 
justifying future improvements, on the other hand the radiation transfer methods that we use 
in this study might, in their turn, be improved as well. 
The principal limitation of our models comes from the restrictions on the hydrodynamical timestep, controlled 
by the standard Courant-Friedrichs-Levy parameter~\citep{mignone_apj_170_2007,migmone_apjs_198_2012}. It 
decreases proportionally to the smallest grid size of the grid mapping the computational domain, and, 
therefore gets smaller as the spatial resolution augments.
The sink cell radius also influences the speed of the time-marching algorithm in our simulations. It 
should be as small as possible since we study the accretion of inward-migrating disc fragments by 
the protostellar surroundings. In this study, we adopt the smallest value ($r_{\rm in}=10\, \rm au$) 
where the simulations are still computationally affordable, which is still smaller than in recent works 
on disc fragmentation. 
Nevertheless, the most critical point of the disc simulations in spherical coordinate systems remains 
the effect of the inertia \textcolor{black}{of the central star to the disc dynamics, which is neglected when the high-mass 
protostar is fixed at the origin of the computational domain, see also discussion in~\citet{meyer_mnras_482_2019}}. The stellar motion in response to the gravitational
force of the disc has been shown not only to reduce the strength of gravitational instability (but not shutting 
it down completely) in the low-mass regime of star formation but 
also to introduce unpleasant boundary effects~\citep{regaly_aa_601_2017}, as it displaces the 
barycenter of the star-disc system from the origin of the domain. 
This is in accordance with the analytic study of~\citet{adam_apj_347_1989} that pointed out the role of 
stellar inertia in the development of asymmetries in discs. The study of~\citet{meyer_mnras_482_2019} demonstrated 
that the disc inertial force does not entirely stabilise discs or shuts off the burst phases of the protostellar 
growth, in accordance with results of~\citet{regaly_aa_601_2017}. 
The assumptions in our numerical simulations of the disc structures are presented and discussed in 
detail in~\citet{meyer_mnras_473_2018} and~\citet{meyer_mnras_482_2019}.

\begin{figure}  
        \centering
        \begin{minipage}[b]{ 0.55\textwidth}
                \centering
                \includegraphics[width=1.0\textwidth]{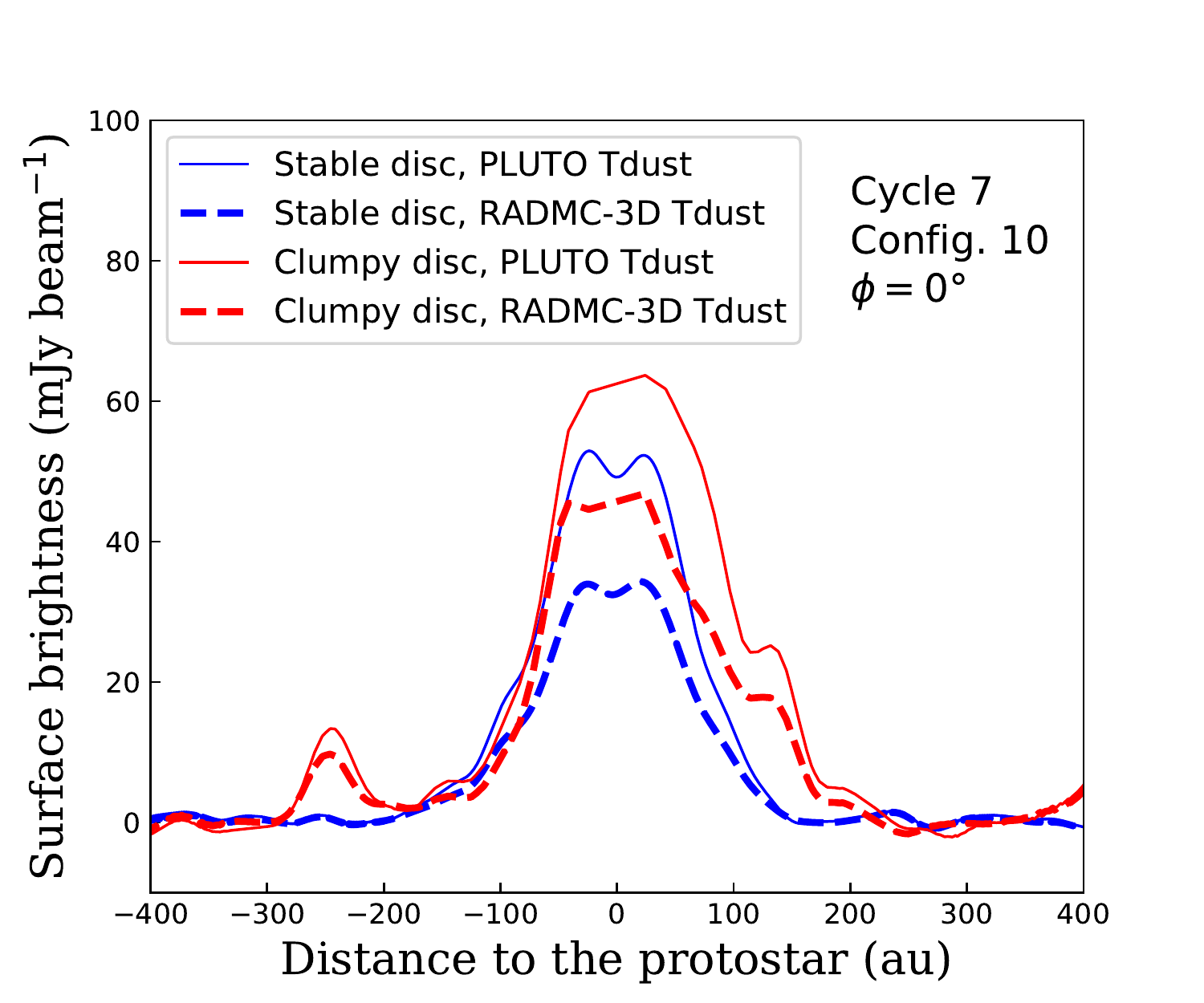}
        \end{minipage} 
        \caption{ 
        	Comparison between the cross-sections of simulated {\sc alma} Cycle 7 band 6 
        	($1.2\, \rm mm$) antenna configuration 10 images of 
        	our snapshots 1 (stable disc model, blue) and 4 (fragmented disc model, red),  
        	in which the dust temperature is either estimated using \textcolor{black}{a radiative transfer calculation with the {\sc radmc-3d} code (thick 
        	dashed lines) or estimated on-the-fly with the hydrodynamical {\sc pluto} code and taking account compressional heating (thin solid lines)}. 
        	Note that the left and right series of images are not plotted on the same physical size, respectively. 
                The source distance is located at $1\, \rm kpc$. 
                 }      
        \label{fig:tdust_cut}  
\end{figure}

The dust temperature is a key parameter governing the brightness of the accretion discs. 
Fig.~\ref{fig:tdust} compares simulated ALMA cycle 6 (1.2 mm) images with an antenna configuration 10 of a 
stable (left) and a fragmented (right) disc model, where the radiation transfer calculations 
are performed with different methods to compute the dust temperature fields: either by Monte-Carlo 
in the {\sc radmc-3d} code (top panels) or using the dust temperatures from the hydrodynamics simulations (bottom 
panels). The images are shown for face-on viewing angle ($\phi=0\degree$). 
The beam size is indicated in the lower left corner of each panel and the source distance 
is $1\, \rm kpc$.
It appears that the synthetic {\sc alma} images of the stable model at time $20.1\, \rm kyr$ 
do not exhibit strong differences if modelled on the basis of either of a Monte-Carlo method or using 
the temperature fields imported from the hydrodynamical simulations (left panels of Fig.~\ref{fig:tdust}). 
Nevertheless, differences are observed with respect to the observability of the clumps of fragmented discs (right panels of Fig.~\ref{fig:tdust}). 
Such changes in the simulated images are due to the fact that the hydrodynamical temperature field results in 
a self-consistent simulation, including the treatment of the mechanical $pdV$ work from gravity forces in 
the disc, i.e. the internal thermodynamics of the clumps are taken into account. 
The cores of these clumps experience an internal gravitational collapse~\citep{meyer_mnras_473_2018}, 
while a Monte-Carlo calculation of the temperature fields is less realistic as it simply ray-traces photon packages 
through the clumps without accounting for their thermal properties. 
This results in brighter $1.2\, \rm mm$ clumps in the spiral arms of the fragmented discs (bottom right panel of 
Fig.~\ref{fig:tdust}), and, consequently, in a higher brightness and better observability at the 
considered sub-millimetre wavelength. 
This effect was also reported in~\citet{dong_apj_823_2016} in the context of fragmented disks around low-mass protostars.

\begin{figure}  
        \centering
        \begin{minipage}[b]{ 0.45\textwidth}   
                \centering
                \includegraphics[width=1.0\textwidth]{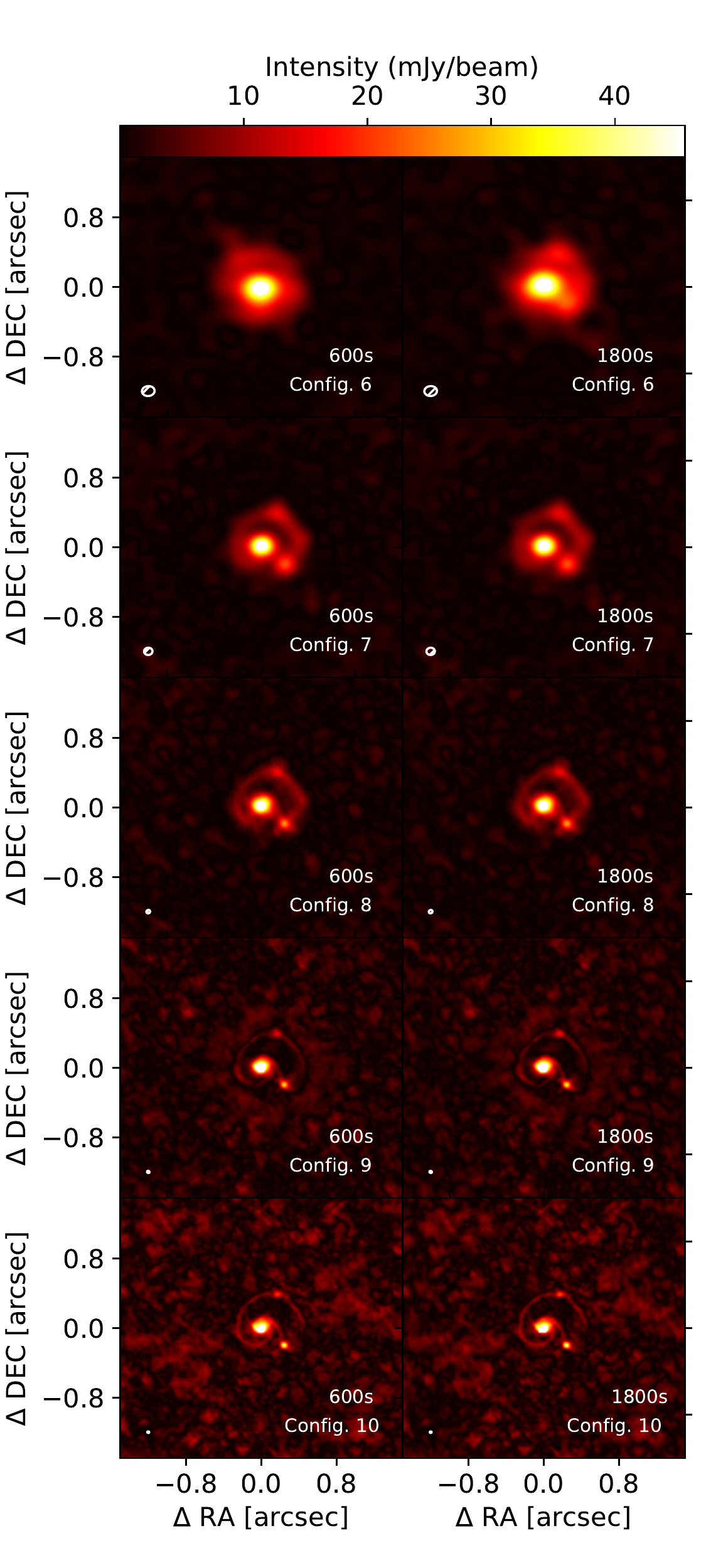}
        \end{minipage}   
        \caption{ 
                 Effects of the antenna configuration and integration time on the simulated 
                 images of our fragmented accretion disc at time $30.7\, \rm kyr$. 
        		 Upper panels are the most compact antenna configuration (Cycle 7 band 6 with antenna configuration 6) and lower 
        		 panels the most extended (Cycle 7 band 6 with antenna configuration 10), respectively. 
        		 The integration time is $600\, \rm s$ (left) and $1800\, \rm s$ (right). 
        		 The images are shown assuming face-on viewing geometry.  
                 The beam size is indicated in the lower left corner of each panel and  
                 the source distance is located at $1\, \rm kpc$. 
                 }      
        \label{fig:config}  
\end{figure}

Fig.~\ref{fig:tdust_cut} compares several of the cross-sections taken through our simulated ALMA 
cycle 6 ($1.2\, \rm mm$) antenna configuration 10 images displayed in Fig.~\ref{fig:tdust}. Two disc models 
are compared: a first one corresponding to a time $20.1\, \rm kyr$ when the disc is stable (blue lines) and another one at time 
$30.7\, \rm kyr$ when the disc fragments (red), in which the dust temperature is estimated 
by Monte-Carlo simulations (thin solid lines) and by thermal equilibrium (thick dashed lines). 
\textcolor{black}{
It further highlights the differences between the two methods for the determination of the disc dust temperature. 
One can see that the simulated images performed with two consecutive Monte-Carlo ray-tracings ({\sc RADMC-3D})  
are slightly fainter in intensity than that carried out on the basis of {\sc PLUTO} temperatures, e.g. 
the central part of the {\sc RADMC-3D} image peaks at $\approx 35\, \rm mJy/beam$ while the {\sc PLUTO} peaks at 
$\approx 50\, \rm mJy/beam$, respectively. 
As another example, the peaks intensities in the clumps calculated with {\sc RADMC-3D} and {\sc PLUTO} temperatures 
are $\approx 10$ and $\approx 14\, \rm mJy/beam$, respectively, which is slightly more noticeable than at the location 
of the central high-mass protostar. 
%
%
%One has also to keep in mind that ray-tracing methods are subject to local, numerical overestimation of the dust 
%temperature resulting in the production of temperatures larger than the sublimation temperature threshold of 
%$\approx 1500\, \rm K$~\citep{kuffmeier_mnras_4754_2018}. 
%
Our simulations with {\sc radmc-3d} tends to underestimate the surface brightness in the synthetic images. }
As a conclusion, we stress the need for the use of self-consistently determined temperatures for the modelling of emission maps of accretion discs of massive protostars.

%\begin{figure}  
%        \begin{minipage}[b]{ 0.5\textwidth}
%               \centering
%                \includegraphics[width=1.0\textwidth]{./cut_convergency_comparison.pdf}
%        \end{minipage} 
%        %
%        \caption{ 
%                 Top panel compares the cross-sections of simulated {\sc alma} cycle 6 
%        		 ($1.2\, \rm mm$) images of a fragmented accretion disc, simulated with 
%        		 various antenna configurations (8 and 10) and different integration 
%        		 times (600s and 1800s).
%        		 %
%        		 Bottom panel compares the intensity of simulated {\sc alma} cycle 6 
%        		 ($1.2\, \rm mm$) when considered with different antenna configurations. 
%        		 %
%                 The source distance is located at $1\, \rm kpc$. 
%                 }      
%        \label{fig:config_cuts}  
%\end{figure}

\subsection{Effect of integration time and angular resolution}    
\label{sect:effects}

Fig.~\ref{fig:config} compares the effects of the antenna configuration and of the integration 
time on the simulated images of our fragmented accretion disc at time $30.7\, \rm kyr$. 
The top panels are simulated with the most compact antenna configuration (cycle 6.6) and lower 
panels are simulated with the most extended configuration (cycle 6.10), respectively. 
The integration time is $600\, \rm s$ 
(left) and $1800\, \rm s$ (right). All images are shown assuming a face-on viewing angle of the discs. 
The beam size is indicated in the lower left corner 
of each panel and the source is located at $1\, \rm kpc$. 
The series of tests highlight that the more extended the antenna configuration, i.e. the smaller 
the beam scanning a region of the sky, the better visible are the substructures inside the protostellar disc. 
If the images performed with $600\, \rm s$ of integration time and an antenna configuration 6 do 
not reveal clumps in the disc, those with antenna configuration $\ge 7$ do. The latter show  
both the distinct features of two clumps which are brighter than the background, and it equivalently reveals 
the main spiral arm of the disc. Note that the best available beam resolution (bottom panels) even shows the 
asymmetric character of some clumps. 
The two series of panels comparing the {\sc alma} integration time of $600\, \rm s$ (right) and $1800\, \rm s$ 
(left) indicate that the more extended the antenna configuration, the faster the convergence of the solution 
in terms of exposure time. Indeed, the more compact antenna configuration 6 (top series of panel) shows that the 
image modeled with $1800\, \rm s$ reveals a clump above the protostar that is not visible with only $600\, \rm s$ 
integration time, while the other images with the antenna configuration $\ge 7$ do not exhibit such 
differences, i.e. the beam resolution is more important than the integration time. 
Similarly, we derive a reasonable exposure time to observe sources located at $2\, \rm kpc$ to be $30\, \rm min$.

%Fig.~\ref{fig:config_cuts} plots cross-sections taken through the synthetic images displayed in 
%Fig.~\ref{fig:config}. The top panel shows the normalised intensity for two different antenna 
%configuration, i.e. 7 (blue lines) and 10 (red lines), and with two different integration times of 
%$600\, \rm s$ (thin solid lines) and $1800\, \rm s$ (thick dashed lines), respectively. 
%The bottom panel compares the cross-sections taken through all the models simulated with 
%$600\, \rm s$ with different antenna configurations (from 6 to 10). 
%
%The top panel of the figure shows that the integration time is converged for images simulated with an antenna 
%configuration 7 or above, as long as the integration time is $600\, \rm s$ or longer. This applies to the 
%fragmented discs exhibiting substructures in the protostellar surroundings. 
%
%\textcolor{black}{
%The bottom panel illustrates that the smaller the beam size (the higher the spatial resolution), the smaller the 
%accumulated intensity, i.e. an exposure time of $600\, \rm s$ with the antenna configuration 6 reaches a 
%surface brightness of $\approx 600\, \rm mJy\, \rm beam^{-1}$ while it is $\le 100\, \rm mJy\, \rm beam^{-1}$ 
%for the antenna configuration 9 or 10 (red dotted lines). On the other hand, the resolution of the disc substructures 
%increases with the beam size. 
%}

\begin{table*}
	\centering
	\caption{
	Adopted properties of the bursting MYSOs S255IR-NIRS3~\citep{caratti_nature_2016} and (b) NGC 6334I-MM1~\citep{hunter_apj_837_2017}.   
	}
	\begin{tabular}{lcccccccr}
	\hline
	${\rm {Name}}$  &  $M_{\star}$ ($\rm M_{\odot}$)    &  $L_{\star}$ ($\rm L_{\odot}$)  &  $R_{\star}$ ($\rm R_{\odot}$)  &  $T_{\rm eff}$ ($\rm K$)  &  $d$ ($\rm kpc$)  & $\phi$ & References\\ 
	\hline    
	S255IR-NIRS3   &  $20$  &  $2.9\times 10^{4}$       &  $~~10$    &  $23800^{a}$   &  $1.8$  &  $80\degree$        & \citet{caratti_nature_2016},~\citet{meyer_mnras_482_2019} \\  
	NGC 6334I-MM1  &  $20$  &  $~~1.5\times 10^{5a}$    &  $350$     &  $6100$        &  $1.3$  &  $~~\phi^{b}$  & \citet{forgan_mnras_463_2016},~\citet{hunter_apj_837_2017} \\           
	\hline    
%	\hline 
	\end{tabular}
\label{tab:mysos}\\
\footnotesize{ ${(a)}$Derived assuming that the MYSO just experienced a strong accretion burst as in Fig.~\ref{fig:star}, ${(b)}$Undetermined. }\\
\end{table*}

\begin{figure}  
        \centering
        \begin{minipage}[b]{ 0.5\textwidth}
                \includegraphics[width=1.0\textwidth]{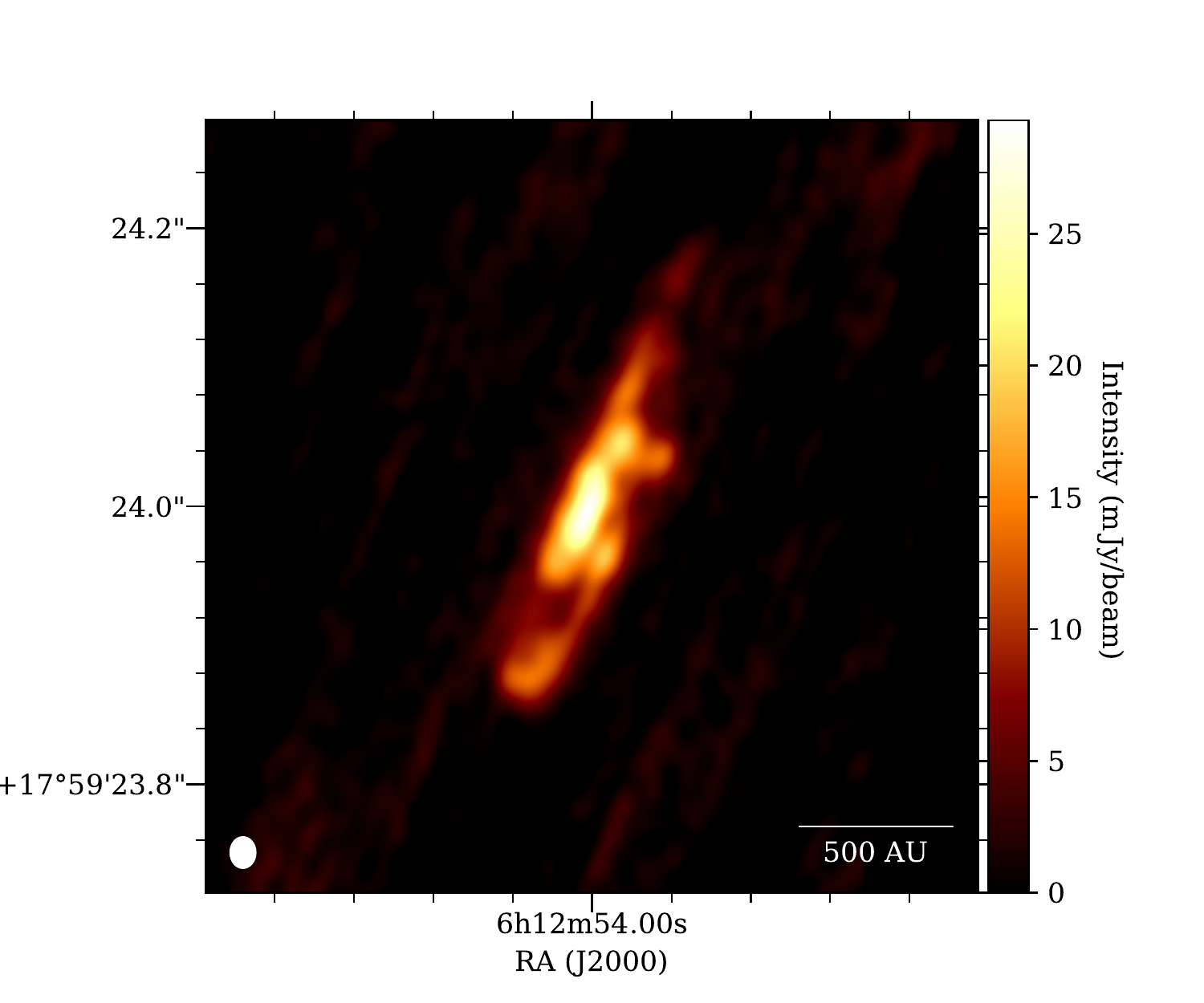}
        \end{minipage}        
        \caption{ 
                 Dust continuum {\sc alma} Cycle 7 C43-10 at band 6 ($1.2\, \rm mm$) predictions for the observability of disc 
                 substructures in the context of the bursting young high-mass star  
                 S255IR-NIRS3~\citep{caratti_nature_2016}.      
                 }      
        \label{fig:predictions}  
\end{figure}

\subsection{Comparison with other studies} 
\label{sect:studies}

A couple of previous works have performed synthetic imaging of both analytic and self-consistent hydrodynamical simulations 
of forming massive stars at the disc scale ($\le 1000\, \rm au$), see~\citet{krumholz_apj_665_2007},~\citet{harries_2017},~\citet{meyer_mnras_473_2018} 
and~\citet{jankovic_mnras_482_2019}. 
First, the paper of~\citet{krumholz_apj_665_2007} investigated synthetic {\sc alma} and {\sc noema} images of disc structures 
forming from collapsing turbulent pre-stellar cores located at $\le 2\, \rm kpc$, in which secondary star formation is 
tracked with the used of Lagrangian sink particles whose caveats are discussed in~\citet{meyer_mnras_473_2018}. Our models therefore provide 
more accurate density and temperature disc structures, as the gaseous clumps and secondary low-mass stars are self-consistently produced and 
their internal thermodynamics is intrinsically taken into account. 
Secondly,~\citet{harries_2017} carried out state-of-art gravito-radiation-chemo-hydrodynamical simulations and 
checked the disc spiral arm observability with simulated $1\, \rm mm$ {\sc alma} interferometric observations. 
This study includes the most accurate treatment of the radiative protostellar feedback to-date, however, its disc 
models neither strongly fragment nor produce the formation of clumps because of a lack of spatial resolution.

In the earlier studies of this series devoted to the burst mode of massive star formation,~\citet{meyer_mnras_473_2018} 
showed that spiral arms can be observed in continuum and molecular emission for a fragmenting 
disc located at a distance and with a inclination corresponding to the disk around the MYSO AFGL-4176~\citep{johnston_apj_813_2015}.  
Last, the work of~\citet{jankovic_mnras_482_2019} predicts the observability of spiral arms and clumps in discs around 
massive stars by computing images of analytic disc structures. They predict their observability up to a 
distance $\le 5\, \rm kpc$ for arbitrary inclinations by direct imaging discs with $\phi \le 50\degree$ and/or by their 
kinematic signature. Our results show that clumps can, under some circumstances, be directly imaged or at least inferred 
for edge-viewed discs. 
These studies highlight the possible observation of circumstellar discs around young massive stars. Our work allows for the 
first time to obtain self-consistent solution from the hydrodynamical, stellar evolution and radiative transfer point 
of view, which permits qualitative predictions relative to the emission properties of disc substructures.

\subsection{Prospect for observing disc fragmentation around MYSOs: the cases of S255IR-NIRS3 and NGC 6334I-MM1} 
\label{sect:prospect}

In this sub-section, we discuss the prospects for the potential observability of signatures of gravitational 
instabilities in the surroundings of outbursting sources. We focus on the two high-mass protostars for which 
potentially accretion-driven bursts have been observed, namely S255IR-NIRS3~\citep{caratti_nature_2016} and 
NGC 6334I-MM1~\citep{hunter_apj_837_2017}. Note that a third burst from a MYSO, V723 Car, has been reported 
in~\citet{tapia_mnras_446_2015}, however, its discovery in archival data did not permit further monitoring of it. 
S255IR-NIRS3 is the very first massive protostar observed experiencing a disc-mediated burst and it is 
therefore a natural target to investigate the possible presence of other clumps in its disc. NGC 6334I-MM1 is 
the second known bursting MYSO and it exhibited a successive quadruple temporal variation in the dust continuum 
emission which suggests that this eruptive event is stronger in intensity than that of S255IR-NIRS3. This makes 
NGC 6334I-MM1 the second most obvious candidate to chase for substructures in its close surroundings. A summary of the recent research 
concerning these two MYSOs is given in Section~5.4 of~\citet{meyer_mnras_482_2019}. 
The predictions for their circumstellar medium are performed following the comparison 
method utilised for the proto-O-type star AFGL 4176 in~\citet{meyer_mnras_473_2018}. We select a simulation snapshot of 
our hydrodynamical model at a time corresponding to the estimated masses of the MYSOs and run radiative transfer 
calculations as above described, with {\sc radmc3d} and {\sc casa}. The dust continuum models are run using the 
values for $L_{\star}$, $R_{\star}$ and $T_{\rm eff}$ either constrained by observations or derived from our stellar 
evolution calculation, and we account for the position of the MYSO on the sky, its distance to the observer and, when possible, 
the inclination angle of the disc with respect to the plane of sky.

\begin{figure}  
        \begin{minipage}[b]{ 0.5\textwidth}
                \includegraphics[width=1.0\textwidth]{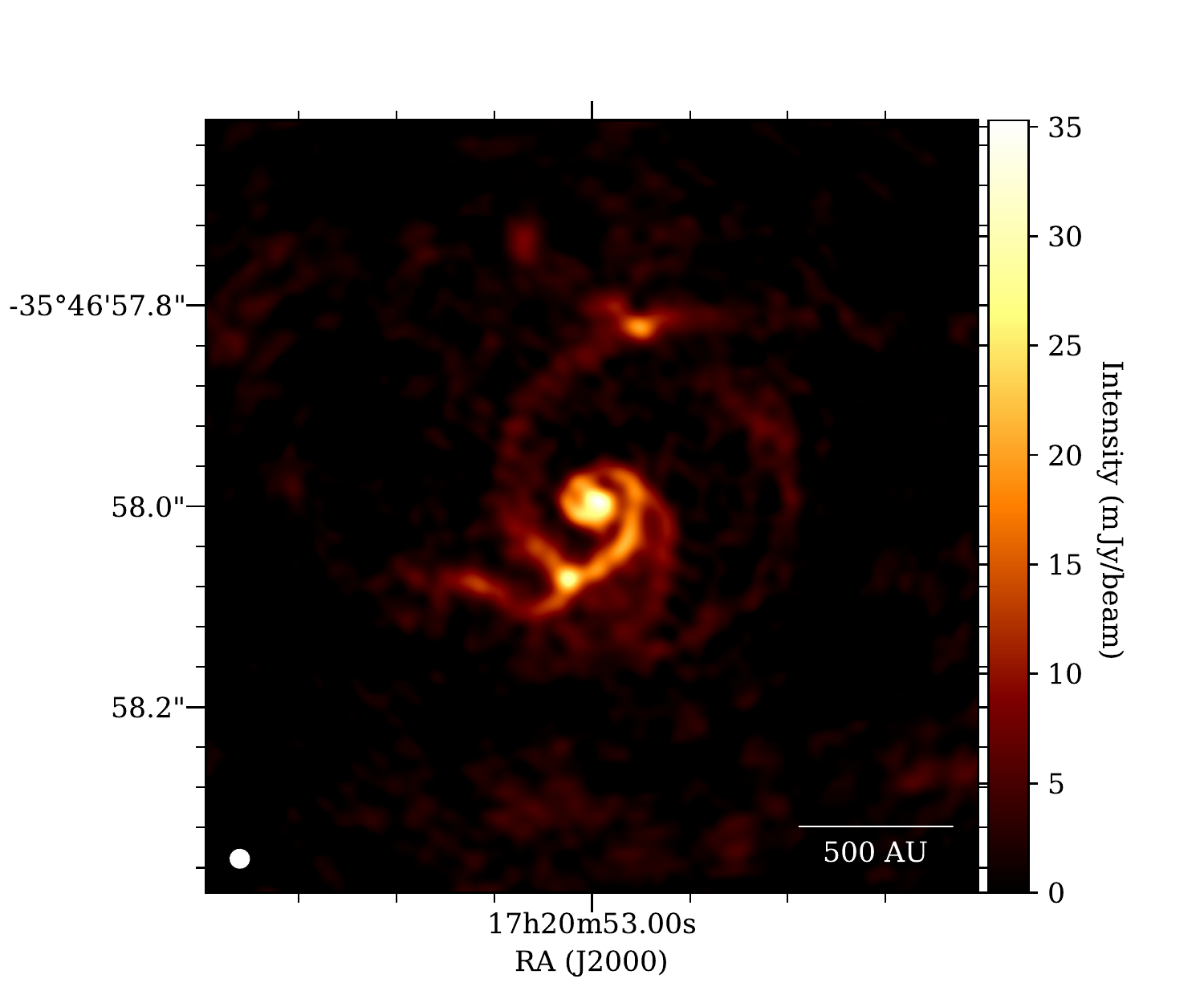}
        \end{minipage} 
        \centering     
        \begin{minipage}[b]{ 0.5\textwidth}
                \includegraphics[width=1.0\textwidth]{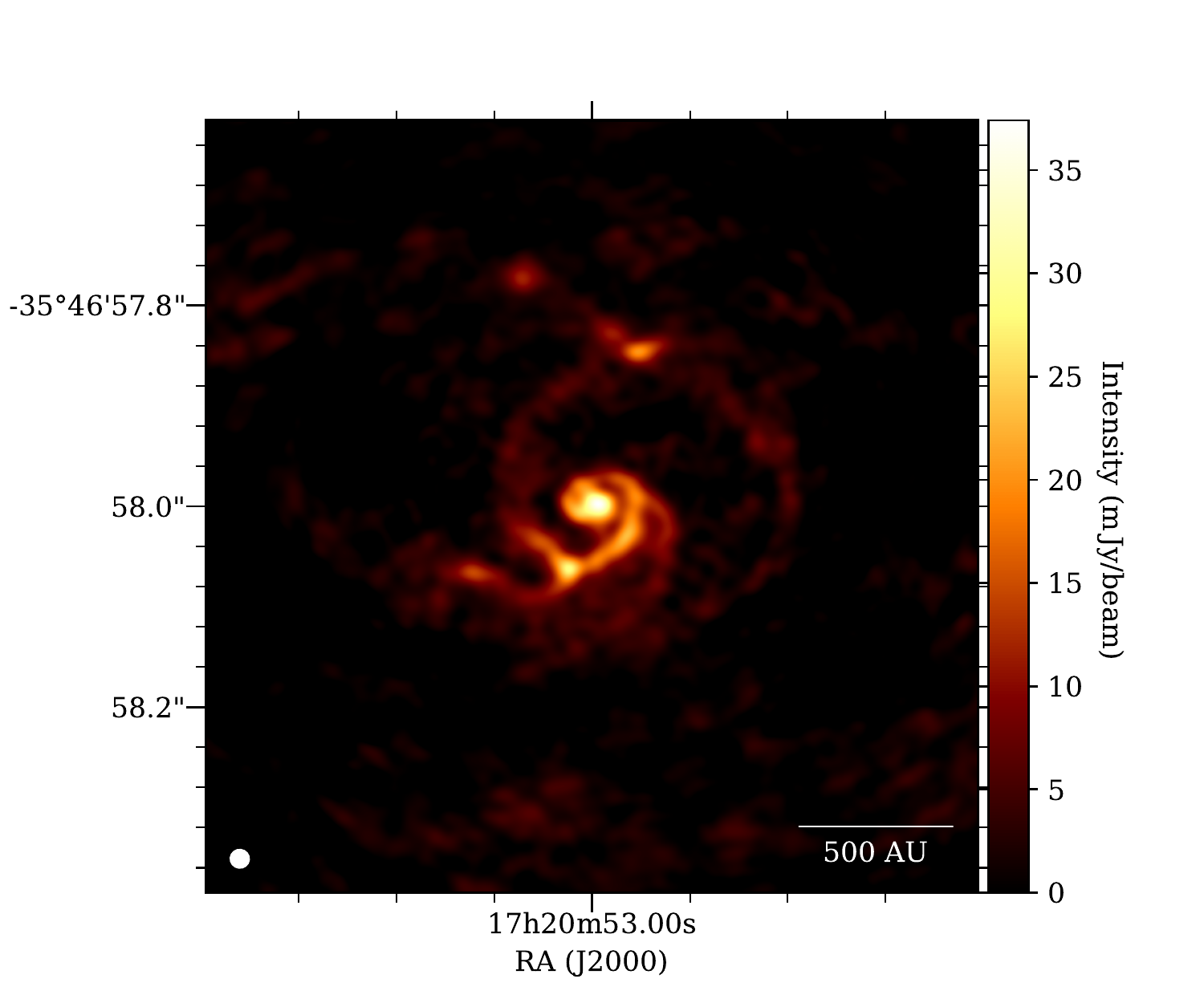}
        \end{minipage}  
        \centering     
        \begin{minipage}[b]{ 0.5\textwidth}
                \includegraphics[width=1.0\textwidth]{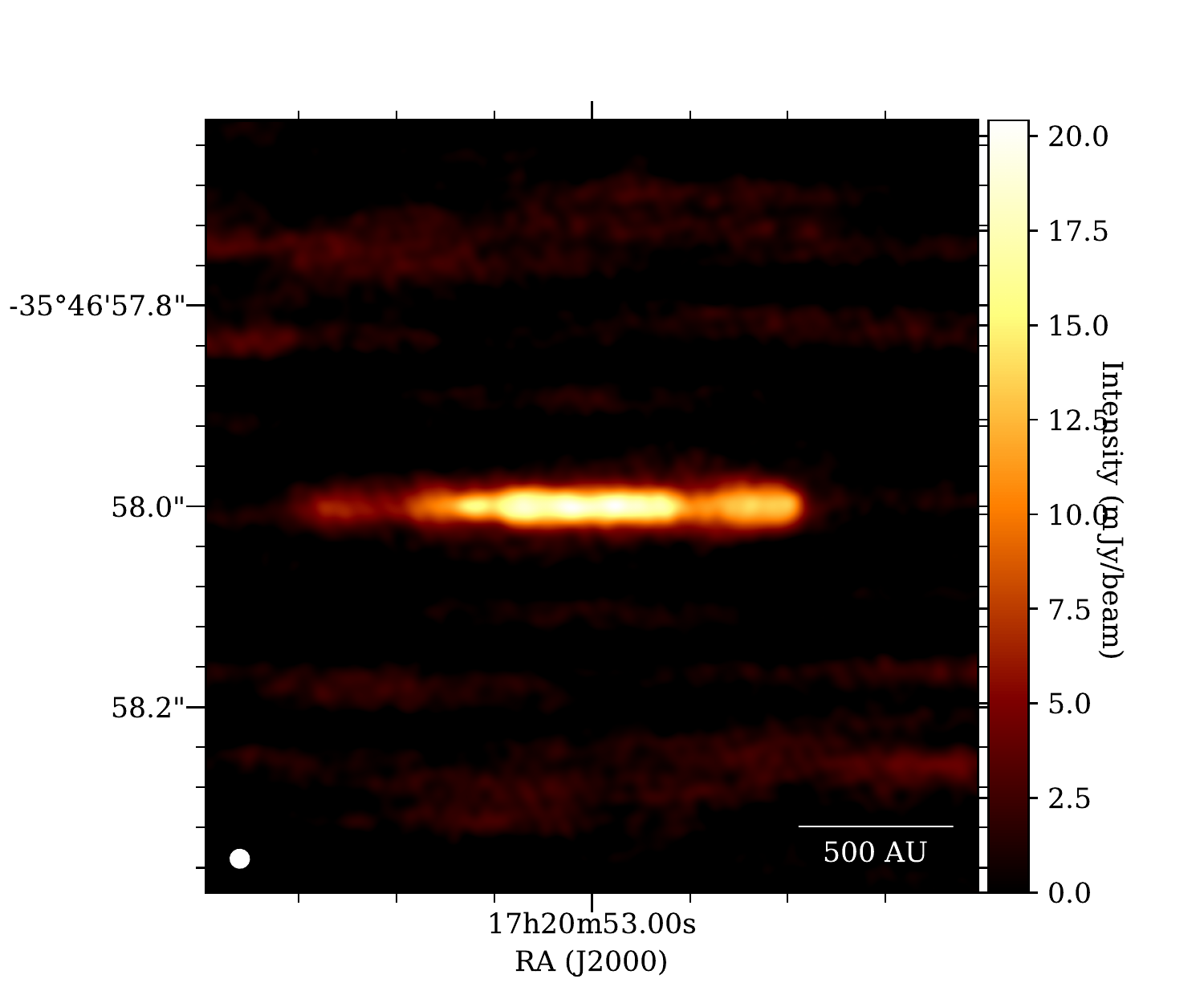}
        \end{minipage}         
        \caption{ 
                 Dust continuum {\sc alma} Cycle 7 C43-10 at band 6 ($1.2\, \rm mm$) predictions for the observability of disc 
                 substructures in the context of the bursting young high-mass star 
                 NGC 6334I-MM1~\citep{hunter_apj_837_2017}. The viewing angle of the disc, not constrained by 
                 observations, is hereby assumed to be $0\degree$ (top), $30\degree$ (middle) and $90\degree$ 
                 (bottom), respectively. 
                 }      
        \label{fig:predictions2}  
\end{figure}

We adopt the stellar parameters of S255IR-NIRS3 and NGC 6334I-MM1 summarised in the 
studies of~\citet{caratti_nature_2016},~\citet{eisloeffel_apss_233_1995},~\citet{bogelund_aa_615_2018} 
and~\citet{hunter_apj_837_2017}. 
For S255IR-NIRS3, we consider that it has recovered the pre-burst quiescent value of $L_{\star}$, although 
our stellar evolution calculation predict a rather slow post-burst luminosity decay lasting $\simeq 10^{2}\, \rm yr$, 
see our Fig.~\ref{fig:star} and~\citet{2019MNRAS.tmp...10M}. 
Our choice to adopt protostellar parameters fainter than that of Run-1-hr is justified because we have classified the flare 
of S255IR-NIRS3 to be a modest 2-mag burst in~\citet{meyer_mnras_482_2019} which can not be responsible for 
long-lasting spectral excursions in the Hertzsprung-Russell diagram as experienced by the MYSO of Run-1-hr, 
see also~\citet{2019MNRAS.tmp...10M}. 
Concerning NGC 6334I-MM1 which experienced a strong burst, one has to keep in mind than the values of $R_{\star}$ 
in~\citet{forgan_mnras_463_2016} do not account for the burst-induced protostellar bloating. However, the values of $T_{\rm eff}$ 
are fully consistent with our stellar evolution calculation. We correct this discrepancy by assuming a larger value of the protostellar radius, 
taken from Run-1-hr, which in its turn also affects $L_{\star}$. Hence, the only unknown is the inclination angle $\phi$ of the possible 
disc of NGC 6334I-MM1 with respect to the plane of sky. 
The various parameters of these two MYSOs are summarised in Tab.~\ref{tab:mysos} and our results are shown 
in Figs.~\ref{fig:predictions} and~\ref{fig:predictions2}.

The Cycle 7 band 6 synthetic observation of S255IR-NIRS3 is plotted in Fig.~\ref{fig:predictions}. The 
disk position angle is chosen to match the direction of the two lobes inside which material is expelled during the 
burst in~\citet{caratti_nature_2016}. It is interesting to see that a few clumps in the disc can still be 
observed despite of the almost edge-on view of the disc around S255IR-NIRS3. 
The Cycle 7 band 6 synthetic observation of NGC 6334I-MM1 is shown in Fig.~\ref{fig:predictions2}. As the 
inclination of its disc is not known, we assumed several angles ($\phi=0\degree$, $30\degree$ and $90\degree$). 
We conclude for all the considered angles $\phi$ that substructures can easily be observed in the disc and its surroundings.  
Hence, these two targets are good candidates for further exploration of substructures in their circumstellar discs. 
Our results also permit to speculate that NGC 6334I-MM1 might be recovering its quiescent, luminous ground state after a 
spectral excursion in the red part of the Hertzsprung-Russell. Consequently, this object is also a candidate to test our 
prediction regarding the intermittency of the \hii region surrounding outbursting MYSOs that was predicted in~\citet{2019MNRAS.tmp...10M}.

One should keep in mind that the two MYSOs that we selected to study the observability of their (possibly substructured) 
accretion discs are the two MYSOs which had a known burst. This was intentional, as the burst mode picture 
of star formation predicts a direct correspondance between the occurrence and properties of the burst 
with the effects of gravitational instability in those discs~\citep{meyer_mnras_464_2017}. 
However, all MYSOs may have an accretion disc and the absence of observed bursts in most of them can simply be interpreted 
as them being in a quiescent phase of accretion, not as the absence of discs or even not as the absence of fragments 
in their discs. Indeed, the numerical models for bursting massive protostars 
of~\citet{meyer_mnras_482_2019} indicate that MYSOs spend $<2\, \%$ of their early $60\, \rm kyr$ being in the   
eruptive state. Consequently, the other young massive stars with candidate discs might also be worth observing 
with future high-resolution {\sc alma} campaigns. 
Possible targets are listed in the observational studies reporting the Red MSW Source survey of Galactic hot massive 
protostars~\citep{lumsden_apjs_208_2013}. The closest ones, located at $\approx\, 1.4\, \rm pc$ are G076.3829-00.6210 (S106) and 
G080.9383-00.1268 (IRAS 20376+4109) while a few others are at a distance of $\approx\, 2.0\, \rm pc$.  
For instance, G015.0357-00.6795 (M17), the W3IRS objects, G192.5843-00.0417 (G192.58) might deserve particular 
attention for the observer observational astronomical concentrating on multi-epoch VLBI (very long baseline interferometry) 
observations of maser emission.

Last, we should also mention that there is a growing suspicion of correlations between the maser emission  
from MYSOs and their recent accretion activity, as maser flares originating from the surroundings of massive 
protostars continue to be monitored~\citep{burns_mnras_460_2016,burns_mnras_467_2017,macloed_mnras_478_2018}. This 
promising research avenue might investigate, amongst others, if maser variability is an outcome of 
episodic accretion. Such results would be of great interest not only in the understanding of massive star formation 
mechanisms but also a huge leap forward in the validity of the burst mode picture of star formation.

%%%%%%%%%%%%%%%%%%%%%%%%%%%%%%%%%%%%%%%%%%%%%%%%%%%%%%%%%%%%%%%%%%%%%%%%%%%%%%%%%%%%%%%%%%%
%%%%%%%%%%%%%%%%%%%%%%%%%%%%%%%%%%%%%%%%%%%%%%%%%%%%%%%%%%%%%%%%%%%%%%%%%%%%%%%%%%%%%%%%%%%
%%%%%%%%%%%%%%%%%%%%%%%%%%%%%%%%%%%%%%%%%%%%%%%%%%%%%%%%%%%%%%%%%%%%%%%%%%%%%%%%%%%%%%%%%%%

\section{Conclusions}
\label{sect:cc}

This work explores the observability of a fragmenting 
accretion disc surrounding massive young stellar objects (MYSOs), as modelled in~\citet{meyer_mnras_464_2017,meyer_mnras_473_2018}, 
using the {\sc alma} interferometer. Such discs are subject 
to efficient gravitational instabilities, generating the formation of dense spiral arms and gaseous clumps, which 
episodically inward-migrate to the inner disc region. Some of them further contract up to become the precursors of 
close/spectroscopic binary companions to the central protostar~\citep{meyer_mnras_473_2018} and/or (simultaneously) 
provoke accretion driven outbursts of various intensities~\citep{meyer_mnras_482_2019}. 
Representative simulation snapshots exhibiting the characteristic formation and evolutionary stages of the disc 
are selected from model run-1-hr of~\citet{meyer_mnras_473_2018} which results from the gravitational 
collapse of a $100\, \rm M_{\odot}$ molecular pre-stellar core. 
We performed stellar evolution and radiative transfer calculations to produce {\sc alma} at band 6 ($1.2\, \rm mm$) emission 
maps of the discs and obtain synthetic images. These images are then used to investigate the observability of spiral arms and clumpy fragments.  
In addition to the disc evolution, we explore how its inclination angle with respect to the plane of sky 
affects the $1.2\, \rm mm$ projected emission signatures. We also look at the effects of various {\sc alma} beam resolutions and/or 
antenna configurations to predict the best facility configuration and the ages of accretion discs from MYSOs that are 
the easiest ones to observe.

Our results demonstrate that not only the spiral arms developing in the disc by gravitational instability, but also the gaseous 
clumps can be observed, as their internal density and temperature make them much brighter than the disc itself. 
For sources located at a distance of $1.0\, \rm kpc$, $10\, \rm min$ exposition time with a $0.015^{\prime\prime}$ beam 
resolution permits to clearly distinguish those nascent stars and reveal the multiplicity in the environment of young massive protostars. 
We predict that in the millimeter waveband (band 6), an antenna configuration 8 (corresponding to a beam resolution of 
$0.029^{\prime\prime}$) is necessary to reach a beam resolution required to identify the disc substructures with a rather 
short $10\, \rm min$ exposure time. However, only antenna configuration 10 (corresponding to a beam resolution of 
$0.015^{\prime\prime}$) might allow us to distinguish the finest disc substructures.
Interestingly, our {\sc alma} images also revealed clumps in inclined discs and they can be inferred for edge-on  
accretion discs, as long as these fragments develop at radii $\ge 600\, \rm au$ from the protostar, i.e. they 
are not fully screened by the irradiation of the central massive protostar. In our models, this corresponds to discs 
older than $\approx 26.7\, \rm kyr$ after the onset of the gravitational collapse, when the MYSO has already reached a mass of 
$\approx 16\, \rm M_{\odot}$. Hence, good candidates for hunting disc substructures are MYSOs of $\simeq 20\, \rm M_{\odot}$. 
Our conclusions regarding the observability of substructures in our disc models are valid up to distances of $2\, \rm kpc$, 
using a $0.015^{\prime\prime}$ beam resolution with the {\sc alma} Cycle 7 antenna configuration 10 and at least 
$30\, \rm min$ exposure time.

The detection probability of the structures increases as the disc grows, i.e. accretion discs around protostars close 
to the ZAMS are more extended, and host more clumps in the outer region of the disc that is not screened by the MYSO irradiation. 
Infalling clumps in the inner few $~10-100\, \rm au$ of the disc might not by observable and the probability to directly 
associate a clump with a burst is rather small. 
Last, our study emphasises the necessity to perform simulated images of discs around massive stars by accounting for the 
gaseous clumps thermodynamics by using temperatures self-consistently determined during the hydrodynamics simulations instead of dust 
temperature fields obtained using steady-state radiative transfer models which neglect $pdV$ work.  
%
%We also notice from our previous hydrodynamical models~\citep{meyer_mnras_473_2018} and stellar evolution 
%calculations~\citep{2019MNRAS.tmp...10M} that the observation probability of the clumps does depend on the current accretion 
%rate onto the MYSOs. Discs of hot protostars in (or returning in) the quiescent phase are more likely to exhibit clumps via 
%{\sc ALMA} than discs of MYSOs entering a burst phase. 
%
%Contrarily, MYSOs developing internal luminosity wave and undergoing spectral excursions in the Hertzsprung-Russell diagram~\citep{2019MNRAS.tmp...10M} 
%seems to be less likely to harbor clumpy discs than in the quiescent phase of accretion. 
%
In addition to demonstrating the possible direct imaging of nascent multiple massive protostellar systems with {\sc alma}, we 
also highlight that gaseous clumps and spiral overdensities can clearly be observed/inferred at distances corresponding to the 
bursting massive protostars S255IR-NIRS3~\citep{caratti_nature_2016} and NGC 6334I-MM1~\citep{hunter_apj_837_2017}. 
Those objects might deserve particular future attention from the point of view of high-resolution observations~\citep{zinchenko_apj_810_2015}. 
Similarly, the Earth's closest sources, at $\approx\, 1.4-2.0\, \rm pc$, of the Red MSW Source survey of Galactic hot massive 
protostars~\citep{lumsden_apjs_208_2013} also constitute valuable potential targets. 
As the models for fragmented discs of~\citet{meyer_mnras_464_2017,meyer_mnras_473_2018} predict a pre-main-sequence outburst 
of massive protostars to be the direct outcome of rapidly inward-migrating clumps, we suggest observers to consider their 
respective circumstellar discs as priority targets to search for the substructures which can be signatures of fragmentation by 
gravitational instability in their accretion disc.

%%%%%%%%%%%%%%%%%%%%%%%%%%%%%%%%%%%%%%%%%%%%%%%%%%%%%%%%%%%%%%%%%%%%%%%%%%%%%%%%%%%%%%%%%%%
%%%%%%%%%%%%%%%%%%%%%%%%%%%%%%%%%%%%%%%%%%%%%%%%%%%%%%%%%%%%%%%%%%%%%%%%%%%%%%%%%%%%%%%%%%%
%%%%%%%%%%%%%%%%%%%%%%%%%%%%%%%%%%%%%%%%%%%%%%%%%%%%%%%%%%%%%%%%%%%%%%%%%%%%%%%%%%%%%%%%%%%

\section*{Acknowledgements}

\textcolor{black}{The authors thank the referee for her/his valuable comments which improved the quality of the manuscript.} 
D.~M.-A.~Meyer acknowledges B.~Magnelli from the German ALMA Regional Center (ARC) Node at 
the Argelander-Institut f\" ur Astronomie of the University of Bonn for his help, and 
C.~Dullemond for his advices in using the {\sc radmc-3d} code. D.~M.-A.~Meyer also thanks 
Dr.~Z.~Reg\' aly from Konkoly Observatory for his advices in using the {\sc casa} software 
and Dr.~K.~G.~Johnston for kindly sharing her explanations with the {\sc matplotlib} python 
package. 
S. Kraus and A. Kreplin acknowledge support from an ERC Starting Grant (Grant Agreement No.\ 639889), 
a STFC Rutherford Fellowship (ST/J004030/1) and an STFC Rutherford Grant (ST/K003445/1).
E. I. Vorobyov acknowledges support from the Russian Science Foundation grant 18-12-00193. 
This work was sponsored by the Swiss National Science Foundation (project number 200020-172505).

%%%%%%%%%%%%%%%%%%%%%%%%%%%%%%%%%%%%%%%%%%%%%%%%%%%%%%%%%%%%%%%%%%%%%%%%%%%%%%%%%%%%%%%%%%%
%%%%%%%%%%%%%%%%%%%%%%%%%%%%%%%%%%%%%%%%%%%%%%%%%%%%%%%%%%%%%%%%%%%%%%%%%%%%%%%%%%%%%%%%%%%
%%%%%%%%%%%%%%%%%%%%%%%%%%%%%%%%%%%%%%%%%%%%%%%%%%%%%%%%%%%%%%%%%%%%%%%%%%%%%%%%%%%%%%%%%%%

% Bibliography style for the bibtex
\bibliographystyle{mn2e}

\footnotesize{
% Create the reference Section using BibTeX:
\bibliography{grid}

\begin{thebibliography}{}

\bibitem[\protect\citeauthoryear{{Adams}, {Ruden} \& {Shu}}{{Adams}
  et~al.}{1989}]{adam_apj_347_1989}
{Adams} F.~C.,  {Ruden} S.~P.,    {Shu} F.~H.,  1989, \apj, 347, 959

\bibitem[\protect\citeauthoryear{{Ahmadi}, {Beuther}, {Mottram}, {Bosco},
  {Linz}, {Henning}, {Winters}, {Kuiper}, {Pudritz}, {S{\'a}nchez-Monge},
  {Keto}, {Beltran}, {Bontemps}, {Cesaroni}, {Csengeri}, {Feng} \&
  {Galvan-Madrid}}{{Ahmadi} et~al.}{2018}]{ahmadi_aa_618_2018}
{Ahmadi} A.,  {Beuther} H.,  {Mottram} J.~C.,  {Bosco} F.,  {Linz} H.,
  {Henning} T.,  {Winters} J.~M.,  {Kuiper} R.,  {Pudritz} R.,
  {S{\'a}nchez-Monge} {\'A}.,  {Keto} E.,  {Beltran} M.,  {Bontemps} S.,
  {Cesaroni} R.,  {Csengeri} T.,  {Feng} S.,    {Galvan-Madrid} R. e.~a.,
  2018, \aap, 618, A46

\bibitem[\protect\citeauthoryear{{Asplund}, {Grevesse} \& {Sauval}}{{Asplund}
  et~al.}{2005}]{asplund_ASPC_2005}
{Asplund} M.,  {Grevesse} N.,    {Sauval} A.~J.,  2005, in {Barnes} III T.~G.,
  {Bash} F.~N.,  eds, Cosmic Abundances as Records of Stellar Evolution and
  Nucleosynthesis Vol.~336 of Astronomical Society of the Pacific Conference
  Series, {The Solar Chemical Composition}.
p.~25

\bibitem[\protect\citeauthoryear{{Behrend} \& {Maeder}}{{Behrend} \&
  {Maeder}}{2001}]{behrend_aa_373_2001}
{Behrend} R.,  {Maeder} A.,  2001, \aap, 373, 190

\bibitem[\protect\citeauthoryear{{Beuther}, {Ahmadi}, {Mottram}, {Linz},
  {Maud}, {Henning}, {Kuiper}, {Walsh}, {Johnston} \& {Longmore}}{{Beuther}
  et~al.}{2019}]{2018arXiv181110245B}
{Beuther} H.,  {Ahmadi} A.,  {Mottram} J.~C.,  {Linz} H.,  {Maud} L.~T.,
  {Henning} T.,  {Kuiper} R.,  {Walsh} A.~J.,  {Johnston} K.~G.,    {Longmore}
  S.~N.,  2019, \aap, 621, A122

\bibitem[\protect\citeauthoryear{{Beuther}, {Soler}, {Vlemmings}, {Linz},
  {Henning}, {Kuiper}, {Rao}, {Smith}, {Sakai}, {Johnston}, {Walsh} \&
  {Feng}}{{Beuther} et~al.}{2018}]{beuther_aa_614_2018}
{Beuther} H.,  {Soler} J.~D.,  {Vlemmings} W.,  {Linz} H.,  {Henning} T.,
  {Kuiper} R.,  {Rao} R.,  {Smith} R.,  {Sakai} T.,  {Johnston} K.,  {Walsh}
  A.,    {Feng} S.,  2018, \aap, 614, A64

\bibitem[\protect\citeauthoryear{{B{\o}gelund}, {McGuire}, {Ligterink},
  {Taquet}, {Brogan}, {Hunter}, {Pearson}, {Hogerheijde} \& {van
  Dishoeck}}{{B{\o}gelund} et~al.}{2018}]{bogelund_aa_615_2018}
{B{\o}gelund} E.~G.,  {McGuire} B.~A.,  {Ligterink} N.~F.~W.,  {Taquet} V.,
  {Brogan} C.~L.,  {Hunter} T.~R.,  {Pearson} J.~C.,  {Hogerheijde} M.~R.,
  {van Dishoeck} E.~F.,  2018, \aap, 615, A88

\bibitem[\protect\citeauthoryear{{Brown}, {Wild} \& {Cunningham}}{{Brown}
  et~al.}{2004}]{brown_AdSpR_2004}
{Brown} R.~L.,  {Wild} W.,    {Cunningham} C.,  2004, Advances in Space
  Research, 34, 555

\bibitem[\protect\citeauthoryear{{Burns}}{{Burns}}{2018}]{burns_iaus_336_2018}
{Burns} R.~A.,  2018, in {Tarchi} A.,  {Reid} M.~J.,   {Castangia} P.,  eds,
  Astrophysical Masers: Unlocking the Mysteries of the Universe Vol.~336 of IAU
  Symposium, {Water masers in bowshocks: Addressing the radiation pressure
  problem of massive star formation}.
pp 263--266

\bibitem[\protect\citeauthoryear{{Burns}, {Handa}, {Imai}, {Nagayama},
  {Omodaka}, {Hirota}, {Motogi}, {van Langevelde} \& {Baan}}{{Burns}
  et~al.}{2017}]{burns_mnras_467_2017}
{Burns} R.~A.,  {Handa} T.,  {Imai} H.,  {Nagayama} T.,  {Omodaka} T.,
  {Hirota} T.,  {Motogi} K.,  {van Langevelde} H.~J.,    {Baan} W.~A.,  2017,
  \mnras, 467, 2367

\bibitem[\protect\citeauthoryear{{Burns}, {Handa}, {Nagayama}, {Sunada} \&
  {Omodaka}}{{Burns} et~al.}{2016}]{burns_mnras_460_2016}
{Burns} R.~A.,  {Handa} T.,  {Nagayama} T.,  {Sunada} K.,    {Omodaka} T.,
  2016, \mnras, 460, 283

\bibitem[\protect\citeauthoryear{{Caratti o Garatti}, {Stecklum}, {Garcia
  Lopez}, {Eisloffel}, {Ray}, {Sanna}, {Cesaroni}, {Walmsley}, {Oudmaijer}, {de
  Wit}, {Moscadelli}, {Greiner}, {Krabbe}, {Fischer}, {Klein} \&
  {Ibanez}}{{Caratti o Garatti} et~al.}{2017}]{caratti_nature_2016}
{Caratti o Garatti} A.,  {Stecklum} B.,  {Garcia Lopez} R.,  {Eisloffel} J.,
  {Ray} T.~P.,  {Sanna} A.,  {Cesaroni} R.,  {Walmsley} C.~M.,  {Oudmaijer}
  R.~D.,  {de Wit} W.~J.,  {Moscadelli} L.,  {Greiner} J.,  {Krabbe} A.,
  {Fischer} C.,  {Klein} R.,    {Ibanez} J.~M.,  2017, Nature Physics, 13, 276

\bibitem[\protect\citeauthoryear{{Caratti o Garatti}, {Stecklum}, {Linz},
  {Garcia Lopez} \& {Sanna}}{{Caratti o Garatti}
  et~al.}{2015}]{caratti_aa_573_2015}
{Caratti o Garatti} A.,  {Stecklum} B.,  {Linz} H.,  {Garcia Lopez} R.,
  {Sanna} A.,  2015, \aap, 573, A82

\bibitem[\protect\citeauthoryear{{Cesaroni}, {Hofner}, {Araya} \&
  {Kurtz}}{{Cesaroni} et~al.}{2010}]{cesaroni_aa_509_2010}
{Cesaroni} R.,  {Hofner} P.,  {Araya} E.,    {Kurtz} S.,  2010, \aap, 509, A50

\bibitem[\protect\citeauthoryear{{Chen}, {Ren}, {Zhang}, {Shen} \&
  {Qiu}}{{Chen} et~al.}{2017}]{chen_apj_835_2017}
{Chen} X.,  {Ren} Z.,  {Zhang} Q.,  {Shen} Z.,    {Qiu} K.,  2017, \apj, 835,
  227

\bibitem[\protect\citeauthoryear{{Chini}, {Hoffmeister}, {Nasseri}, {Stahl} \&
  {Zinnecker}}{{Chini} et~al.}{2012}]{chini_424_mnras_2012}
{Chini} R.,  {Hoffmeister} V.~H.,  {Nasseri} A.,  {Stahl} O.,    {Zinnecker}
  H.,  2012, \mnras, 424, 1925

\bibitem[\protect\citeauthoryear{{Cunha}, {Hubeny} \& {Lanz}}{{Cunha}
  et~al.}{2006}]{cunha_apj_647_2006}
{Cunha} K.,  {Hubeny} I.,    {Lanz} T.,  2006, \apjl, 647, L143

\bibitem[\protect\citeauthoryear{{Cunningham}, {Moeckel} \&
  {Bally}}{{Cunningham} et~al.}{2009}]{Cunningham_apj_692_2009}
{Cunningham} N.~J.,  {Moeckel} N.,    {Bally} J.,  2009, \apj, 692, 943

\bibitem[\protect\citeauthoryear{{Dong}, {Vorobyov}, {Pavlyuchenkov}, {Chiang}
  \& {Liu}}{{Dong} et~al.}{2016}]{dong_apj_823_2016}
{Dong} R.,  {Vorobyov} E.,  {Pavlyuchenkov} Y.,  {Chiang} E.,    {Liu} H.~B.,
  2016, \apj, 823, 141

\bibitem[\protect\citeauthoryear{{Dullemond}}{{Dullemond}}{2012}]{dullemond_2012}
{Dullemond} C.~P., , 2012, {RADMC-3D: A multi-purpose radiative transfer tool},
  Astrophysics Source Code Library

\bibitem[\protect\citeauthoryear{{Eggenberger}, {Meynet}, {Maeder}, {Hirschi},
  {Charbonnel}, {Talon} \& {Ekstr{\"o}m}}{{Eggenberger}
  et~al.}{2008}]{eggenberger_apss_316_2008}
{Eggenberger} P.,  {Meynet} G.,  {Maeder} A.,  {Hirschi} R.,  {Charbonnel} C.,
  {Talon} S.,    {Ekstr{\"o}m} S.,  2008, \apss, 316, 43

\bibitem[\protect\citeauthoryear{{Eisl{\"o}ffel} \& {Davis}}{{Eisl{\"o}ffel} \&
  {Davis}}{1995}]{eisloeffel_apss_233_1995}
{Eisl{\"o}ffel} J.,  {Davis} C.~J.,  1995, \apss, 233, 59

\bibitem[\protect\citeauthoryear{{Ekstr{\"o}m}, {Georgy}, {Eggenberger},
  {Meynet}, {Mowlavi}, {Wyttenbach}, {Granada}, {Decressin}, {Hirschi},
  {Frischknecht}, {Charbonnel} \& {Maeder}}{{Ekstr{\"o}m}
  et~al.}{2012}]{ekstroem_aa_537_2012}
{Ekstr{\"o}m} S.,  {Georgy} C.,  {Eggenberger} P.,  {Meynet} G.,  {Mowlavi} N.,
   {Wyttenbach} A.,  {Granada} A.,  {Decressin} T.,  {Hirschi} R.,
  {Frischknecht} U.,  {Charbonnel} C.,    {Maeder} A.,  2012, \aap, 537, A146

\bibitem[\protect\citeauthoryear{{Elbakyan}, {Vorobyov}, {Rab}, {Meyer},
  {G{\"u}del}, {Hosokawa} \& {Yorke}}{{Elbakyan}
  et~al.}{2019}]{elbakyan_mnras_484_2019}
{Elbakyan} V.~G.,  {Vorobyov} E.~I.,  {Rab} C.,  {Meyer} D.~M.-A.,  {G{\"u}del}
  M.,  {Hosokawa} T.,    {Yorke} H.,  2019, \mnras, 484, 146

\bibitem[\protect\citeauthoryear{{Forgan}, {Ilee}, {Cyganowski}, {Brogan} \&
  {Hunter}}{{Forgan} et~al.}{2016}]{forgan_mnras_463_2016}
{Forgan} D.~H.,  {Ilee} J.~D.,  {Cyganowski} C.~J.,  {Brogan} C.~L.,
  {Hunter} T.~R.,  2016, \mnras, 463, 957

\bibitem[\protect\citeauthoryear{{Fujii} \& {Portegies Zwart}}{{Fujii} \&
  {Portegies Zwart}}{2011}]{fujii_sci_334_2011}
{Fujii} M.~S.,  {Portegies Zwart} S.,  2011, Science, 334, 1380

\bibitem[\protect\citeauthoryear{{Ginsburg}, {Bally}, {Goddi}, {Plambeck} \&
  {Wright}}{{Ginsburg} et~al.}{2018}]{2018arXiv180410622G}
{Ginsburg} A.,  {Bally} J.,  {Goddi} C.,  {Plambeck} R.,    {Wright} M.,  2018,
  \apj, 860, 119

\bibitem[\protect\citeauthoryear{{Greif}, {Bromm}, {Clark}, {Glover}, {Smith},
  {Klessen}, {Yoshida} \& {Springel}}{{Greif}
  et~al.}{2012}]{greif_mnras_424_2012}
{Greif} T.~H.,  {Bromm} V.,  {Clark} P.~C.,  {Glover} S.~C.~O.,  {Smith} R.~J.,
   {Klessen} R.~S.,  {Yoshida} N.,    {Springel} V.,  2012, \mnras, 424, 399

\bibitem[\protect\citeauthoryear{{Haemmerl{\'e}}}{{Haemmerl{\'e}}}{2014}]{haemmerle_phd_2014}
{Haemmerl{\'e}} L.,  2014, PhD thesis, University of Geneva

\bibitem[\protect\citeauthoryear{{Haemmerl{\'e}}, {Eggenberger}, {Meynet},
  {Maeder} \& {Charbonnel}}{{Haemmerl{\'e}}
  et~al.}{2016}]{haemmerle_585_aa_2016}
{Haemmerl{\'e}} L.,  {Eggenberger} P.,  {Meynet} G.,  {Maeder} A.,
  {Charbonnel} C.,  2016, \aap, 585, A65

\bibitem[\protect\citeauthoryear{{Haemmerl{\'e}} \& {Peters}}{{Haemmerl{\'e}}
  \& {Peters}}{2016}]{haemmerle_458_mnras_2016}
{Haemmerl{\'e}} L.,  {Peters} T.,  2016, \mnras, 458, 3299

\bibitem[\protect\citeauthoryear{{Harries}}{{Harries}}{2015}]{harries_mnras_448_2015}
{Harries} T.~J.,  2015, \mnras, 448, 3156

\bibitem[\protect\citeauthoryear{{Harries}, {Douglas} \& {Ali}}{{Harries}
  et~al.}{2017}]{harries_2017}
{Harries} T.~J.,  {Douglas} T.~A.,    {Ali} A.,  2017, \mnras, 471, 4111

\bibitem[\protect\citeauthoryear{{Hennebelle}, {Commer{\c c}on}, {Chabrier} \&
  {Marchand}}{{Hennebelle} et~al.}{2016}]{hennebelle_apj_830_2016}
{Hennebelle} P.,  {Commer{\c c}on} B.,  {Chabrier} G.,    {Marchand} P.,  2016,
  \apjl, 830, L8

\bibitem[\protect\citeauthoryear{{Hosokawa} \& {Omukai}}{{Hosokawa} \&
  {Omukai}}{2009}]{hosokawa_apj_691_2009}
{Hosokawa} T.,  {Omukai} K.,  2009, \apj, 691, 823

\bibitem[\protect\citeauthoryear{{Hosokawa}, {Yorke} \& {Omukai}}{{Hosokawa}
  et~al.}{2010}]{hosokawa_apj_721_2010}
{Hosokawa} T.,  {Yorke} H.~W.,    {Omukai} K.,  2010, \apj, 721, 478

\bibitem[\protect\citeauthoryear{{Hunter}, {Brogan}, {MacLeod}, {Cyganowski},
  {Chandler}, {Chibueze}, {Friesen}, {Indebetouw}, {Thesner} \&
  {Young}}{{Hunter} et~al.}{2017}]{hunter_apj_837_2017}
{Hunter} T.~R.,  {Brogan} C.~L.,  {MacLeod} G.,  {Cyganowski} C.~J.,
  {Chandler} C.~J.,  {Chibueze} J.~O.,  {Friesen} R.,  {Indebetouw} R.,
  {Thesner} C.,    {Young} K.~H.,  2017, \apjl, 837, L29

\bibitem[\protect\citeauthoryear{{Ilee}, {Cyganowski}, {Brogan}, {Hunter},
  {Forgan}, {Haworth}, {Clarke} \& {Harries}}{{Ilee}
  et~al.}{2018}]{2018ApJ...869L..24I}
{Ilee} J.~D.,  {Cyganowski} C.~J.,  {Brogan} C.~L.,  {Hunter} T.~R.,  {Forgan}
  D.~H.,  {Haworth} T.~J.,  {Clarke} C.~J.,    {Harries} T.~J.,  2018, \apjl,
  869, L24

\bibitem[\protect\citeauthoryear{{Ilee}, {Cyganowski}, {Nazari}, {Hunter},
  {Brogan}, {Forgan} \& {Zhang}}{{Ilee} et~al.}{2016}]{ilee_mnras_462_2016}
{Ilee} J.~D.,  {Cyganowski} C.~J.,  {Nazari} P.,  {Hunter} T.~R.,  {Brogan}
  C.~L.,  {Forgan} D.~H.,    {Zhang} Q.,  2016, \mnras, 462, 4386

\bibitem[\protect\citeauthoryear{{Jankovic}, {Haworth}, {Ilee}, {Forgan},
  {Cyganowski}, {Walsh}, {Brogan}, {Hunter} \& {Mohanty}}{{Jankovic}
  et~al.}{2019}]{jankovic_mnras_482_2019}
{Jankovic} M.~R.,  {Haworth} T.~J.,  {Ilee} J.~D.,  {Forgan} D.~H.,
  {Cyganowski} C.~J.,  {Walsh} C.,  {Brogan} C.~L.,  {Hunter} T.~R.,
  {Mohanty} S.,  2019, \mnras, 482, 4673

\bibitem[\protect\citeauthoryear{{Johnston}, {Robitaille}, {Beuther}, {Linz},
  {Boley}, {Kuiper}, {Keto}, {Hoare} \& {van Boekel}}{{Johnston}
  et~al.}{2015}]{johnston_apj_813_2015}
{Johnston} K.~G.,  {Robitaille} T.~P.,  {Beuther} H.,  {Linz} H.,  {Boley} P.,
  {Kuiper} R.,  {Keto} E.,  {Hoare} M.~G.,    {van Boekel} R.,  2015, \apjl,
  813, L19

\bibitem[\protect\citeauthoryear{{Keto} \& {Wood}}{{Keto} \&
  {Wood}}{2006}]{keto_apj_637_2006}
{Keto} E.,  {Wood} K.,  2006, \apj, 637, 850

\bibitem[\protect\citeauthoryear{{Klassen}, {Pudritz}, {Kuiper}, {Peters} \&
  {Banerjee}}{{Klassen} et~al.}{2016}]{klassen_apj_823_2016}
{Klassen} M.,  {Pudritz} R.~E.,  {Kuiper} R.,  {Peters} T.,    {Banerjee} R.,
  2016, \apj, 823, 28

\bibitem[\protect\citeauthoryear{{Kobulnicky}, {Kiminki}, {Lundquist}, {Burke},
  {Chapman}, {Keller}, {Lester}, {Rolen}, {Topel}, {Bhattacharjee}, {Smullen},
  {Vargas {\'A}lvarez}, {Runnoe}, {Dale} \& {Brotherton}}{{Kobulnicky}
  et~al.}{2014}]{2014ApJS..213...34K}
{Kobulnicky} H.~A.,  {Kiminki} D.~C.,  {Lundquist} M.~J.,  {Burke} J.,
  {Chapman} J.,  {Keller} E.,  {Lester} K.,  {Rolen} E.~K.,  {Topel} E.,
  {Bhattacharjee} A.,  {Smullen} R.~A.,  {Vargas {\'A}lvarez} C.~A.,  {Runnoe}
  J.~C.,  {Dale} D.~A.,    {Brotherton} M.~M.,  2014, \apjs, 213, 34

\bibitem[\protect\citeauthoryear{{Kolb}, {Stute}, {Kley} \& {Mignone}}{{Kolb}
  et~al.}{2013}]{kolb_aa_559_2013}
{Kolb} S.~M.,  {Stute} M.,  {Kley} W.,    {Mignone} A.,  2013, \aap, 559, A80

\bibitem[\protect\citeauthoryear{{Kraus}, {Kluska}, {Kreplin}, {Bate},
  {Harries}, {Hofmann}, {Hone}, {Monnier}, {Weigelt}, {Anugu}, {de Wit} \&
  {Wittkowski}}{{Kraus} et~al.}{2017}]{kraus_apj_835_2017}
{Kraus} S.,  {Kluska} J.,  {Kreplin} A.,  {Bate} M.,  {Harries} T.~J.,
  {Hofmann} K.-H.,  {Hone} E.,  {Monnier} J.~D.,  {Weigelt} G.,  {Anugu} A.,
  {de Wit} W.~J.,    {Wittkowski} M.,  2017, \apjl, 835, L5

\bibitem[\protect\citeauthoryear{{Krumholz}, {Klein} \& {McKee}}{{Krumholz}
  et~al.}{2007}]{krumholz_apj_665_2007}
{Krumholz} M.~R.,  {Klein} R.~I.,    {McKee} C.~F.,  2007, \apj, 665, 478

\bibitem[\protect\citeauthoryear{{Kuffmeier}, {Frimann}, {Jensen} \&
  {Haugb{\o}lle}}{{Kuffmeier} et~al.}{2018}]{kuffmeier_mnras_4754_2018}
{Kuffmeier} M.,  {Frimann} S.,  {Jensen} S.~S.,    {Haugb{\o}lle} T.,  2018,
  \mnras, 475, 2642

\bibitem[\protect\citeauthoryear{{Langer}}{{Langer}}{2012}]{langer_araa_50_2012}
{Langer} N.,  2012, \araa, 50, 107

\bibitem[\protect\citeauthoryear{{Laor} \& {Draine}}{{Laor} \&
  {Draine}}{1993}]{laor_apj_402_1993}
{Laor} A.,  {Draine} B.~T.,  1993, \apj, 402, 441

\bibitem[\protect\citeauthoryear{{Lumsden}, {Hoare}, {Urquhart}, {Oudmaijer},
  {Davies}, {Mottram}, {Cooper} \& {Moore}}{{Lumsden}
  et~al.}{2013}]{lumsden_apjs_208_2013}
{Lumsden} S.~L.,  {Hoare} M.~G.,  {Urquhart} J.~S.,  {Oudmaijer} R.~D.,
  {Davies} B.,  {Mottram} J.~C.,  {Cooper} H.~D.~B.,    {Moore} T.~J.~T.,
  2013, \apjs, 208, 11

\bibitem[\protect\citeauthoryear{{MacLeod}, {Smits}, {Goedhart}, {Hunter},
  {Brogan}, {Chibueze}, {van den Heever}, {Thesner}, {Banda} \&
  {Paulsen}}{{MacLeod} et~al.}{2018}]{macloed_mnras_478_2018}
{MacLeod} G.~C.,  {Smits} D.~P.,  {Goedhart} S.,  {Hunter} T.~R.,  {Brogan}
  C.~L.,  {Chibueze} J.~O.,  {van den Heever} S.~P.,  {Thesner} C.~J.,  {Banda}
  P.~J.,    {Paulsen} J.~D.,  2018, \mnras, 478, 1077

\bibitem[\protect\citeauthoryear{{Mahy}, {Rauw}, {De Becker}, {Eenens} \&
  {Flores}}{{Mahy} et~al.}{2013}]{2013A&A...550A..27M}
{Mahy} L.,  {Rauw} G.,  {De Becker} M.,  {Eenens} P.,    {Flores} C.~A.,  2013,
  \aap, 550, A27

\bibitem[\protect\citeauthoryear{{Maud}, {Cesaroni}, {Kumar}, {van der Tak},
  {Allen}, {Hoare}, {Klaassen}, {Harsono} \& {Hogerheijde}}{{Maud}
  et~al.}{2018}]{maud_aa_620_2018}
{Maud} L.~T.,  {Cesaroni} R.,  {Kumar} M.~S.~N.,  {van der Tak} F.~F.~S.,
  {Allen} V.,  {Hoare} M.~G.,  {Klaassen} P.~D.,  {Harsono} D.,
  {Hogerheijde} M.~R. e.~a.,  2018, \aap, 620, A31

\bibitem[\protect\citeauthoryear{{Maud}, {Hoare}, {Galv{\'a}n-Madrid}, {Zhang},
  {de Wit}, {Keto}, {Johnston} \& {Pineda}}{{Maud}
  et~al.}{2017}]{maud_467_mnras_2017}
{Maud} L.~T.,  {Hoare} M.~G.,  {Galv{\'a}n-Madrid} R.,  {Zhang} Q.,  {de Wit}
  W.~J.,  {Keto} E.,  {Johnston} K.~G.,    {Pineda} J.~E.,  2017, \mnras, 467,
  L120

\bibitem[\protect\citeauthoryear{{McMullin}, {Waters}, {Schiebel}, {Young} \&
  {Golap}}{{McMullin} et~al.}{2007}]{McMullin_aspc_376_2007}
{McMullin} J.~P.,  {Waters} B.,  {Schiebel} D.,  {Young} W.,    {Golap} K.,
  2007, in {Shaw} R.~A.,  {Hill} F.,   {Bell} D.~J.,  eds, Astronomical Data
  Analysis Software and Systems XVI Vol.~376 of Astronomical Society of the
  Pacific Conference Series, {CASA Architecture and Applications}.
p.~127

\bibitem[\protect\citeauthoryear{{Meyer}, {Haemmerl{\'e}} \&
  {Vorobyov}}{{Meyer} et~al.}{2019}]{2019MNRAS.tmp...10M}
{Meyer} D.~M.-A.,  {Haemmerl{\'e}} L.,    {Vorobyov} E.~I.,  2019, \mnras, 484,
  2482

\bibitem[\protect\citeauthoryear{{Meyer}, {Kuiper}, {Kley}, {Johnston} \&
  {Vorobyov}}{{Meyer} et~al.}{2018}]{meyer_mnras_473_2018}
{Meyer} D.~M.-A.,  {Kuiper} R.,  {Kley} W.,  {Johnston} K.~G.,    {Vorobyov}
  E.,  2018, \mnras, 473, 3615

\bibitem[\protect\citeauthoryear{{Meyer}, {van Marle}, {Kuiper} \&
  {Kley}}{{Meyer} et~al.}{2016}]{meyer_459_mnras_2016}
{Meyer} D.~M.-A.,  {van Marle} A.-J.,  {Kuiper} R.,    {Kley} W.,  2016,
  \mnras, 459, 1146

\bibitem[\protect\citeauthoryear{{Meyer}, {Vorobyov}, {Elbakyan}, {Stecklum},
  {Eisl{\"o}ffel} \& {Sobolev}}{{Meyer} et~al.}{2019}]{meyer_mnras_482_2019}
{Meyer} D.~M.-A.,  {Vorobyov} E.~I.,  {Elbakyan} V.~G.,  {Stecklum} B.,
  {Eisl{\"o}ffel} J.,    {Sobolev} A.~M.,  2019, \mnras, 482, 5459

\bibitem[\protect\citeauthoryear{{Meyer}, {Vorobyov}, {Kuiper} \&
  {Kley}}{{Meyer} et~al.}{2017}]{meyer_mnras_464_2017}
{Meyer} D.~M.-A.,  {Vorobyov} E.~I.,  {Kuiper} R.,    {Kley} W.,  2017, \mnras,
  464, L90

\bibitem[\protect\citeauthoryear{{Mignone}, {Bodo}, {Massaglia}, {Matsakos},
  {Tesileanu}, {Zanni} \& {Ferrari}}{{Mignone}
  et~al.}{2007}]{mignone_apj_170_2007}
{Mignone} A.,  {Bodo} G.,  {Massaglia} S.,  {Matsakos} T.,  {Tesileanu} O.,
  {Zanni} C.,    {Ferrari} A.,  2007, \apjs, 170, 228

\bibitem[\protect\citeauthoryear{{Mignone}, {Zanni}, {Tzeferacos}, {van
  Straalen}, {Colella} \& {Bodo}}{{Mignone}
  et~al.}{2012}]{migmone_apjs_198_2012}
{Mignone} A.,  {Zanni} C.,  {Tzeferacos} P.,  {van Straalen} B.,  {Colella} P.,
     {Bodo} G.,  2012, \apjs, 198, 7

\bibitem[\protect\citeauthoryear{{Norberg} \& {Maeder}}{{Norberg} \&
  {Maeder}}{2000}]{norberg_aa_159_2000}
{Norberg} P.,  {Maeder} A.,  2000, \aap, 359, 1025

\bibitem[\protect\citeauthoryear{{Palla} \& {Stahler}}{{Palla} \&
  {Stahler}}{1992}]{palla_apj_392_1992}
{Palla} F.,  {Stahler} S.~W.,  1992, \apj, 392, 667

\bibitem[\protect\citeauthoryear{{Purser}, {Lumsden}, {Hoare} \&
  {Cunningham}}{{Purser} et~al.}{2018}]{purser_mnras_475_2018}
{Purser} S.~J.~D.,  {Lumsden} S.~L.,  {Hoare} M.~G.,    {Cunningham} N.,  2018,
  \mnras, 475, 2

\bibitem[\protect\citeauthoryear{{Purser}, {Lumsden}, {Hoare}, {Urquhart},
  {Cunningham}, {Purcell}, {Brooks}, {Garay}, {G{\'u}zman} \&
  {Voronkov}}{{Purser} et~al.}{2016}]{purser_mnras_460_2016}
{Purser} S.~J.~D.,  {Lumsden} S.~L.,  {Hoare} M.~G.,  {Urquhart} J.~S.,
  {Cunningham} N.,  {Purcell} C.~R.,  {Brooks} K.~J.,  {Garay} G.,
  {G{\'u}zman} A.~E.,    {Voronkov} M.~A.,  2016, \mnras, 460, 1039

\bibitem[\protect\citeauthoryear{{Reg{\'a}ly} \& {Vorobyov}}{{Reg{\'a}ly} \&
  {Vorobyov}}{2017}]{regaly_aa_601_2017}
{Reg{\'a}ly} Z.,  {Vorobyov} E.,  2017, \aap, 601, A24

\bibitem[\protect\citeauthoryear{{Reid}, {Menten}, {Zheng}, {Brunthaler},
  {Moscadelli}, {Xu}, {Zhang}, {Sato}, {Honma}, {Hirota}, {Hachisuka}, {Choi},
  {Moellenbrock} \& {Bartkiewicz}}{{Reid} et~al.}{2009}]{reid_apj_700_2009}
{Reid} M.~J.,  {Menten} K.~M.,  {Zheng} X.~W.,  {Brunthaler} A.,  {Moscadelli}
  L.,  {Xu} Y.,  {Zhang} B.,  {Sato} M.,  {Honma} M.,  {Hirota} T.,
  {Hachisuka} K.,  {Choi} Y.~K.,  {Moellenbrock} G.~A.,    {Bartkiewicz} A.,
  2009, \apj, 700, 137

\bibitem[\protect\citeauthoryear{{Reiter}, {Kiminki}, {Smith} \&
  {Bally}}{{Reiter} et~al.}{2017}]{reiter_mnras_470_2017}
{Reiter} M.,  {Kiminki} M.~M.,  {Smith} N.,    {Bally} J.,  2017, \mnras, 470,
  4671

\bibitem[\protect\citeauthoryear{{Samal}, {Chen}, {Takami}, {Jose} \&
  {Froebrich}}{{Samal} et~al.}{2018}]{samal_mnras_477_2018}
{Samal} M.~R.,  {Chen} W.~P.,  {Takami} M.,  {Jose} J.,    {Froebrich} D.,
  2018, \mnras, 477, 4577

\bibitem[\protect\citeauthoryear{{Sana}, {de Mink}, {de Koter}, {Langer},
  {Evans}, {Gieles}, {Gosset}, {Izzard}, {Le Bouquin} \& {Schneider}}{{Sana}
  et~al.}{2012}]{sana_sci_337_2012}
{Sana} H.,  {de Mink} S.~E.,  {de Koter} A.,  {Langer} N.,  {Evans} C.~J.,
  {Gieles} M.,  {Gosset} E.,  {Izzard} R.~G.,  {Le Bouquin} J.-B.,
  {Schneider} F.~R.~N.,  2012, Science, 337, 444

\bibitem[\protect\citeauthoryear{{Sanna}, {Koelligan}, {Moscadelli}, {Kuiper},
  {Cesaroni}, {Pillai}, {Menten}, {Zhang}, {Garatti}, {Goddi}, {Leurini} \&
  {Carrasco-Gonzalez}}{{Sanna} et~al.}{2018}]{2018arXiv180509842S}
{Sanna} A.,  {Koelligan} A.,  {Moscadelli} L.,  {Kuiper} R.,  {Cesaroni} R.,
  {Pillai} T.,  {Menten} K.~M.,  {Zhang} Q.,  {Garatti} A.~C.~o.,  {Goddi} C.,
  {Leurini} S.,    {Carrasco-Gonzalez} C.,  2018, arXiv e-prints

\bibitem[\protect\citeauthoryear{{Seifried}, {Banerjee}, {Klessen}, {Duffin} \&
  {Pudritz}}{{Seifried} et~al.}{2011}]{seifried_mnras_417_2011}
{Seifried} D.,  {Banerjee} R.,  {Klessen} R.~S.,  {Duffin} D.,    {Pudritz}
  R.~E.,  2011, \mnras, 417, 1054

\bibitem[\protect\citeauthoryear{{Seifried}, {Pudritz}, {Banerjee}, {Duffin} \&
  {Klessen}}{{Seifried} et~al.}{2012}]{seifried_mnras_422_2012}
{Seifried} D.,  {Pudritz} R.~E.,  {Banerjee} R.,  {Duffin} D.,    {Klessen}
  R.~S.,  2012, \mnras, 422, 347

\bibitem[\protect\citeauthoryear{{Seifried}, {S{\'a}nchez-Monge}, {Walch} \&
  {Banerjee}}{{Seifried} et~al.}{2016}]{seifried_mnras_571_2016}
{Seifried} D.,  {S{\'a}nchez-Monge} {\'A}.,  {Walch} S.,    {Banerjee} R.,
  2016, \mnras

\bibitem[\protect\citeauthoryear{{Stecklum}, {Heese}, {Wolf}, {Garatti},
  {Ibanez} \& {Linz}}{{Stecklum} et~al.}{2017}]{stecklum_2017a}
{Stecklum} B.,  {Heese} S.,  {Wolf} S.,  {Garatti} A.~C.~o.,  {Ibanez} J.~M.,
   {Linz} H.,  2017, ArXiv e-prints

\bibitem[\protect\citeauthoryear{{Tapia}, {Roth} \& {Persi}}{{Tapia}
  et~al.}{2015}]{tapia_mnras_446_2015}
{Tapia} M.,  {Roth} M.,    {Persi} P.,  2015, \mnras, 446, 4088

\bibitem[\protect\citeauthoryear{{Vaidya}, {Fendt}, {Beuther} \&
  {Porth}}{{Vaidya} et~al.}{2011}]{vaidya_apj_742_2011}
{Vaidya} B.,  {Fendt} C.,  {Beuther} H.,    {Porth} O.,  2011, \apj, 742, 56

\bibitem[\protect\citeauthoryear{{Vink}, {de Koter} \& {Lamers}}{{Vink}
  et~al.}{2000}]{vink_aa_362_2000}
{Vink} J.~S.,  {de Koter} A.,    {Lamers} H.~J.~G.~L.~M.,  2000, \aap, 362, 295

\bibitem[\protect\citeauthoryear{{Vorobyov} \& {Basu}}{{Vorobyov} \&
  {Basu}}{2010}]{vorobyov_apj_719_2010}
{Vorobyov} E.~I.,  {Basu} S.,  2010, \apj, 719, 1896

\bibitem[\protect\citeauthoryear{{Vorobyov}, {DeSouza} \& {Basu}}{{Vorobyov}
  et~al.}{2013}]{vorobyov_apj_768_2013}
{Vorobyov} E.~I.,  {DeSouza} A.~L.,    {Basu} S.,  2013, \apj, 768, 131

\bibitem[\protect\citeauthoryear{{Vorobyov}, {Zakhozhay} \&
  {Dunham}}{{Vorobyov} et~al.}{2013}]{vorobyov_mnras_433_2013}
{Vorobyov} E.~I.,  {Zakhozhay} O.~V.,    {Dunham} M.~M.,  2013, \mnras, 433,
  3256

\bibitem[\protect\citeauthoryear{{Wootten} \& {Thompson}}{{Wootten} \&
  {Thompson}}{2009}]{wootten_IEEEP_2009}
{Wootten} A.,  {Thompson} A.~R.,  2009, IEEE Proceedings, 97, 1463

\bibitem[\protect\citeauthoryear{{Zinchenko}, {Liu}, {Su}, {Salii}, {Sobolev},
  {Zemlyanukha}, {Beuther}, {Ojha}, {Samal} \& {Wang}}{{Zinchenko}
  et~al.}{2015}]{zinchenko_apj_810_2015}
{Zinchenko} I.,  {Liu} S.-Y.,  {Su} Y.-N.,  {Salii} S.~V.,  {Sobolev} A.~M.,
  {Zemlyanukha} P.,  {Beuther} H.,  {Ojha} D.~K.,  {Samal} M.~R.,    {Wang} Y.,
   2015, \apj, 810, 10

\bibitem[\protect\citeauthoryear{{Zinnecker} \& {Yorke}}{{Zinnecker} \&
  {Yorke}}{2007}]{zinnecker_araa_45_2007}
{Zinnecker} H.,  {Yorke} H.~W.,  2007, \araa, 45, 481

\end{thebibliography}
}

%%%%%%%%%%%%%%%%%%%%%%%%%%%%%%%%%%%%%%%%%%%%%%%%%%%%%%%%%%%%%%%%%%%%%%%%%%%%%%%%%%%%%%%%%%%
%%%%%%%%%%%%%%%%%%%%%%%%%%%%%%%%%%%%%%%%%%%%%%%%%%%%%%%%%%%%%%%%%%%%%%%%%%%%%%%%%%%%%%%%%%%
%%%%%%%%%%%%%%%%%%%%%%%%%%%%%%%%%%%%%%%%%%%%%%%%%%%%%%%%%%%%%%%%%%%%%%%%%%%%%%%%%%%%%%%%%%%

\end{document}